\documentclass[prl,showkeys,showpacs,superscriptaddress,nofootinbib,nobibnotes,twocolumn]{revtex4-2}

\usepackage[T1]{fontenc}
\usepackage[utf8]{inputenc}

\usepackage[caption=false]{subfig}
\usepackage{graphicx}
\usepackage{epstopdf}
\usepackage{array}
\usepackage{verbatim}
\usepackage{amsmath,amsfonts,amssymb,amscd,mathtools}
\usepackage{dsfont}
\usepackage{physics}
\usepackage{tabularx}
\usepackage{stmaryrd}
\usepackage{enumerate}
\usepackage[utf8]{inputenc}

 \usepackage[colorlinks=true,        
            allcolors = black,  
            citecolor=DarkBlue, 
            linkcolor=DarkBlue,
            urlcolor=DarkBlue
        ]{hyperref}

\usepackage{wasysym}
\usepackage{braket}
\usepackage{mathrsfs}
\usepackage{microtype}
\usepackage{hyperref}
\usepackage[dvipsnames]{xcolor}

\hypersetup{
    bookmarksnumbered=true, 
    unicode=false, 
    pdfstartview={FitH}, 
    pdftitle={}, 
    pdfauthor={}, 
    pdfsubject={}, 
    pdfcreator={}, 
    pdfproducer={}, 
    pdfkeywords={}, 
    pdfnewwindow=true, 
    colorlinks=true, 
    linkcolor=NavyBlue, 
    citecolor=NavyBlue, 
    filecolor=NavyBlue, 
    urlcolor=NavyBlue 
}
\usepackage{diagbox}
\usepackage{tikz}
\usepackage{newtxtext,newtxmath}
\usetikzlibrary{decorations.pathreplacing}

\usepackage{algorithm}
\usepackage{algorithmicx}
\usepackage{algpseudocode}
\usepackage{amsthm}



\usepackage{quantikz}
\usepackage{physics}
\usepackage{amssymb}
\usepackage{amsthm}
\usepackage{bm}
\usepackage{rotating} 
\usepackage{atbegshi} 

\usepackage{hyperref}
\hypersetup{%
	unicode=true,
	breaklinks=true,
	colorlinks,
	urlcolor={black},
	pdfpagemode=UseOutlines,
	pdfborder={0 0 0}
}

\usepackage{comment}

\usepackage{adjustbox} 

\usepackage{pgfplots}
\pgfplotsset{compat=1.18}
\usepackage{tikz}
\usetikzlibrary{quantikz2}

\newcommand{\free}[0]{\text{free}}
\newcommand{\clamped}[0]{\text{clamped}}

\newcommand{\final}[0]{\text{final}}

\newcommand{\prop}[0]{\text{prop}}

\newtheorem{theorem}{Theorem}
\newtheorem{lemma}[theorem]{Lemma}

\begin{document}

\title{Equilibrium Learning of Quantum Circuits}

\author{Frank Barrows$^{\S}$}
\affiliation{Theoretical Division, Los Alamos National Laboratory, Los Alamos, NM USA}

\author{Touheed Atif$^{\S}$}
\affiliation{Computing and Artificial Intelligence Division, Los Alamos National Laboratory, Los Alamos, NM, USA}

\author{Yi\u{g}it Suba\c{s}\i}
\affiliation{Computing and Artificial Intelligence Division, Los Alamos National Laboratory, Los Alamos, NM, USA}

\author{Tyler Volkoff}
\affiliation{Theoretical Division, Los Alamos National Laboratory, Los Alamos, NM USA}

\begin{abstract}
Equilibrium learning trains a physical system using observables of its relaxed states. We extend this principle to gate-based quantum circuits by embedding a parametrized circuit in a Feynman-Kitaev (FK) Hamiltonian and imposing the compilation target either through an output clamping term or through a target-dependent cyclic condition on the clock register. The ground state energies and the overlaps of the corresponding ground states with the free history state define faithful cost functions for unitary and state compilation. We show that these costs are nonlinear functions of the  compilation infidelity and therefore reshape the optimization landscape. We show that starting from the free history state, quantum phase estimation can prepare the relevant ground states and estimate the cost functions with polynomial resources. In two-qubit benchmarks, optimization with the equilibrium cost functions reaches a fixed fidelity threshold in fewer sweeps than standard compilation infidelity cost function for many instances. The results demonstrate the potential for encoding quantum learning problems as computation of equilibrium observables.
\end{abstract}

\maketitle

\begingroup
\renewcommand{\thefootnote}{*}
\footnotetext{
fbarrows@lanl.gov; tatif@lanl.gov; ysubasi@lanl.gov; volkoff@lanl.gov
}
\endgroup
\begingroup
\renewcommand{\thefootnote}{\S}
\footnotetext{These authors contributed equally to this work.}
\endgroup


\textit{\textbf{Introduction and motivation.}} Classical electrical networks provide a familiar setting in which computation is organized around an equilibrium boundary-value problem. Kirchhoff's law, Ohm's law, and source/load models determine the network state as the stationary point of a power functional \cite{dorfler2018electrical}. In trained resistor networks, the edge conductances are adjusted so that the equilibrium response to prescribed boundary conditions---for example, output voltages generated by input voltage bias---approaches a desired input-output relation. Because the network physically relaxes to its equilibrium state, evaluation of this response is performed directly by the physical system.

Methods such as equilibrium propagation and coupled learning perform this training by comparing a free equilibrium with a second equilibrium in which the output is weakly nudged or clamped toward its desired value \cite{scellier2017equilibrium,kendall2020training,dillavou2024machine}. The resulting changes in local observables, such as dissipated powers at each edge, provide local gradient information for updating the trainable conductances \cite{lin2026train}. 
Thus, physical relaxation determines the free and clamped network states,
while the contrast between local observables measured at these equilibria
provides the gradient information used to update the trainable parameters $\theta$ by descending the energy,
\begin{equation}
\dot{\theta}\propto-\partial_\theta E .
\label{eq:EnergyUpdate}
\end{equation}
Thus, encoding a target cost function as an energy provides a natural way by which physical systems evolve and minimize loss.

Quantum equilibrium propagation has extended this paradigm to tunable quantum Hamiltonians, including in quantum annealers~\cite{Laydevant_NatComm_2024,Zhang_CommPhys_2025}, and through the use of Onsager reciprocity to extract training gradients from linear-response probes
\cite{doi_10_1038_s41467_025_61665_6}. 
That framework concerns the equilibrium training of analog quantum systems rather than compilation in the gate-based circuit model, leaving open whether an analogous equilibrium formulation can be constructed for gate-based quantum compiling.


Quantum compiling refers to algorithms aimed at efficiently implementing a quantum operation of interest on a quantum computer. 
A common framework seeks to approximate a target unitary $V$ up to a global phase $\varphi \in [0,2\pi)$ by a
parametrized circuit $U(\theta):=U_{T}(\theta_{T})\cdots U_{1}(\theta_{1})$ subject to architectural and gate-alphabet constraints:
\begin{equation}
    \operatorname*{arg\,min}_{\theta} \min_\varphi\Vert U(\theta) - e^{i \varphi}V\Vert \, .
    \label{eq:compiling_goal}
\end{equation}

Compiling an arbitrary $n$-qubit unitary is a difficult global optimization problem. This is reflected in the QMA-completeness of closely related circuit-verification problems such as \textsc{Non-Identity Check} \cite{idcheck}, and cautions against expecting a generic compiling procedure to be efficient in the worst case. 
Quantum-assisted quantum compiling (QAQC) uses Hilbert-Schmidt and local Hilbert-Schmidt tests to compare a target unitary with a trainable unitary on a quantum processor \cite{khatri2019quantum}.  Related work has studied the noise resilience of such variational compiling objectives~\cite{Sharma2020noise}, local or subsystem compiling costs~ \cite{mituza2022local}, and reinforcement-learning methods for searching over gate sequences or circuit structures~\cite{Moro2021quantum}.  At the same time, the trainability of parametrized quantum circuits is strongly cost-function dependent, and global objectives can exhibit barren plateaus even for shallow ans{\"a}tze~\cite{Cerezo2021}.

Our aim in this work is not to circumvent all difficulties associated with quantum compiling, but to formulate the compiling problem in an equilibrium language. To this end, we use the circuit-to-Hamiltonian construction~\cite{kitaevbook}, which maps a gate sequence to a local Hamiltonian on clock and system registers: propagation terms penalize inconsistencies between successive circuit steps, while boundary terms penalize disagreement with the prescribed input and target states. 
From the ground state energies and overlaps, we derive faithful cost functions that are nonlinear functions of the ordinary compilation infidelity. These cost functions change the optimization landscape without changing the compilation target.
Our main results consist of gradient-alignment theorems that indicate the shape of the cost function relative to the compilation infidelity, depth-independent lower bounds on the overlaps between the ground states and the free history state, and inverse-quadratic scaling of the spectral gaps with circuit depth. These results allow the cost functions to be estimated with polynomial quantum resources. In idealized, noiseless two-qubit benchmarks, optimization using these cost functions reaches a fixed fidelity threshold in fewer sweeps than direct QAQC for many instances, although this comparison does not account for the quantum resources required per sweep.

\begin{figure}[t]
\centering
\includegraphics[scale=.39]{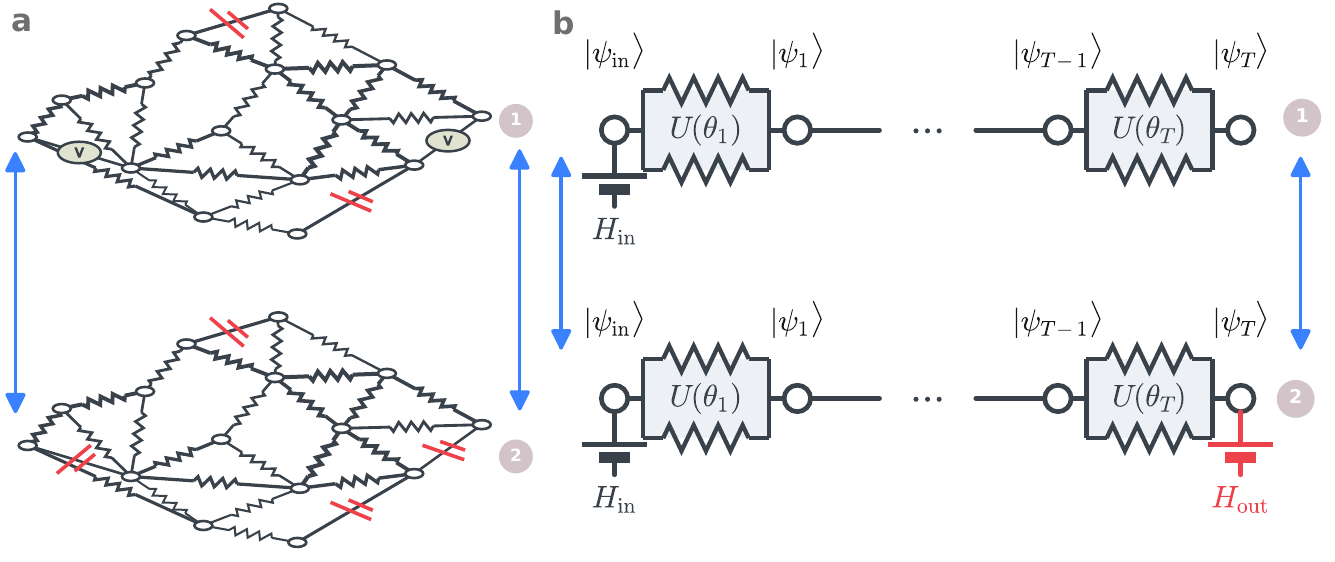}
\caption{Schematic of equilibrium learning in resistor networks and FK clock networks. (a)~A resistor network relaxes to a free equilibrium under input boundary conditions (1); clamping the outputs toward their targets defines a second equilibrium (2), and the contrast between the two provides local gradient information for the trainable conductances. (b)~FK analogue: the clock network of a parametrized circuit with input term $H_{\rm in}$ has the history state as its free equilibrium (1); adding the output clamp $H_{\rm out}$ defines the clamped equilibrium (2) used for training.}
\label{fig:schematic}
\end{figure}

\textbf{\textit{From resistor networks to FK Hamiltonians.}} The common mathematical object behind the electrical and quantum constructions is a Laplacian energy. In a resistor network with node voltages $v_a$ and edge conductances $g_e>0$, with input source nodes held at fixed voltages $v_i=x_i$, the free node voltages are determined by minimizing the dissipated-power functional
\begin{equation}
    D_{\rm res}(v)
    =
    \frac{1}{2}\sum_{e:a\to b} g_e\,|v_a-v_b|^2 ,
    \label{eq:resistor_dirichlet}
\end{equation}
whose stationary point over those voltages is Kirchhoff's law,~\cite{dorfler2018electrical}. 
Expressed in the node potentials this relaxation is an unconstrained minimization of $D_{\rm res}$, whereas the dual description in edge currents is bound by current conservation~\cite{Meixner_JMathPhys_1963,Meixner_1973,BraytonMoser1964}. 
Training such a system adds a clamp that biases the outputs toward targets $y_o$,
\begin{equation}
    D_\beta(v)
    =
    D_{\rm res}(v)
    +
    \frac{\beta}{2}\sum_{o\in\mathcal O}(v_o-y_o)^2 ,
    \label{eq:resistor_clamped_energy}
\end{equation}
so that $\beta$ interpolates between the free equilibrium ($\beta=0$) and the clamped one. The two equilibria are stationary states of the same energy function at different clamp strengths, and their difference is exactly the local gradient information of Eq.~\eqref{eq:EnergyUpdate}: descending $D_\beta$ in the conductances lowers the dissipation consistent with the target, realizing the desired input-output relation  (Fig.~\ref{fig:schematic}a)~\cite{scellier2017equilibrium,kendall2020training,dillavou2024machine,lin2026train}.

Now consider a quantum circuit in which an initial state $\ket{\psi_\text{in}}$ is acted on by $T$ unitaries $U_1,\dots,U_T$. The FK Hamiltonian associated with this circuit acts on the composite Hilbert space of a clock register and a system register, $\mathcal{H}_\text{C}\otimes\mathcal{H}_\text{S}$, and is given by $H_\text{free}
    = J_\text{in}H_\text{in} + J_\text{prop}H_\text{prop}$, $J_\text{in},J_\text{prop}>0$ where
\begin{align}
    H_\text{in} &= \ketbra{0}{0} \otimes \left(\mathbb{I} -\ketbra{\psi_\text{in}}{\psi_\text{in}} \right)\, , \\
    H_\text{prop} &= \frac{1}{2} \sum_{t=1}^{T}\left(\vphantom{U^{\dagger}}\ket{t}\otimes U_t-\ket{t-1}\otimes\mathbb{I} \right)\left(\bra{t}\otimes U_t^\dagger-\bra{t-1}\otimes\mathbb{I}\right).
\end{align}
For a state written in the clock basis as $\ket{\Psi}=\sum_{t=0}^{T}\ket{t}\otimes\ket{\phi_t}$, the gate $U_t(\theta_t)$ transports the amplitude at clock time $t-1$. 
Thus the classical scalar voltage drop $v_b-v_a$ is replaced by the difference between neighboring clock states after parallel transport 
\begin{equation}
     \delta_U\phi_t:=\phi_t-U_t(\theta_t)\phi_{t-1},
    \qquad t=1,\ldots,T,
    \label{eq:clock_edge_mismatch}
\end{equation}
and the propagation energy is
\begin{align}
    \bra{\Psi}H_\text{prop}\ket{\Psi} &=\frac{1}{2}\sum_{t=1}^{T}\left\|\ket{\phi_t}-U_t\ket{\phi_{t-1}}\right\|^2 
    =
    \frac{1}{2}\sum_{t=1}^{T}\|\delta_U\phi_t\|^2.
    \label{eq:FK_energy}
\end{align}
This is the analogue of Eq.~\eqref{eq:resistor_dirichlet}. Thus the circuit-to-Hamiltonian construction provides a vector-valued analogue of the resistor-network problem. The scalar voltage drops are replaced by covariant edge difference, Eq.~\eqref{eq:clock_edge_mismatch}, and   $H_\text{prop}$ is the corresponding unitary Laplacian on the clock path;  see Supplemental Material~\ref{sec:graph-laplacian}. Finally, the term $H_\text{in}$ is a soft implementation of the input boundary condition.

The unique ground state of $H_{\rm free}$ is the ``history state''
\begin{equation}
    \ket{\Psi_\mathrm{free}}=\frac{1}{\sqrt{T+1}}\sum_{t=0}^{T}\ket{t}\otimes U_t\cdots U_1\ket{\psi_\text{in}},
\end{equation}
where the $t=0$ term in the sum is $\ket{0}\otimes \ket{\psi_\text{in}}$.
Each clock basis state $\ket{t}$ 
labels the system state after the first $t$ gates of the circuit.

\textbf{\textit{Compiling using FK Hamiltonians.}}  To approach the state compilation problem in the spirit of equilibrium learning, we add a term to the Hamiltonian that penalizes disagreement of the history state at final clock slice and a target state $\ket{\psi_{f}}$ or disagreement with the variational circuit $U_{T}\cdots U_{1}$ with a target unitary $V$.
This is done in a way analogous to clamping in resistor networks (Fig.~\ref{fig:schematic}b), namely by introducing a soft clamp at the final clock state or a soft twist connecting the first and final clock states:
\begin{align}
    H_\text{out} &= \ketbra{T}{T}
    \otimes \left(\mathbb I-\ketbra{\psi_f}{\psi_f}\right), \\
    H_{\text{periodic}}&= {1\over 2}\left( \vphantom{\sum} \ket{0}\otimes V^\dagger - \ket{T}\otimes \mathbb{I}  \right)\left( \vphantom{\sum} \bra{0}\otimes V - \bra{T}\otimes \mathbb{I} \right)
\end{align}
and defining the clamped Hamiltonian $H_{\rm clamped}:= H_{\rm free}+J_{\text{out}}H_{\text{out}}$ with $0<J_{\text{out}}\le J_{\text{in}}$ and $T\ge 1$, and the cyclic Hamiltonian $H_{\text{cyclic}}:=H_\text{free} +J_{\text{cyc}}H_{\text{periodic}}$ with $J_\text{cyc}>0$ and $T\ge 2$. The domain $J_{\text{out}}\le J_{\text{in}}$ is required so that the history state is in the lowest-energy subspace of $\mathcal{H}_{C}\otimes \mathcal{H}_{S}$ in the limit $\epsilon \rightarrow 1$. 
One observes that $H_{\rm clamped}$ biases the final clock state toward the target and  is the direct analogue of the output clamp in Eq.~\eqref{eq:resistor_clamped_energy}~\footnote{The analogy is not perfect: the dissipated power of the resistor network can be nonzero even if the output voltages match the target. }.  In the free history state the  differences $\delta_U\phi_t$ vanish, whereas an incompatible output clamp introduces a tradeoff between matching the target state at the final time and exact propagation through the circuit, 
so these differences generally become nonzero. In $H_{\rm cyclic}$, tunneling between the $\ket{T}$ and $\ket{0}$ clock states comes with an action of $V^{\dagger}$ on the system, which allows the history state to be a ground state if $ \bra{\psi_{\text{in}}}V^{\dagger}U_{T}\cdots U_{1}\ket{\psi_{\text{in}}}=1$ (the converse result is part of Theorem \ref{thm:ttt} below). A similar cyclic clock register dynamics was recently used to construct a perfectly complete verifier circuit for QMA \cite{grewal2026qma}. Because $H_{\rm cyclic}$ does not generally possess a dynamical subspace, we will actually implement $PH_{\rm cyclic}P$ for a projection $P$ that can be implemented using the dynamics of $H_{\rm free}$ only \cite{supp}.  The energy-based cost functions $E_\text{clamped}(\theta)$ and $E_{\text{cyclic}}(\theta)$ are defined as the smallest eigenvalues of $H_{\text{clamped}}$ and $PH_{\text{cyclic}}P$, respectively. 
These cost functions are applied to the unitary compilation task by ampliating the system register to two copies of the system on which the $U_{t}$ act and taking $\ket{\psi_{\text{in}}}$ to consist of maximally entangled states between each register.
The overlap cost functions $C_\text{clamped}(\theta)$ and $C_{\text{cyclic}}(\theta)$ are defined as the infidelity between the respective ground states of $H_{\text{clamped}}$ and $H_{\text{cyclic}}$ and the history state,
    $C (\theta):= 1-\big|\langle\Psi |\Psi_\mathrm{free}\rangle\big|^2.$
\\

The history state is efficiently preparable given the circuit description. For example, with a unary clock one first prepares the uniform superposition of clock states and then applies each $U_t$ controlled on the $t$-th clock qubit, requiring $O(T)$ controlled gates \cite{supp}. The energy cost can be estimated by applying high-confidence quantum phase estimation (QPE)~\cite{patel2026optimal} to the input state $\ket{\Psi_{\rm free}}$, using an efficient Hamiltonian simulation or qubitized representation of $H_{\rm clamped}$ or $PH_{\rm cyclic}P$. 
This follows from our result that the squared overlap between the history state and the ground state $\ket{\Psi}$ is bounded below by a constant independent of $T$, so that QPE obtains the ground-energy component with constant probability. Thus, the energy cost can be estimated to inverse-polynomial accuracy with polynomial quantum resources.

For the overlap cost function, QPE must additionally resolve the ground state eigenspace from the excited spectrum. Applied to $\ket{\Psi_{\rm free}}$, the probability of the ground state outcome directly gives $|\langle\Psi_{\rm free}|\Psi\rangle|^2$; alternatively, a coherent version of this ground state test can be combined with quantum amplitude estimation to obtain the usual quadratic improvement in precision \cite{Brassard2002amplitude}. The required spectral resolution is controlled by the gap $\Delta$, which we show satisfies $\Delta=\Theta(T^{-2})$ for both Hamiltonians. Consequently, both the energy and overlap cost functions can be estimated efficently for inverse-polynomial accuracy.  Whereas lower bounds on the spectral gap of modified FK Hamiltonians are often proved by leveraging Cheeger's inequality for ergodic time-reversible Markov chains, which involves the conductance of weighted graphs \cite{SINCLAIR198993,aharonov,Bohdanowicz2019GoodApproximate}, our exact spectral gap results are based on analytic properties of the characteristic equation of the Hamiltonian and may be of interest in other applications of FK Hamiltonians, such as properties of local quantum lattice models \cite{tvny-gtzp}. 

We now present the main cost function results which capture the essential features of the proposed approach and results in simple explicit expressions. For general parameter domains, the results are given in full generality \cite{supp}. Our results characterize the cost functions while noting their faithfulness to the compilation error, and determine the alignment of their gradients. We define the compilation error $\epsilon := 1-|\bra{\psi_f}U_T\cdots U_1\ket{\psi_\text{in}}|^2$ for the clamped Hamiltonian and $\epsilon:=1-\vert \langle \psi_{\text{in}}\vert V^{\dagger}U_{T}\cdots U_{1}\ket{\psi_{\text{in}}}\vert^{2}$ for the cyclic Hamiltonian, omitting the clamped/cyclic label as the cost function setting will always be clear.

\begin{theorem}[Cost functions of $H_\mathrm{clamped}$]
\label{thm:ooo}
For $J_{\mathrm{in}}=J_{\mathrm{out}}=1$, the ground-state energy $E_{\rm clamped}(\epsilon)$ and the
ground-state overlap cost function $C_{\rm clamped}(\epsilon)$ associated with $H_{\textnormal{clamped}}$ are given by $E_{\rm clamped}(\epsilon) = 1-\cos\vartheta^*$ and \begin{align}
    C_{\rm clamped}(\epsilon)
    &=
    1-
    \frac{\epsilon}
    {4(T+1)^2
    \sin^2\left(
    \frac{\vartheta^*}{2}
    \right)},
    \label{eq:overlapJin1Jout1}
    \end{align}
where
$\vartheta^*(\epsilon)
    =
    \frac{1}{T+1}\sin^{-1}\sqrt{\epsilon}$.
 The cost functions are faithful, strictly increasing functions of $\epsilon$.
\end{theorem}

The domain $J_{\rm out}<J_{\rm in}$ possesses a potentially useful property that is not covered by Theorem \ref{thm:ooo}: the overlap cost function $C_{\rm clamped}(\epsilon)$ has negative derivative with respect to $\epsilon$ above a critical value, and therefore has anti-parallel gradient to $\epsilon(\theta)$ in neighborhoods of critical points above this value. No such anti-parallel gradient behavior occurs in the case of the $H_{\rm cyclic}$ overlap cost function in its general domain $J_{\rm cyc}>0$. Both results are stated rigorously and proven in \cite{supp}.

\begin{theorem}[Cost functions of $H_{\mathrm{cyclic}}$]
\label{thm:ttt}
For $J_{\mathrm{cyc}}>0$, let
$\vartheta^*(\epsilon)$ denote the smallest non-negative solution of the characteristic
equation
\begin{equation}
    \frac{
        \cos\!\left(\frac{(T+2)\vartheta}{2}\right)
    }{
        \cos\!\left(\frac{T\vartheta}{2}\right)
    }
    = 1-J_{\rm cyc}\left( 1 - \sqrt{1-\epsilon}\right).
    \label{eq:secEquationFloquent}
\end{equation}
Then the ground state energy and the ground-state-overlap cost function of
$PH_{\mathrm{cyclic}}P$ are faithful and strictly
increasing functions of $\epsilon \in (0,1)$ given by $E_{\textnormal{cyclic}}(\epsilon)
    =
    1-\cos\vartheta^*$ and
\begin{align}
C_{\textnormal{cyclic}}(\epsilon)
    &=
    1- \frac{
        2\sin^2\left(\frac{(T+1)\vartheta^*}{2}\right)}
        {(T+1)\sin^2\left(\frac{\vartheta^*}{2}\right)
        \left(
            T+1+\frac{\sin((T+1)\vartheta^*)}{\sin(\vartheta^*)}
        \right)}.
\end{align}
\end{theorem}

Theorems \ref{thm:ooo} and {\ref{thm:ttt}} are proved making use of the theory of rank-2 modifications of Hermitian matrices~\cite{arbenz1988restricted}. Beyond having the same global minimum as $\epsilon(\theta)$, it is a remarkable consequence of analyticity of $H_{\clamped}$ and $PH_{\text{cyclic}}P$ with respect to $\epsilon\in (0,1)$ \cite[Theorem 1.10]{Kato1995} that all critical sets of the cost functions are the same as those of $\epsilon(\theta)$. Therefore, the only differences between the compilation error and the equilibrium-based cost functions are in their overall magnitude and shapes. For example, it is clear from Theorem \ref{thm:ooo} that the norm of the gradient of $E_{\rm clamped}(\epsilon(\theta))$ approaches infinity as $\epsilon \rightarrow 1$, which implies smaller level surfaces of neighborhoods of local minima that compile poorly ($\epsilon \approx 1$). We now explore the consequences of this geometric observation on cost function optimization.

\textbf{\textit{Benchmarking.}} 
Our benchmarks probe the training dynamics of the equilibrium costs in an idealized setting, noiseless simulation with exact gradients. 
All trainable circuits are products of Pauli rotations $U_j(\theta_j)=\exp\!\big({-\tfrac{i}{2}\theta_jG_j}\big)$ with $G_j^{2}=I$. 
Each cost $C$ is minimized by fixed-learning-rate coordinate-gradient descent, $\theta_j \leftarrow \theta_j-\eta\,\partial_{\theta_j}C(\theta)$, with one sweep defined as one pass through the $p$ trainable parameters; the compared methods differ only in the cost and in how its gradient is evaluated. For the Pauli-rotation ansätze used here,  QAQC trains  the infidelity $C_{\rm QAQC}:=\epsilon$ with the ordinary parameter-shift gradient $\partial_{\theta_j}C_{\rm QAQC} = \tfrac{1}{2}\big(\epsilon(\theta_j+\tfrac{\pi}{2})-\epsilon(\theta_j-\tfrac{\pi}{2})\big)$ \cite{khatri2019quantum}. The equilibrium methods realize the update of Eq.~\eqref{eq:EnergyUpdate} by descending the ground-state energies $E_{\rm clamped}$ and $E_{\rm cyclic}$, or the overlap cost $C_{\rm clamped}$. 
The energy gradients are Hellmann-Feynman derivatives,
\begin{equation}
    \partial_{\theta_j} E_x(\theta) =     \bra{\Psi_{x}(\theta)}         \partial_{\theta_j}H_x(\theta) \ket{\Psi_{x}(\theta)},
    \; x\in\{{\rm clamped},{\rm cyclic}\},
    \label{eq:HF_gradient}
\end{equation}
$C_{\rm clamped}$ instead depends on the ground-state eigenvectors, so no Hellmann-Feynman identity applies and its gradient is estimated by central finite differences of the cost. 
The operator $\partial_{\theta_j}H_x$ is supported on the single clock edge $(j-1,j)$ and follows from a $\pi$-shift identity, $\partial_{\theta_j}U_j(\theta_j)=\tfrac{1}{2}U_j(\theta_j+\pi)$. 
The ordinary $\pm\pi/2$ parameter-shift rule does not apply to the equilibrium costs, which are evaluated on equilibrium states that change with the same Hamiltonian being differentiated. 

\begin{figure}[t]
\centering
\includegraphics[width=\columnwidth]{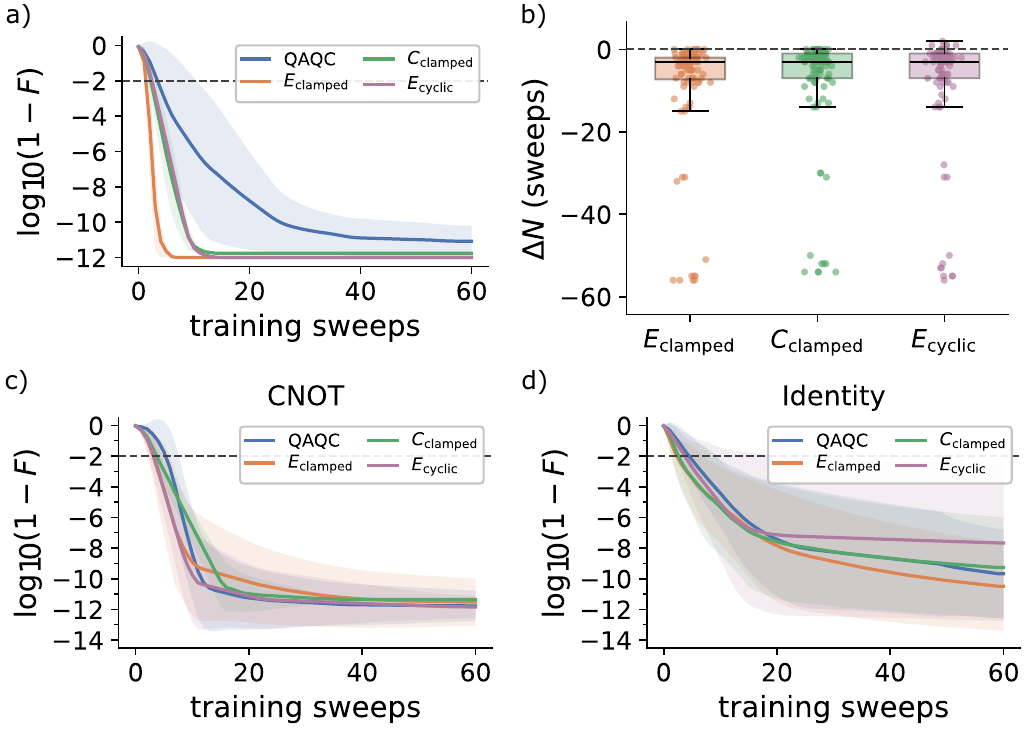}
\caption{ (a,b)~Random-generator identity compiling: (a)~geometric-mean infidelity $1-F$ over paired initializations, (b)~paired sweep differences $\Delta N$ relative to tuned QAQC, with $\Delta N<0$ indicating convergence in fewer sweeps. (c,d)~Geometric-mean infidelity for unitary compiling: $V=\mathrm{CNOT}_{12}$ (c) and $V=I$ (d).}
\label{fig:Fig1_Compiling}
\end{figure}

We benchmark the equilibrium methods against direct QAQC on two suites of two-qubit unitary-compiling tasks, a random-generator identity compiling task (Fig.~\ref{fig:Fig1_Compiling}~a,b) and fixed-ansatz \texttt{CNOT} and identity compiling tasks (Fig.~\ref{fig:Fig1_Compiling}~c,d), which share a common structure: the active register $A$ is entangled with a reference register $R$ in the maximally entangled state $\ket{\Omega}=2^{-1}\sum_{x\in\{0,1\}^2}\ket{x}_A\ket{x}_R$, the trainable circuit acts on $A$ alone, and the objective is the Choi-state fidelity $\vert\langle\Omega_V\vert\Omega_{U(\theta)}\rangle\vert^{2}$, where $\ket{\Omega_{W}}:=(W\otimes I_R)\ket{\Omega}$, which measures agreement between $U(\theta)$ and $V$ on the full two-qubit input space and coincides with the Hilbert-Schmidt-test cost of QAQC. In the FK Hamiltonians these tasks correspond to the boundary states $\ket{\psi_{\rm in}}=\ket{\Omega}$ and $\ket{\psi_f}=\ket{\Omega_{V}}$. All benchmarks were run in noiseless state-vector simulation with fixed input states, randomness entering only through the initial parameters $\theta_j(0)\sim\mathrm{Unif}[-\pi,\pi]$, with identical initialization seeds across methods; learning rates were calibrated independently for each method and task on held-out initializations \cite{supp}. Convergence is defined as reaching fidelity $F=1-\epsilon(\theta)\geq 0.99$, and methods are compared through the 
difference in sweeps required to reach this threshold $\Delta N=N_{\rm method}-N_{\rm QAQC}$, with $\Delta N<0$ indicating convergence in fewer sweeps than tuned QAQC.

In the random-generator setting, 
 $U(\theta)$ is a product of $T=4$ Pauli rotations with  two-qubit generators $G_t$ sampled and then held fixed, and the target is the identity, 
so the optimizer must tune a short coherent circuit to implement the trivial gate ($5$ generator instances, $20$ initializations each). 
In Figure~\ref{fig:Fig1_Compiling}(a,b), all three equilibrium methods reach threshold in fewer median sweeps than QAQC, with a tail of instances gaining tens of sweeps ($\Delta N \lesssim -50$).

In the fixed-ansatz suite, $U(\theta)$ is a fifteen-parameter two-qubit block with three entangling gates (so $T=18$) \cite{supp}, and the targets are $V=\mathrm{CNOT}_{12}$ and $V=I$ ($20$ initializations each). Figure~\ref{fig:Fig1_Compiling}(c,d) shows the corresponding training curves; the paired $\Delta N$ distributions~\cite{supp} give median $\Delta N$ negative for \texttt{CNOT} compiling and near zero for identity compiling.

The sweep reduction reflects improved conditioning of the objective, consistent with the enhanced equilibrium-cost gradients near $\epsilon\approx1$; per-sweep requirements, including  parallel evaluation of clock-edge gradients, are discussed in \cite{supp}.

\textbf{\textit{Discussion.}} In this work, we have not attempted a complete resource analysis in terms of gate, query, and sampling complexities. Our goal has been to introduce the equilibrium-learning approach and establish that the quantum algorithm can be implemented with polynomial resources.
The precise cost of each optimization step depends on how the  ground state of the full Hamiltonian is prepared and how the corresponding cost functions and gradient are estimated. Alternatively, because the cost functions derived here depend only on the compilation infidelity, they could be evaluated by estimating the fidelity directly and then applying the corresponding function classically. The relevant comparison would then depend on the precision with which the fidelity and its gradient must be estimated to train effectively with the different cost functions. We leave a detailed comparison of these implementation-dependent costs for future work and focus here on the number of optimization sweeps, which directly probes how the choice of cost function affects trainability. A reduction in sweep count should therefore be interpreted as an improvement in optimization performance, rather than by itself as an end-to-end quantum-resource advantage.

The differences between the clamped and cyclic Hamiltonians have algorithmic consequences. For small compilation fidelity $F=1-\epsilon$, the clamped gap scales as  $\Theta(\sqrt{F}/T^2)$, while the cyclic gap remains $\Theta(1/T^2)$. Thus, when the initial fidelity is very small, as can occur in high-dimensional compiling problems, the cyclic Hamiltonian may have an advantage for ground state preparation. In contrast, for  short-depth examples we provide an analytic guarantee via the Polyak-Lojasiewicz constant that the clamped energy cost has larger convergence rate than the cyclic energy cost \cite{karimi,supp}. These results do not establish a general ordering between the two approaches, but they suggest hybrid strategies in which one cost function is used at low fidelity and another closer to the solution. Coupling strengths can also serve as hyperparameters that can be changed over the course of optimization. In particular, for $J_{\rm out}<J_{\rm in}$, the clamped overlap cost has a high-infidelity regime in which its gradient becomes anti-aligned with the gradient of infidelity cost function. This raises the possibility that varying the clamp strengths during training could help an optimizer avoid becoming trapped in a local minima.

The question remains open whether one can construct and sample a faithful cost function directly from local observables. A naive construction based on differences of free and clamped overlaps of neighboring clock slices is not faithful. For $PH_\text{cyclic}P$, normalizing nearest-neighbor clock coherences by their values in the free history state yields a faithful cost function built from local observables, but we found its optimization to be slow relative to our other methods. 
Despite this, the construction illustrates the flexibility of the equilibrium-learning framework to generate new faithful cost functions with different evaluation procedures and different optimization landscapes, while encoding the same underlying compilation task. Extension of our framework could allow constraints to be imposed at intermediate clock times, allowing partial information about the desired computation, such as known symmetries or conserved quantities, to be incorporated into the training objective.

\textbf{\textit{Acknowledgements.}} We thank Eddie Schoute for discussions on low-rank modifications of Hermitian matrices. This work was primarily led and supported by the U.S. Department of Energy, Office of Science, National
Quantum Information Science Research Centers, Quantum Science Center (QSC). TAA, YS, and TV were supported by QSC. 
FB acknowledges support by the DOE Advanced Scientific Computing Research (ASCR) program  Award No. DE-SCL0000118 and by the U.S.\ Department of  Energy (DOE), Office of Science, Basic Energy Sciences program (award No.\ DE-SCL0000157). 
This work was initiated at Aspen Center for Physics, which is supported by National Science Foundation grant PHY-2210452. 
LLMs were used to generate figures, edit text, and in some simplification steps in proofs, which the authors have independently verified. Conceptualization, definitions of the models, and writing the manuscript were done by the authors.

\bibliography{references}

\clearpage
\twocolumngrid 
\widetext  
\begin{center}
  \textbf{\large Supplemental Material}
\end{center}
\vspace{0.5cm}

\setcounter{theorem}{0}
\renewcommand{\thetheorem}{S\arabic{theorem}}

\setcounter{equation}{0}
\renewcommand{\theequation}{S\arabic{equation}}

\setcounter{table}{0}
\renewcommand{\thetable}{S\arabic{table}}

\setcounter{secnumdepth}{3}

\setcounter{section}{0}
\setcounter{subsection}{0}
\setcounter{subsubsection}{0}

\renewcommand{\thesection}{S\Roman{section}}

\renewcommand{\thesubsection}{\thesection.\arabic{subsection}}

\renewcommand{\thesubsubsection}{\thesubsection.\Alph{subsubsection}}

\makeatletter
\renewcommand{\p@subsection}{}
\renewcommand{\p@subsubsection}{}
\makeatother

\setcounter{tocdepth}{2}
\tableofcontents
\vspace{1em}

\section{FK Hamiltonians, ground states, and rank-2 modifications of Hermitian matrices}
\label{sec:first}

\begin{table}[h]
    \centering
    \begin{tabular}{|c||c|c||c|}
        \hline
        &\multicolumn{2}{c|}{$H_{\rm clamped}$} 
        & $H_{\rm cyclic}$ \\
        \hline
        \hline
        Parameters
        &\multicolumn{2}{c|}{Thm. \ref{thm:clampedparams}}
        & L-Thm. \ref{thm:cyclicparams}
        \\
        \hline
        & $0< J_{\mathrm{out}} = J_{\mathrm{in}}=J$
        & $0< J_{\mathrm{out}} < J_{\mathrm{in}}$
        & $  J_\mathrm{in} \gtrsim 2J_{\rm cyc} > 0 $\\
        \hline
        
        Secular eqn, faithfulness
        &
        & Thm. \ref{thm:clampedsec}
        & Thm. \ref{thm:cyclicsec}
         \\
        \hline

        \textbf{GSE cost function, Gap}
        &
        & L-Thm. \ref{thm:gap_gen_j_in_j_out}
        & L-Thm. \ref{thm:cycliclargeTgap}
         \\
        \hline

        GSE Gradient
        &
        & Thm. \ref{thm:GSEgradientClampedGeneralJ}    & Thm. \ref{thm:GSE_H_cyclic_gradient}    \\
        \hline

        Ground state
        &
        & Thm. \ref{thm:clampedgs}
        & Thm. \ref{thm:cyclicgs}
         \\
        \hline

        \textbf{Overlap cost function}
        &
        Lem. \ref{lem:llll}
        & (\ref{eq:OverlapCost_Hclampd_GenJ})
        & Thm. \ref{thm:overlapcostcyclic}
         \\
        \hline

        Overlap Gradient
        & Thm. \ref{thm:jin_jout_parallel}
        & L-Thm. \ref{thm:ggoo}
        & Thm. \ref{thm:GSO_H_cyclic2}
         \\
        \hline

        Min Overlap
        &
        & $1\over 4$ - Thm. \ref{thm:LowerBoundOverlapLargeTClampedGeneralJ}
        & ${1\over 3}$ - Thm. \ref{thm:worstOverlapCyclicGeneral}
         \\
        \hline
    \end{tabular}
    \caption{Summary of results. GSE denotes the ground state energy, and Gradient refers to a result that determines the sign of the derivative of the cost function with respect to $\epsilon$. The L- prefix indicates that the result is in the limit of large $T$ (large variational circuit depth). Empty boxes in the $J_{\rm out}=J_{\rm in}=J$ column mean that the $0<J_{\rm out}<J_{\rm in}$ theorem holds for $J_{\rm out}\le J_{\rm in}$. The minimum overlap results are stated explicitly.}
    \label{tab:summaryResults}
\end{table}

Theorem 1 and of the main text is the $J_{\rm in}=J_{\rm out}=1$ special case of our more general results for $H_{\rm clamped}$ in Theorems (S2)-(S5), (S11) and the special form of $\ket{\Psi_{\rm clamped}}$ in Eq.(\ref{eqn:specialclampedgs}). Theorem 2 of the main text follows from the detailed statements of Theorems (S13)-(S18). We prove the general results and demonstrate how they reduce to the formulas and claims in Theorems 1 and 2. We first show the matrices of $H_{\rm clamped}$ and $P_{\mathcal{D}}H_{\rm cyclic}P_{\mathcal{D}}$  and derive their respective ground states and ground state energies. The coupling constant $J_{\rm prop}$ of the Feynman-Kitaev propagation Hamiltonian sets the energy scale and is taken to be 1 throughout the analysis. The clock register $C$ is spanned by $T+1$ orthonormal states $\ket{0},\ldots,\ket{T}$ and the system register $S$ is a register of $n$ qubits. Defining an orthogonal state $\ket{\psi_{\rm in}^{\perp}}$ orthogonal to $\ket{\psi_{\rm in}}$, the target state $\ket{\psi_{f}}_{S}$ is decomposed into its component on $\ket{\psi_{T}}_{S}:=\prod_{j=1}^{T}U_{j}(\theta_{j})\ket{\psi_{\rm in}}_{S}$ and an orthogonal state $\ket{\psi_{T}^{\perp}}:= U_{T}^{\dagger}\cdots U_{1}^{\dagger}\ket{\psi_{\rm in}^{\perp}}$ according to
\begin{equation}
    \ket{\psi_{f}}=\sqrt{1-\epsilon(\theta)}e^{i\Theta_{1}}\ket{\psi_{T}} + \sqrt{\epsilon(\theta)}e^{i\Theta_{2}}\ket{\psi_{T}^{\perp}}.
\end{equation}
By taking $\ket{\tilde{T}}:=e^{i\Theta_{1}}\ket{T}$ and redefining $\ket{\psi_{\rm in}^{\perp}}$ so that $e^{i\Theta_{2}}\ket{T}\ket{\psi_{T}^{\perp}} = \ket{\tilde{T}}\ket{\psi_{T}^{\perp}}$, one can write
\begin{equation}
    \ket{T}\ket{\psi_{f}}=\sqrt{1-\epsilon(\theta)}\ket{\tilde{T}}\ket{\psi_{T}} + \sqrt{\epsilon(\theta)}\ket{\tilde{T}}\ket{\psi_{T}^{\perp}}.
\end{equation}
Introducing the ordered orthonormal set $\mathcal{B}$
\begin{equation}
\ket{\gamma_{0}}_{CS}:=\ket{0}_{C}\ket{\psi_{\text{in}}}_{S},\ldots,\ket{\gamma_{T}}_{CS}:=\ket{\tilde{T}}_{C}\ket{\psi_{T}}_{S},\ket{\eta_{T}}_{CS}:=\ket{\tilde{T}}_{C}\ket{\psi_{T}^{\perp}}_{S},\cdots , \ket{\eta_{0}}_{CS}=\ket{0}_{C}\ket{\psi_{\text{in}}^{\perp}}_{S},
    \label{eqn:spsp}
\end{equation}
where $\ket{\psi_{t}}:=  U_t \dots U_1 \ket{\psi_\text{in}}$ and $\ket{\psi_{t}^{\perp}}:=  U_t^{\dagger} \dots U_1^{\dagger} \ket{\psi_\text{in}^{\perp}}$, the matrix of $H_{\rm clamped}$ in the low energy subspace is given by the  $2(T+1)\times 2(T+1)$ tridiagonal matrix
\begin{align}
H_{\rm clamped}\big\vert_{\text{span}\mathcal{B}}=\begin{pmatrix}
{1\over 2} &-{1\over 2} &            &             &                           &                            &            &               & \\
-{1\over 2}&  1         &-{1\over 2} &             &                           &                            &            &               &\\
           &            &\ddots      &             &                           &                            &            &              & \\
           &            & -{1\over 2}& 1           & -{1\over 2}               &                            &            &               &\\
           &            &            &-{1\over 2} & {1\over 2}+J_\text{out}\epsilon        &-J_\text{out}\sqrt{\epsilon(1-\epsilon)}&            &               &\\
           &            &            &            &-J_\text{out}\sqrt{\epsilon(1-\epsilon)}& J_\text{out}(1-\epsilon)+{1\over 2}       & -{1\over 2}&              &\\
           &            &            &            &                            & -{1\over 2}                & 1          & -{1\over 2} & \\
           &            &            &            &                           &                            &             & \ddots       & \\
            &            &            &            &                           &                            &             & -{1\over 2} & 1 & -{1\over 2}     \\
                        &            &            &            &                           &                            &             &  & -{1\over 2} & {1\over 2}+J_\text{in}     \\
\end{pmatrix},
\label{eqn:matmat}
\end{align}
where $\epsilon(\theta) := 1-F(\theta)$ with \begin{equation}
    F(\theta)
    :=
    \Big|
        \bra{\psi_f}
        \prod_{t=1}^{T}U_t(\theta_t)
        \ket{\psi_{\rm in}}
    \Big|^2=|\bra{\psi_f}{\psi_T}\rangle|^{2}.
    \label{eq:fidelity_infidelity}
\end{equation}
Because of our freedom in choosing $\ket{\tilde{T}}$ and $\ket{\psi_{\rm in}^{\perp}}$ when analyzing a dynamical subspace at circuit parameters $\theta$, we will dispense with the tilde in the $\ket{\tilde{T}}$ from here on and always define $\ket{T}$ and $\ket{\psi_{\rm in}^{\perp}}$ to get rid of the phases, so that we can write
\begin{equation}
    \ket{\psi_{f}}=\sqrt{1-\epsilon(\theta)}\ket{\psi_{T}} + \sqrt{\epsilon(\theta)}\ket{\psi_{T}^{\perp}}
\end{equation}
without loss of generality.

Note that the Feynman-Kitaev history state can be written \begin{equation}
    \ket{\Psi_{\rm free}}=(T+1)^{-1/2}\sum_{t=0}^{T}\ket{\gamma_{t}}_{CS}
    \label{eq:historystate}
\end{equation} and that it is in the linear span of $\mathcal{D}\subset \mathcal{B}$ where $\mathcal{D}:=\lbrace \ket{\gamma_{0}}_{CS},\ldots,\ket{\gamma_{T}}_{CS}\rbrace$. If $J_{\text{out}}>J_{\text{in}}$, the amplitude of the ground state of $H_{\rm clamped}\big\vert_{\text{span}\mathcal{B}}$ in $\text{span}\mathcal{D}$ goes to 0 as $\epsilon$ goes to 1. Because a constant fidelity of the ground state with $\ket{\Psi_{\rm free}}$ is desired for all $\epsilon \in [0,1]$, we therefore take $J_{\rm out} \le J_{\rm in}$.  See Section \ref{sec:leds} for an example result showing the energy gap of $H_{\rm clamped}$ between $\text{span}\mathcal{B}$ and its complement in $CS$.

Because $\mathcal{D}$ is not a dynamical subspace of $H_{\rm cyclic}$, we consider $P_{\mathcal{D}}H_{\rm cyclic}P_{\mathcal{D}}$ which is a $(T+1)\times (T+1)$ tridiagonal matrix, where we use $V\ket{\psi_{\rm in}}$ instead of $\ket{\psi_{f}}$ above
\begin{equation}
P_{\mathcal{D}}H_{\text{cyclic}}P_{\mathcal{D}}=\begin{pmatrix}
{1\over 2}+{J_{\text{cyc}}\over 2} &-{1\over 2} &     0        &     \cdots        &         \cdots          & 0 & -{J_{\text{cyc}}\over 2}\sqrt{1-\epsilon} \\
-{1\over 2}&  1         &-{1\over 2}&  &             &                    &  0\\
      0     &-{1\over 2} &1 &-{1\over 2}&            &                    &  \vdots \\
      \vdots     &            &  &\ddots     &            &                    &  \vdots  \\
       \vdots    &            &  &-{1\over 2}& 1          &    -{1\over 2}     &   0  \\
        0   &            &  &           &-{1\over 2} & 1 &-{1\over 2} \\
      -{J_{\text{cyc}}\over 2}\sqrt{1-\epsilon}     &      0      &  \cdots &    \cdots       &       0     &-{1\over 2}& {1\over 2}+{J_{\text{cyc}}\over 2} \\
\end{pmatrix}.
\end{equation}
Note that this matrix does not depend on $J_{\text{in}}$ because we have restricted to a Hilbert subspace defined by the kernel of $H_{\rm in}$. Despite the fact that $\text{span}\mathcal{D}$ gets no penalty from $J_{\text{in}}$, it is not necessarily the lowest energy $(T+1)$-dimensional subspace spanned by $\lbrace \ket{t}_{C}\otimes U_{t}\cdots U_{1}\ket{\psi}_{S}\rbrace_{t=0}^{T}$ for some system state $\ket{\psi}$. For example, it possible that the circuit parameters $\theta$ are chosen such that compiling in the $\text{span}\mathcal{D}$ fails, whereas compiling in an orthogonal subspace of the above form succeeds, compensating for the initialization energy penalty. `` For sufficiently large $T$, Theorem \ref{thm:cyclicparams} guarantees that $J_{\rm in} > 2J_{\rm cyc}$ suffices. 

Before analyzing the two Hamiltonians in more detail, we emphasize that the time dynamics of $P_{\mathcal{D}}H_{\text{cyclic}}P_{\mathcal{D}}$ can be simulated using the time dynamics of the local Hamiltonian $H_{\text{cyclic}}$ and $H_{\free}$. This point, shown in Section \ref{sec:qpesupp} allows us to implement the QPE algorithm discussed in the main text for calculating the cost functions in the cyclic setting. 

Both Hamiltonians $H_{\clamped}\big\vert_{\text{span}\mathcal{B}}$ and $P_{\mathcal{D}}H_{\text{cyclic}}P_{\mathcal{D}}$ are rank-2 modifications of tridiagonal matrices in the sense that each can be written
\begin{align}
    \label{eq:modified_H}
    H = H_0 + \sum_{\ell=1}^L a_\ell \ketbra{\xi_\ell} \; .
\end{align}
where $H_{0}$ can be diagonalized in closed form. Secular equations and the associated eigenvector formulas for finite-rank matrix modifications are well established~\cite{arbenz1988restricted}. For completeness, we rederive the results needed here in Dirac notation and then specialize them to rank-2 modifications.
An eigenvector $\ket{\varphi}$ of this Hamiltonian with eigenvalue $\lambda$ satisfies
\begin{align}
    \left( H_0 + \sum_{\ell=1}^L a_\ell \ketbra{\xi_\ell}-\lambda \right) \ket{\varphi} &=0 .
\end{align}
We define the resolvent of the unmodified Hamiltonian as $R_0(\lambda) := (H_0-\lambda)^{-1}$. In what follows we will suppress the $\lambda$ dependence of $R_0(\lambda)$ and other quantities to avoid clutter and reinsert it only when we want to draw attention to it. Applying the resolvent to the above equation we obtain
\begin{align}
    \ket{\varphi} + \sum_{\ell=1}^L a_\ell \langle \xi_\ell \vert \varphi\rangle R_{0}(\lambda)\ket{\xi_\ell} &= 0. 
\end{align}
Contracting this equation with $\bra{\xi_k}$ and defining $c_\ell = \langle \xi_\ell\vert \varphi\rangle$ and $r_{k\ell}(\lambda) = \bra{\xi_k} R_0(\lambda) \ket{\xi_\ell}$, we obtain
\begin{align}
    c_k + \sum_{\ell=1}^L a_\ell   r_{k\ell} c_{\ell} &= 0
    \label{eqn:eigeq}
\end{align}
Collecting these equations over $k=1,\ldots, L$, the linear system is
\begin{align}
    \label{eq:c}
    M(\lambda)\vec{c} = 0   
\end{align}
where $M(\lambda)$ is a matrix with entries $M(\lambda)_{ij}=\delta_{ij}+  a_j r_{ij}(\lambda)$. 
Since $\vec{c}$ is generally nonzero, this equation holds only if 
\begin{align}
    \label{eq:eigenvalue_equation}
    \text{det}M(\lambda) = 0,
\end{align}
which is the characteristic equation for $\lambda$. The eigenstates are determined by $\vec{c}$ solving \eqref{eq:c} for $\lambda$ values that satisfy \eqref{eq:eigenvalue_equation}: 
\begin{align}
    \ket{\varphi} &\propto R_0  \sum_{\ell=1}^L c_\ell a_\ell \ket{\xi_\ell}
    \label{eqn:r2p}
\end{align}

Specializing in the case of rank-2 modifications, i.e. $L=2$, we get the eigenvalue equation (i.e., secular equation, characteristic equation):
\begin{align}
    \label{eq:eig_equation}
    g = (1+a_1 r_{11}(\lambda))(1+a_2r_{22}(\lambda)) - a_1 a_2 r_{12}(\lambda)r_{21}(\lambda) = 0 
\end{align}
The eigenstates can be computed explicitly as well resulting in
\begin{align}
    \ket{\varphi} &\propto R_{0}(c_{1}a_{1}\ket{\xi_{1}}+ c_{2}a_{2}\ket{\xi_{2}})\nonumber \\
    &\propto R_0 \left( \ket{\xi_1} + A \ket{\xi_2} \right)
    \label{eqn:resolvprop}
\end{align}
where the second proportionality eliminates the $c_{1}$, $c_{2}$ with
\begin{align}
    A &= - \frac{1+a_1 r_{11}}{a_1r_{12}} \\
      &= - \frac{a_2 r_{21}}{1+a_2r_{22}}
\end{align}
These two expressions are equal because $\lambda$ is chosen to satisfy Eq.~\eqref{eq:eigenvalue_equation}. 

\subsection{Analysis of the clamped energy cost function $E_{\rm clamped}$}

We first apply the rank-2 modification formalism to analyze the cost functions associated with $H_{\rm clamped}$. This is made easier by first introducing a shifted version of $H_{\rm prop}$.

\begin{lemma}
\label{lem:clem}
    Define the matrix $C$ as the $(T+1)\times (T+1)$ matrix
\begin{equation}
    C=\begin{pmatrix}
\frac{1}{2} & \frac{1}{2} & 0 & \cdots & 0 \\
\frac{1}{2} & 0 & \frac{1}{2} & \ddots & \vdots \\
0 & \frac{1}{2} & 0 & \ddots & 0 \\
\vdots & \ddots & \ddots & 0 & \frac{1}{2} \\
0 & \cdots & 0 & \frac{1}{2} & \frac{1}{2}
\end{pmatrix}
\label{eqn:cmat}
\end{equation}
with basis indexing from $0,\ldots, T$, and define the resolvent $R_{C}(\cos\vartheta):= \left( \cos\vartheta - C\right)^{-1}$. One has matrix elements
\begin{align}
    R_{C}(\cos\vartheta)_{0,T} &= -{\cot\left( {\vartheta\over 2} \right) \over \sin((T+1)\vartheta)} \\
    R_{C}(\cos\vartheta)_{0,0} &=R_{C}(\cos\vartheta)_{T,T} = -1-\cot\left({\vartheta\over 2}\right)\cot((T+1)\vartheta). 
\end{align}
\end{lemma}
\begin{proof}
    The reason for the ``backward'' convention for the resolvent $R_{C}$ is that we will find it useful to write the ground state eigenvalue $\lambda$ of the full matrix $H$ in (\ref{eq:modified_H}) as $\lambda = 1-\cos\vartheta$, which will be a valid transformation for all $H$ considered in this work. Therefore, if $C:= I - H_{0}$, we have that $R_{0}(1-\cos\vartheta) = \left( \cos\vartheta - C\right)^{-1}$. Note that in the orthonormal basis $\mathcal{D}=\lbrace \ket{\gamma_{0}},\ldots,\ket{\gamma_{T}}\rbrace$, the matrix $C$ in (\ref{eqn:cmat}) is $C=I_{T+1}-H_{\rm \prop}$. The eigenvalues of $H_{\rm prop}$ in this basis are given by \cite{Losonczi1992}
\begin{align} 
    \omega_{k}:= 1-\cos\left( {k\pi \over T+1}\right).
    \label{eqn:h0eigs}
\end{align}
From the identity $\left( \cos\vartheta - C \right)R_{C}(\cos\vartheta)\ket{\gamma_{T}} = \ket{\gamma_{T}}$ one has the recurrence 
\begin{align}
R_{C}(\cos \vartheta)_{1,T}&= (2\cos\vartheta -1)R_{C}(\cos \vartheta)_{0,T}\nonumber \\
     R_{C}(\cos \vartheta)_{t+1,T} &=    2\cos\vartheta R_{C}(\cos \vartheta)_{t,T}-R_{C}(\cos \vartheta)_{t-1,T}  \, , \, t=1,\ldots,T-2\nonumber \\
    R_{C}(\cos\vartheta)_{T,T}&= {2+R_{C}(\cos \vartheta)_{T-1,T} \over 2\cos\vartheta -1},
    \label{eqn:resolverecurrence}
\end{align}
holds, which has the solution \begin{equation}R_{C}(\cos\vartheta)_{t,T}=-\frac{\cos \left(\left(t+\tfrac{1}{2}\right)\vartheta\right)}
{\sin\left(\frac{\vartheta}{2}\right)\,\sin\left((T+1)\vartheta\right)}.
\end{equation}The fact that $R_{C}(\cos\vartheta)_{0,0}=R_{C}(\cos\vartheta)_{T,T}$ follows is a consequence of $C$ being a centrosymmetric matrix.
\end{proof}

The matrix $C$ allows to evaluate the matrix elements of the resolvent of $H_{\rm prop}$ in closed form, which leads to a closed form characteristic equation for $H_{\rm clamped}$.

\begin{theorem}
\label{thm:clampedsec}
    For $0<J_{\rm out}\le J_{\rm in}$, the ground state energy of $H_{\rm clamped}$ is given by $E_{\rm clamped}(\epsilon)=1-\cos\vartheta^*(\epsilon)$, where $\vartheta^*$ is the smallest non-negative solution of $g(\epsilon,\vartheta)=0$ where
\begin{align}\label{eq:SecGenJinJout}
    g(\epsilon,\vartheta) &:= \left(1+J_{\rm out}\left(-1-\cot\left({\vartheta\over 2}\right)\cot((T+1)\vartheta)\right)\right)\left(1+J_{\rm in}\left(-1-\cot\left({\vartheta\over 2}\right)\cot((T+1)\vartheta)\right)\right) \nonumber \\
    &{}-J_{\rm out}J_{\rm in}(1-\epsilon)\left(\frac{\cot(\frac{\vartheta}{2})}{\sin((T+1)\vartheta)}\right)^{2}.
\end{align}
$E_{\rm clamped}(\epsilon)$  takes values in $\mathbb{R}_{\ge 0}$ and is equal to zero if and only if the circuit parameters $\theta$ are such that $\epsilon(\theta) =0$.
\end{theorem}

\begin{proof}
    In the basis $\mathcal{B}=\lbrace \ket{\gamma_{0}},\ldots,\ket{\gamma_{T}},\ket{\eta_{T}},\ldots,\ket{\eta_{0}}\rbrace$, $H_{\rm prop}$ is the sum of two identical tridiagonal matrices, whose eigenvectors are given by
\begin{align}
    \ket{\Psi_{k}} &= \sum_{t=0}^{T}h_{kt} \ket{\gamma_{t}}\, ,  \\ 
    \ket{\Psi_{k}^\perp} &= \sum_{t=0}^{T}h_{kt}\ket{\eta_{T-t}}\, ,
\end{align}
for $k=0,\ldots, T$ where
\begin{align}
    h_{kt} =
\begin{cases}
\sqrt{\frac{1}{T+1}} &  k=0 \, ,\\
\sqrt{\frac{2}{T+1}} \cos \left( {k(2t+1)\pi\over 2(T+1)} \right) &  k\ge 1.
\end{cases}
\end{align}
Note that the $\ket{\Psi_{k}}$ are also the eigenvectors of $C$ in Lemma \ref{lem:clem}. Both sets of eigenvectors have the same eigenvalues given by 
\begin{align} 
    \omega_{k}:= 1-\cos\left( {k\pi \over T+1}\right).
\end{align}
Working in $\text{span}\mathcal{B}$, one finds that $H_{\rm clamped}$ has the form of $H$ in (\ref{eq:modified_H}) with
\begin{align}
    H_0 &= H_\prop \nonumber \\
    a_1 &= J_{\rm out} \nonumber \\
    \ket{\xi_1} &=  \sqrt{\epsilon}\ket{\gamma_{T}} -\sqrt{1-\epsilon} \ket{\eta_{T}}  \nonumber \\
    a_2 &= J_{\rm in} \nonumber \\
    \ket{\xi_2} &= \ket{\eta_{0}}.
    \label{eqn:idents}
\end{align}
The resolvent is 
\begin{align}
    R_{0}(\lambda) &= \sum_{k=0}^T \frac{\ketbra{\Psi_k} + \ketbra{\Psi_k^\perp}}{\omega_k - \lambda}
    \label{eqn:res0}
\end{align}
with $1,1$ matrix element given by Lemma \ref{lem:clem} (we rewrite $\lambda = 1-\cos\vartheta$, $\vartheta \in (0,\pi)$)
\begin{align}
    r_{11} &= \epsilon\sum_{k=0}^{T} \frac{h_{kT}^2}{\omega_k - \lambda} +(1-\epsilon)\sum_{k=0}^{T} \frac{h_{k0}^2}{\omega_k - \lambda} \\
    &= \epsilon R_{C}(\cos\vartheta)_{T,T} + (1-\epsilon)R_{C}(\cos\vartheta)_{0,0}\nonumber \\
    &= R_{C}(\cos\vartheta)_{0,0} \nonumber \\
    &= -1-\cot\left({\vartheta\over 2}\right)\cot((T+1)\vartheta).\nonumber
    \end{align}
     The 2,2 and 1,2 matrix elements are also given by Lemma \ref{lem:clem}
    and
    \begin{align}
    r_{22} &= \sum_{k=0}^{T} \frac{h_{k0}^2}{\omega_k - \lambda} = r_{11} \\
    r_{12} &= r_{21} = - \sqrt{1-\epsilon}\sum_{k=0}^{T} \frac{h_{kT}h_{k0}}{\omega_k - \lambda} \nonumber \\
    &= - \sqrt{1-\epsilon} \left( -{1\over (T+1)\lambda} + {2\over T+1}\sum_{k=1}^{T}{\cos\left({k(2T+1)\pi \over 2(T+1)}\right) \cos \left( {k\pi \over 2(T+1)}\right)\over \omega_{k}-\lambda} \right)\nonumber \\
    &= - \sqrt{1-\epsilon} R_{C}(\cos\vartheta)_{0,T} \nonumber \\
    &= {\sqrt{1-\epsilon}\cot\left( {\vartheta\over 2} \right) \over \sin((T+1)\vartheta)}
    \label{eqn:uuuu}
\end{align}
The characteristic equation (\ref{eq:eigenvalue_equation}) of $H_{\rm clamped}$ is obtained from (\ref{eq:eig_equation}) as
\begin{align}
    \label{eq:eig_equation2}
    g = \left(1+J_{\rm out} \sum_{k=0}^{T+1} \frac{h_{k0}^2}{\omega_k - \lambda}\right)\left(1+J_{\rm in} \sum_{k=0}^{T+1} \frac{h_{kT}^2}{\omega_k - \lambda}\right) - J_{\rm in} J_{\rm out} (1-\epsilon) \left(\sum_{k=0}^{T+1} \frac{h_{kT}h_{k0}}{\omega_k - \lambda}\right)^2 = 0 .
\end{align}
Using the resolvent matrix elements derived above, this simplifies as
\begin{align}
    g(\epsilon,\vartheta) & = \left(1+J_\mathrm{in}\left(-1-\cot\left({\vartheta\over 2}\right)\cot((T+1)\vartheta)\right)\right)\left(1+J_\mathrm{out}\left(-1-\cot\left({\vartheta\over 2}\right)\cot((T+1)\vartheta)\right)\right) \nonumber \\
    & \hspace{1in}-J_\mathrm{in}J_\mathrm{out}(1-\epsilon)\left(\frac{\cot(\frac{\vartheta}{2})}{\sin((T+1)\vartheta)}\right)^2.
    \label{eqn:genjinjout}
\end{align}
The secular equation $g(\epsilon,\vartheta)=0$ is valid on $\epsilon\in (0,1)$. For $\epsilon = 0,1$ the Hamiltonian $H_{\clamped}$ decouples into a direct sum with the ground state in the $\ket{\gamma_{t}}$ sector when $J_{\text{out}}< J_{\text{in}}$.
For $J_{\rm in} = J_{\rm out}=J$, we get
\begin{align}
    g(\epsilon,\vartheta) = \left(1+J\left(-1-\cot\left({\vartheta\over 2}\right)\cot((T+1)\vartheta)\right)\right)^2 -J^2(1-\epsilon)\left(\frac{\cot(\frac{\vartheta}{2})}{\sin((T+1)\vartheta)}\right)^2.
    \label{eqn:genjsecular}
\end{align}
 For general $J_{\rm out}$ and $J_{\rm in}$, it does not have a closed form solution.

 Note that  $E_{\rm clamped}(\epsilon(\theta)) = \langle \Psi_{\clamped}(\theta)\vert H_{\clamped}(\theta)\vert \Psi_{\clamped}(\theta)\rangle = 1-\cos \vartheta^{*}(\epsilon(\theta)) \ge 0$ due to the positive semidefiniteness of $H_\clamped$. For the faithfulness statement in the theorem, one can argue directly from the characteristic equation, but we provide a more physical proof here. $H_{\final}$ is a positive semidefinite perturbation of $H_{\text{free}}$, with kernel $\mathbb{C}\ket{T}\ket{\psi_{f}}$. If $\vert \langle \psi_f \vert U_{T}(\theta_{T})\cdots U_{1}(\theta_{1}) \ket{\psi_{\rm in}} \vert =1$, the Feynman-Kitaev history state $\ket{\psi_{\rm free} (\theta)}$ is a zero eigenvector of $H_\clamped(\theta)$. Conversely, if $H_\clamped (\theta)\ket{\psi_\clamped (\theta)} = 0$, then according to Corollary 4.3.12 of \cite{hj} there is a vector $\ket{x}\in \mathbb{C}^{T+1}\otimes \mathbb{C}^{d}$ such that $H_{\rm free} \ket{x}=0$ and $H_{\rm final}\ket{x}=0$. The first condition implies that $\ket{x}$ must have the Feynman-Kitaev history state form (\ref{eq:historystate}) with initial state $\ket{\psi_{\rm in}}$. The second condition implies that $\langle T \vert x\rangle \propto \ket{\psi_f}$. Combining these implications, one reaches the conclusion that $U_{T}(\theta_T)\cdots U_{1}(\theta_{1})\ket{\psi_{\rm in}} = \ket{\psi_f}$ up to a phase. 
\end{proof}

For $J_{\rm out}=J_{\rm in}=1$, (\ref{eqn:genjsecular}) becomes \begin{equation}
    \tan^{2}\left( (T+1)\vartheta \right) = {\epsilon \over 1-\epsilon},
    \label{eqn:jjj1}
\end{equation}
which has smallest non-negative solution $\vartheta^{*}$ with
\begin{equation}
    \vartheta^{*} = {1\over T+1}\tan^{-1}\sqrt{\epsilon \over 1-\epsilon}={1\over T+1}\sin^{-1}\sqrt{\epsilon},
    \label{eqn:jjj2}
\end{equation}
proving the form of $E_{\rm clamped}(\epsilon)$ in Theorem 1 of the main text. The fact that $E_{\rm clamped}(\epsilon(\theta))$ has parallel gradient to $\epsilon(\theta)$ is a consequence of the chain rule and the following theorem.
\begin{theorem}
\label{thm:GSEgradientClampedGeneralJ}
    For any $J_{\rm out},J_{\rm in} >0$ and $\epsilon \in (0,1)$, let $\vartheta(\epsilon)$ be the smallest solution on $(0,{\pi\over T+1})$ of $g(\epsilon,\vartheta)=0$. Then
    \begin{equation}
        {d\vartheta \over d\epsilon}\ge 0.
        \end{equation}
\end{theorem}
\begin{proof}
    The secular equation $g(\epsilon,\vartheta)=0$ simplifies to
    \begin{equation}
        q_{\text{in}}(\vartheta)q_{\text{out}}(\vartheta)=1-\epsilon \label{eq:secGenJproduct}
    \end{equation}
    for $\vartheta \in (0,{\pi\over T+1})$, where
    \begin{align}
        q_{\rm in}(\vartheta)
&:=
\cos((T+1)\vartheta)
+
\frac{J_{\rm in}-1}{J_{\rm in}}
\tan\left(\frac{\vartheta}{2}\right)
\sin((T+1)\vartheta) \nonumber \\
q_{\rm out}(\vartheta)
&:=
\cos((T+1)\vartheta)
+
\frac{J_{\rm out}-1}{J_{\rm out}}
\tan\left(\frac{\vartheta}{2}\right)
\sin((T+1)\vartheta).
\label{eqn:qinqout}
    \end{align}
    Each of these functions is 1 for $\vartheta=0$ and -1 for $\vartheta={\pi \over T+1}$, so the smallest solution of the equation is in the interval $(0,{\pi\over T+1})$ by intermediate value theorem. Call the product of the functions by $Q(\vartheta)$ and note that $Q(0)=1$. Take $0<\epsilon_{1}<\epsilon_{2}$ and let $\vartheta_{1}$ and $\vartheta_{2}$ be their corresponding smallest solutions. If $\vartheta_{2}<\vartheta_{1}$, then the intermediate value theorem implies that there must be another solution to $Q(\vartheta)=1-\epsilon_{1} > 1-\epsilon_{2}$ on the interval $(0,\vartheta_{2})$, contradicting the assumption that $\vartheta_{1}$ is the smallest solution. This establishes monotonicity of $\vartheta(\epsilon)$ and the fact that $\vartheta(\epsilon)$ is differentiable follows from the Hellmann-Feynman theorem and the fact that the lowest eigenvalue of $H_{\clamped}$ is non-degenerate (Corollary 4.3.12 of Ref. \cite{hj}, using $H_{0}$ as the unperturbed Hamiltonian). Therefore, $\vartheta(\epsilon)$ is a smooth function of $\epsilon$ for $0<\epsilon<1$.
\end{proof}

We now proceed to analyze the spectral gap of $H_{\rm clamped}$ using methods from analysis instead of Markov chain properties such as Cheeger's inequality.

\begin{theorem}
\label{thm:gap_gen_j_in_j_out}
    Fix $\epsilon\in(0,1)$ and $0< J_\mathrm{out} \leq J_{\mathrm{in}}$, and define
\begin{align}
    n &:= T+1, \quad
    a_y := \frac{J_{y}-1}{J_{y}} \colon y \in \{\mathrm{in},\mathrm{out}\}, \quad a = \frac{a_\mathrm{in}+ a_\mathrm{out}}{2}, \quad 
    x_0 := \sin^{-1}\sqrt{\epsilon}.
\end{align}
Let $\vartheta^{*}$ and $\vartheta_{1}^{*}$ denote the two smallest positive solutions of the characteristic equation defined by \eqref{eqn:genjinjout}, then
as $n\to\infty$,
\begin{align}
    \vartheta^{*}
    & =  \frac{\sin^{-1}\sqrt{\epsilon}}{n} + \frac{a\sin^{-1}\sqrt\epsilon}{2n^2} + O(n^{-3}) \\
    \vartheta_{1}^{*} &= \frac{\pi-\sin^{-1}\sqrt{\epsilon}}{n} + \frac{a(\pi-\sin^{-1}\sqrt{\epsilon})}{2n^2} +O(n^{-3}).
\end{align}
Consequently, for $E_{\rm clamped}(\vartheta)=1-\cos\vartheta$, the spectral gap between the two smallest positive energy levels satisfies
\begin{align}
    \Delta
    &:=
    E(\vartheta_{1}^{*})-E(\vartheta^{*}) = \frac{\pi(\pi-2x_0)}{2n^2} + \frac{a\pi(\pi-2x_0)}{2n^3} + O(n^{-4}).
\end{align}
\end{theorem}

\begin{proof}
     The secular equation is given by \eqref{eq:secGenJproduct}. The secular equation does not have a factor that can impose the scaling with respect to $n$ of a perturbation of a limit solution. Therefore, we introduce a scale parameter $\lambda$ which allows to interpolate between a secular equation we can solve in the limit $\lambda=0$ and the target secular equation at $\lambda={1\over n}$. Specifically, take $x:= n\vartheta$ and define
\begin{align}
    Q(\lambda,x):=q_{\rm in}(\lambda,x)q_{\rm out}(\lambda,x)-(1-\epsilon),
\end{align}
where
\begin{align}
    q_{Y}(\lambda,x)=\cos x+a_Y\tan\left(\frac{\lambda x}{2}\right)\sin x,\quad Y\in\{\mathrm{in},\mathrm{out}\}.
\end{align}
The smallest positive solution of $Q(0,x)=0$ is $x_{0}:=\sin^{-1}\sqrt{\epsilon}$, and we seek the smallest positive solution of $Q\left( {1\over n} , x\right)=0$.
$Q$ is smooth in a neighborhood of $(0,x_0)$ and one finds that $Q(0,x_{0}) = 0$ and 
\begin{align}
    \frac{\partial Q}{\partial x}\bigg|_{(0,x_0)} = -2\cos(x_0)\sin(x_0) = -2\sqrt{\epsilon(1-\epsilon)} \neq 0,
\end{align}
for $\epsilon \in (0,1)$. Hence, by the implicit function theorem, there is a unique smooth function $x(\lambda)$ such that $x(0)=x_{0}$ and $Q(\lambda,x(\lambda))=0$ on a neighborhood of $0$. For large $n$, this neighborhood contains $\lambda = {1\over n}$, and smoothness of $x(\lambda)$ gives that $x({1\over n}) = x_{0}+{1\over n}x'(0) + O(n^{-2})$ where
\begin{align}
    x'(0)&= -{\partial_{\lambda}Q \over \partial_{x}Q} \big\vert_{(0,x_{0})} \nonumber \\
    &= {x_{0}(a_{\text{in}}+a_{\text{out}})\over 4}.
\end{align}
Therefore,
\begin{align}
    \vartheta = \frac{x({1\over n})}{n} = \frac{x_0}{n} + \frac{ax_0}{2n^2} + O(n^{-3}) = \frac{\sin^{-1}(\sqrt{\epsilon})}{n} + \frac{a\sin^{-1}(\sqrt\epsilon)}{2n^2} + O(n^{-3}).
\end{align}
The function $x(\lambda)$ is the smallest positive solution of $Q(\lambda,x)=0$ for sufficiently small $\lambda$. Indeed, $x(\lambda)$ gives the unique solution for an interval of $x$ centered at $x_{0}$. Call the left boundary of this interval $\delta(\lambda)$, and note that $Q(0, x)$ is also bounded away from zero on $[0,\delta(\lambda)]$. Since $Q(\lambda,x)\rightarrow Q(0,x)$ uniformly, just take $\lambda$ small enough that $Q(\lambda,x)$ is bounded away from 0 on $[0,\delta(\lambda)]$.

For the next root, we start from the next-smallest solution of $Q(0,x)=0$, namely $x_1 = \pi - x_0$, and we want the next-smallest solution of $Q(\frac{1}{n},x) = 0$. Since $Q$ is smooth in the neighborhood of $(0,x_1)$ and $Q(0,x_1) = 0$ and $\partial_x Q|_{(0,x_1)} = 2\sqrt{\epsilon(1-\epsilon)} \neq 0$ for $\epsilon\in (0,1)$, the implicit function theorem guarantees that for sufficiently large $n$, there is a unique solution $x_{1}({1\over n})$ of $Q(\frac{1}{n},x)=0$, where $x_{1}(\lambda)$ is the unique function mapping a neighborhood of $0$ containing ${1\over n}$ to a neighborhood of $x_{1}$, with $x_{1}(0)=x_{1}$.  The expansion
$x_1(\frac{1}{n}) = x_1 + \frac{1}{n}x_1'(0) + O(n^{-2})$ holds, where
\begin{align}
    x_1'(0) = -\frac{\partial_\lambda Q}{\partial_x Q}\bigg|_{(0,x_1)} = \frac{x_1 (a_\mathrm{in}+a_\mathrm{out})}{4},
\end{align}
and therefore
\begin{align}
    \vartheta_1 = \frac{x_1(\frac{1}{n})}{n} = \frac{\pi-x_0}{n} + \frac{a(\pi-x_0)}{2n^2} +O(n^{-3}).
\end{align}
$x_1(\frac{1}{n})$ is the next-smallest solution after $x({1\over n})$ in the previous paragraph by the same uniform convergence argument.

Since the corresponding energies are $ E(\vartheta^{*})=1-\cos\vartheta^{*},$
then the large-\(T\) spectral gap is
\begin{align}
       \Delta &:= E(\vartheta_{1}^{*})-E(\vartheta^{*}) = \frac{\pi(\pi-2x_0)}{2n^2} + \frac{a\pi(\pi-2x_0)}{2n^3} + O(n^{-4}) \nonumber \\
    & = \frac{\pi\left(\pi-2\sin^{-1}\sqrt{\epsilon}\right)} {2(T+1)^2}
    \left[1+\left(\frac{J_\mathrm{in}-1}{J_\mathrm{in}}+ \frac{J_\mathrm{out}-1}{J_\mathrm{out}}\right)\frac{1}{2(T+1)}\right]+O(T^{-4}).
\end{align}
\end{proof}

\subsection{Analysis of the clamped overlap cost function $C_{\rm clamped}$}

We start by deriving the $\vartheta^{*}$ dependence of the ground state of $H_{\rm clamped}$.
\begin{theorem}
\label{thm:clampedgs}
    The ground state of $H_{\rm clamped}$ is a $CS$-valued function of the circuit parameters $\theta$ through its dependence on $\epsilon(\theta)$, and is given by
    \begin{align}
    \ket{\Psi_{\rm clamped}}
    &\propto -\sqrt{\epsilon}\sum_{t=0}^{T}  \cos\left( (t+{1\over 2})\vartheta^{*} \right)  \ket{\gamma_{t}} +\sqrt{1-\epsilon}\sum_{t=0}^{T} \cos\left( (t+{1\over 2})\vartheta^{*} \right)\ket{\eta_{t}} \nonumber \\
    &+ {J_{\rm in}\sqrt{1-\epsilon}\cot\left( {\vartheta^{*}\over 2} \right) \over 1-J_{\rm in}\left(1+\cot\left( {\vartheta^{*}\over 2} \right)\cot\left( (T+1)\vartheta^{*} \right) \right) } \sum_{t=0}^{T} {\cos\left( (T-t+{1\over 2})\vartheta^{*} \right)\over \sin \left( (T+1)\vartheta^{*} \right) }\ket{\eta_{t}} .\label{eq:GS_general_Jin_Jout}
\end{align}
\end{theorem}
\begin{proof}
    The ground state of $H_{\clamped}$ is obtained from (\ref{eqn:r2p}) using the same identifications as (\ref{eqn:idents})
\begin{align}
    a_{2} &= J_{\rm in} \nonumber \\
    \ket{\xi_{2}}&=\ket{\eta_{0}} \nonumber \\
    a_{1} &= J_{\final} \nonumber \\
    \ket{\xi_{1}}&=\sqrt{\epsilon}\ket{\gamma_{T}} - \sqrt{1-\epsilon}\ket{\eta_{T}}.
\end{align}
The resolvent matrix elements $r_{ij}:= \langle \xi_{i}\vert R_{0}(\lambda)\vert \xi_{j}\rangle$ are derived in the proof of Theorem \ref{thm:clampedsec}. We will again utilize the matrix $C$ in Lemma \ref{lem:clem}
which has eigenvalues $\cos\left( {k\pi \over T+1}\right)$, and will again use the ``backwards'' notation for its resolvent $R_{C}(\cos\vartheta) := (\cos \vartheta - C)^{-1}$. 

First note that due to the basis ordering $\ket{\gamma_{0}},\ldots,\ket{\gamma_{T}},\ket{\eta_{T}},\ldots,\ket{\eta_{0}}$, the unperturbed eigenvectors in (\ref{eqn:res0}) are
\begin{align}
    \ket{\Psi_{k}} &= \sum_{t=0}^{T}h_{kt}\ket{\gamma_{t}} \nonumber \\
    \ket{\Psi_{k}^{\perp}} &= \sum_{t=0}^{T}h_{kt}\ket{\eta_{T-t}}.
\end{align}
One obtains that
\begin{align}
    R_{0}\ket{\xi_{2}}&=\sum_{k=0}^{T}{h_{kT}\over \omega_{k}-\lambda} \ket{\Psi^{\perp}_{k}} \label{eqn:r0xi2} \\
   \bra{\eta_{t}}R_{0}\ket{\xi_{2}} &= \sum_{k=0}^{T}{h_{k\, T}h_{k\, T-t}\over \omega_{k}-\lambda} \nonumber \\
   &= R_{C}(\cos\vartheta)_{T-t,T} \nonumber \\
   &= {-\cos\left( (T-t+{1\over 2})\vartheta \right)\over \sin\left( {\vartheta\over 2} \right)\sin \left( (T+1)\vartheta \right) }.
\end{align}
We eliminate the $c_{1}$, $c_{2}$ appearing in (\ref{eqn:r2p}) by using (\ref{eq:c})
\begin{equation}
    \ket{\Psi^{\clamped}}\propto R_{0}\left(   \ket{\xi_{1}} + A\ket{\xi_{2}} \right)
    \label{eqn:gengs}
\end{equation}
with $A$ having dependence on $J_{\text{in}}$ so that $\epsilon$ does not appear in the denominator
\begin{align}
    A&= -{a_{2}r_{12} \over 1+a_{2}r_{22}} \nonumber \\
    &= - {J_{\text{in}}r_{12}\over 1+J_{\text{in}}r_{22}} \nonumber \\
    &= -{J_{\text{in}}\sqrt{1-\epsilon}\cot\left( {\vartheta\over 2} \right) \over \sin\left((T+1)\vartheta\right)\left(1-J_{\text{in}}\left(1+\cot\left( {\vartheta\over 2} \right)\cot\left( (T+1)\vartheta \right) \right) \right)}.
\end{align}
So,
\begin{equation}
    AR_{0}\ket{\xi_{2}} = {J_{\text{in}}\sqrt{1-\epsilon}\cot\left( {\vartheta\over 2} \right) \over \sin\left((T+1)\vartheta\right)\left(1-J_{\text{in}}\left(1+\cot\left( {\vartheta\over 2} \right)\cot\left( (T+1)\vartheta \right) \right) \right)} \sum_{t=0}^{T} {\cos\left( (T-t+{1\over 2})\vartheta \right)\over \sin\left( {\vartheta\over 2} \right)\sin \left( (T+1)\vartheta \right) }\ket{\eta_{t}}.
\end{equation}
Now we compute $R_{0}\ket{\xi_{1}}$ and write the final unnormalized ground state.
\begin{align}
    R_{0}\ket{\xi_{1}}&= R_{0}\left( \sqrt{\epsilon}\ket{\gamma_{T}}-\sqrt{1-\epsilon} \ket{\eta_{T}}\right) \label{eqn:r0xi1} \\
    &= \sqrt{\epsilon}\sum_{k=0}^{T} {h_{kT}\over \omega_{k}-\lambda} \ket{\Psi_{k}} - \sqrt{1-\epsilon}\sum_{k=0}^{T}{h_{k0}\over \omega_{k}-\lambda}\ket{\Psi^{\perp}_{k}}.
\end{align}
But we see that
\begin{align}
    \langle \gamma_{t}\vert R_{0}\ket{\xi_{1}} &= \sqrt{\epsilon}\sum_{k=0}^{T}{h_{kt}h_{kT}\over \omega_{k}-\lambda} \nonumber \\
    &= \sqrt{\epsilon}R_{C}(\cos\vartheta)_{t,T}\nonumber \\
    &= -\sqrt{\epsilon}{ \cos\left( (t+{1\over 2})\vartheta \right)\over \sin\left( {\vartheta\over 2} \right)\sin \left( (T+1)\vartheta \right) } \\
    \langle \eta_{t}\vert R_{0}\ket{\xi_{1}} &= -\sqrt{1-\epsilon}\sum_{k=0}^{T}{h_{k0}h_{k\, T-t}\over \omega_{k}-\lambda} \nonumber \\
    &= -\sqrt{1-\epsilon}R_{C}(\cos\vartheta)_{T-t,0}\nonumber \\
    &= \sqrt{1-\epsilon}{ \cos\left( (t+{1\over 2})\vartheta \right)\over \sin\left( {\vartheta\over 2} \right)\sin \left( (T+1)\vartheta \right) }.
\end{align}
Now we are done
\begin{align}
    \ket{\Psi_{\clamped}}&\propto R_{0}(1-\cos\vartheta^{*})\left(   \ket{\xi_{1}} + A\ket{\xi_{2}} \right) 
\end{align}
which is equal to (\ref{eq:GS_general_Jin_Jout}).
\end{proof}

Applying Theorem \ref{thm:clampedgs} to the case of $J_{\rm out}=J_{\rm in}=1$ produces a ground state that is a superposition of rotated states in each $\text{span}\lbrace \ket{\gamma_{t}}, \ket{\eta_{t}}\rbrace$. Recall that the secular equation for this case was shown in (\ref{eqn:jjj1}), which can be rewritten $\cos^{2}((T+1)\vartheta) = 1-\epsilon$ or $\sin((T+1)\vartheta)=\sqrt{\epsilon}$. Using these in (\ref{eq:GS_general_Jin_Jout}) gives the ground state
\begin{align}
    \ket{\Psi_{\rm clamped}}&\propto -\sin((T+1)\vartheta^{*})\sum_{t=0}^{T}\cos\left( (t+{1\over 2})\vartheta^{*}\right)\ket{\gamma_{t}} \nonumber \\
    &+ \sum_{t=0}^{T}\left(  \cos((T+1)\vartheta^{*})\cos\left( (t+{1\over 2})\vartheta^{*}\right) - \cos \left( (T-t+{1\over 2})\vartheta^{*}\right)\right)\ket{\eta_{t}}\nonumber \\
    &\propto \sum_{t=0}^{T}\cos\left( (t+{1\over 2})\vartheta^{*}\right)\ket{\gamma_{t}} + \sin\left( (t+{1\over 2})\vartheta^{*}\right)\ket{\eta_{t}}.
    \label{eqn:specialclampedgs}
\end{align}
The normalization factor is ${1\over \sqrt{T+1}}$ and it becomes a matter of arithmetic to obtain the expression for $C_{\rm clamped}(\epsilon)$ shown in Theorem 1 of the main text.

We now consider the overlap cost function for $0<J_{\text{out}} \le J_{\text{in}}$. Consider the ground state \eqref{eq:GS_general_Jin_Jout}.
Let $c_t := \cos((t+\frac12)\vartheta)$ and 
\begin{align}
    b := {J_{\text{in}}\sqrt{1-\epsilon}\cot\left( {\vartheta\over 2} \right) \over \sin((T+1)\vartheta)\left[1-J_{\text{in}}\left(1+\cot\left( {\vartheta\over 2} \right)\cot\left( (T+1)\vartheta \right) \right) \right]} = 
    -\frac{\sqrt{1-\epsilon}}{\left(\frac{J_\mathrm{in}-1}{J_\mathrm{in}}\right)\tan(\frac{\vartheta}{2})\sin((T+1)\vartheta) +\cos((T+1)\vartheta)} \label{eq:bdef}
\end{align} The normalization of the ground state \eqref{eq:GS_general_Jin_Jout} is 
\begin{align}
    \mathcal{N} 
    & = \frac{(1+b^2)}{2}\left(T+1 + \frac{\sin(2(T+1)\vartheta)}{2\sin(\vartheta)}\right) + b\sqrt{1-\epsilon} \left[(T+1)\cos((T+1)\vartheta)+\frac{\sin((T+1)\vartheta)}{\sin(\vartheta)}\right] \label{eq:normalizationGeneralJ}.
\end{align}
With $n:=T+1$, this gives the overlap cost function as
\begin{align}
    C_{\rm clamped}(\epsilon,\vartheta^{*}) & = 1 - \frac{1}{\mathcal{N}}\frac{\epsilon}{n} \left(\sum_{t=0}^T \cos((t+\frac{1}{2})\vartheta^{*})\right)^2  = 1 - \frac{1}{\mathcal{N}}\frac{\epsilon}{n} \frac{\sin^2(n\vartheta^{*})}{4\sin^2({\vartheta^{*}\over 2})} \label{eq:OverlapCost_Hclampd_GenJ}\nonumber \\
        &= 1- {\epsilon \sin^2(n\vartheta^{*})\over 4n\sin^{2}\left( {\vartheta^{*}\over 2}\right)} {1\over {q_{\rm in}(\vartheta^{*})+q_{\rm out}(\vartheta^{*})\over 2q_{\rm in}(\vartheta^{*})}\left( n + {\sin(2n\vartheta^{*})\over 2\sin\vartheta^{*}} \right) -q_{\rm out}(\vartheta^{*})\left( n\cos(n\vartheta^{*}) + {\sin(n\vartheta^{*})\over \sin\vartheta^{*}} \right)}\nonumber \\
    &= 1-{\epsilon \sin^2(n\vartheta^{*})\over 4n\sin^{2}\left( {\vartheta^{*}\over 2}\right)}{1\over {q_{\rm in}(\vartheta^{*})+q_{\rm out}(\vartheta^{*})\over 2q_{\rm in}(\vartheta^{*})}\left( n + T_{n}U_{n-1} \right) - q_{\rm out}(\vartheta^{*})\left(nT_{n}+U_{n-1}\right)}
\end{align}
with $U_{n-1}={\sin(n\vartheta^{*})\over \vartheta}$, $T_{n}=\cos(n\vartheta^{*})$ the Chebyshev polynomials at $\vartheta^{*}$ and $q_{\rm in(out)}(\vartheta)$ coming from (\ref{eqn:qinqout}).

We now examine the derivative  ${dC_{\clamped}\over d\epsilon}$ in a left neighborhood of $\epsilon =1$ when $0<J_{\rm out}<J_{\rm in}$, which will reveal that $\epsilon$ and $C_{\rm clamped}$ have anti-parallel gradients in this neighborhood. At $\epsilon=1$ and $0<J_{\rm out}<J_{\rm in}$, the smallest positive root of the secular equation is the solution of $q_{\rm out}(\vartheta(1)) = 0$ because ${\cot(n\vartheta)\over \tan\left( {\vartheta\over 2}\right)}$ monotonically decreases from $\infty$ to $-\infty$ as $\vartheta$ varies from $0$ to $\pi\over n$. Considering $H_{\rm clamped}(s)$ as a function of $s:=\sqrt{1-\epsilon}$, one writes
\begin{equation}
    H_{\rm clamped}(s)=H_{\rm clamped}(0)+sV_{1} + s^{2}V_{2}+O(s^{3})
\end{equation}
with
\begin{align}
V_{1} &= -J_{\mathrm{out}}
\left(
|\gamma_T\rangle\langle\eta_T|
+ |\eta_T\rangle\langle\gamma_T|
\right), \\
V_{2} &= J_{\mathrm{out}}
\left(
|\eta_T\rangle\langle\eta_T|
- |\gamma_T\rangle\langle\gamma_T|
\right).
\end{align}
The map $H_{\rm clamped}(s)\mapsto H_{\rm clamped}(-s)$ is achieved by multiplying the $\ket{\eta_{t}}$ sector by $-1$, which does not affect $C_{\rm clamped}(\epsilon)=C_{\rm clamped}(1-s^{2})$. In the following theorem, we will show analyticity of the overlap cost function in a left neighborhood of $\epsilon=1$, and specifically that
\begin{equation}
    C_{\rm clamped}(1-s^{2}) = C_{\rm clamped}(1) + s^{2}K_{n}+O(s^{4})
    \label{eqn:nqnq}
\end{equation}
with $C_{\rm clamped}(1)$ determined from (\ref{eqn:overlapexpl}). Positivity of $K_{n}$ will then be used to conclude that $C_{\clamped}$ and $\epsilon$ have anti-parallel gradients near $\epsilon =1$ in the large-$T$ limit.
\begin{theorem}
    \label{thm:ggoo}
    Let $0<J_{\rm out}<J_{\rm in}$. For sufficiently large $n$, $K_{n}>0$ and there is a left neighborhood of $1$ such that ${dC_{\rm clamped}\over d\epsilon}<0$.
\end{theorem}

\begin{proof}
    Define $\delta = 1-\epsilon$  and take $d := a_{\mathrm{in}} - a_{\mathrm{out}}
= \frac{1}{J_{\mathrm{out}}}
- \frac{1}{J_{\mathrm{in}}} > 0$. With $q_{\rm in}$, $q_{\rm out}$ in (\ref{eqn:qinqout}) now considered as functions of $x(\delta):=n\vartheta(\delta)$, one has $q_{\rm out}(\overline{x})=0$ where $\overline{x}:= x(0)$, and 
    \begin{equation}
        {dx\over d\delta}\big\vert_{\delta=0} = \frac{1}{p_n q'_{\mathrm{out}}(\bar{x})}
    \end{equation}
    where
    \begin{align}
        p_{n}&:= q_{\rm in}(\overline{x})\nonumber \\
        &= d \tan\left(\frac{\bar{x}}{2n}\right)\sin\bar{x} > 0.
    \end{align}
    Expanding the solution $\overline{x}$ about the $n\rightarrow \infty$ solution $\pi\over 2$, one finds that
    \begin{equation}
        \overline{x}={\pi\over 2}+ \frac{a_{\mathrm{out}}\pi}{4n} + O(n^{-2})
    \end{equation}
    and
    \begin{align}
p_n &= \frac{\pi d}{4n}
\left(1 + \frac{a_{\mathrm{out}}}{2n} + O(n^{-2})\right), \\
q'_{\mathrm{out}}(\bar{x})
&= -1 + \frac{a_{\mathrm{out}}}{2n} + O(n^{-2}), \\
{dx\over d\delta}\big\vert_{\delta =0}
&= -\frac{4n}{\pi d} + O(1).
\end{align}
    In the rescaled variable, we again use
    $q_Y(x)=\cos x+a_Y\tan\left(\frac{x}{2n}\right)\sin x$, $Y\in\{\mathrm{in},\mathrm{out}\}$ 
    with primes denoting differentiation with respect to $x$.
    Define
    \begin{align}
        c_t(x)&:=\cos\left(\frac{t+\frac12}{n}x\right),
        \qquad t=0,\ldots,n-1, \\
        M_n(x)&:=\sum_{t=0}^{n-1}c_t(x)
        =\frac{\sin x}{2\sin({x\over 2n})}, \\
        S_n(x)&:=\sum_{t=0}^{n-1}c_t(x)^2
        =\frac{n}{2}+\frac{\sin(2x)}{4\sin({x\over n})}, \\
        R_n(x)&:=\sum_{t=0}^{n-1}c_t(x)c_{n-1-t}(x)
        =\frac12\left(n\cos x+\frac{\sin x}{\sin({x\over n})}\right).
        \label{eqn:sumssums}
    \end{align}
    For $x=x(\delta)$, the unnormalized ground state was found in Theorem \ref{thm:clampedgs} to be
    \begin{align}
        |\widetilde{\Psi}_{\delta}\rangle
        ={}&-\sqrt{1-\delta}
        \sum_{t=0}^{n-1}c_t(x)|\gamma_t\rangle +\sqrt{\delta}\sum_{t=0}^{n-1}
        \left(
            c_t(x)-\frac{c_{n-1-t}(x)}{q_{\rm in}(x)}
        \right)|\eta_t\rangle.
    \end{align}
    Orthogonality of the $\gamma$ and $\eta$ sectors gives $\ket{\Psi_{\rm clamped}}={ \ket{\tilde{\Psi}_{\delta}}\over \sqrt{\mathcal{N}_{n}(\delta)}}$ where
    \begin{align}
        \mathcal N_n(\delta)
        &:=\langle\widetilde{\Psi}_{\delta}
        |\widetilde{\Psi}_{\delta}\rangle \nonumber\\
        &=(1-\delta)S_n(x)
        +\delta\left(
            S_n(x)-\frac{2R_n(x)}{q_{\rm in}(x)}
            +\frac{S_n(x)}{q_{\rm in}(x)^2}
        \right) \nonumber\\
        &=S_n(x)\left(1+\frac{\delta}{q_{\rm in}(x)^2}\right)
        -\frac{2\delta R_n(x)}{q_{\rm in}(x)}.
    \end{align}
    Since
    \[
        |\Psi_{\rm free}\rangle
        =\frac{1}{\sqrt n}\sum_{t=0}^{n-1}|\gamma_t\rangle,
    \]
    we also have
    \[
        \langle\Psi_{\rm free}|\widetilde{\Psi}_{\delta}\rangle
        =-\sqrt{\frac{1-\delta}{n}}\,M_n(x).
    \]
    Consequently, defining
    \[
        \mathcal F_n(x):=\frac{M_n(x)^2}{nS_n(x)},
    \]
    the overlap cost function satisfies
    \begin{equation}
        1-C_{\rm clamped}(1-\delta)
        =
        \frac{(1-\delta)\mathcal F_n(x)}
        {1+\delta\left[
            \frac{1}{q_{\rm in}(x)^2}
            -\frac{2R_n(x)}{S_n(x)q_{\rm in}(x)}
        \right]},
        \qquad x=x(\delta).
        \label{eqn:cdelt}
    \end{equation}

    At $\delta=0$, one has that $q_{\rm in}(\bar{x})=p_n>0$ and
    $p_nq'_{\rm out}(\bar{x})\ne0$. Thus the implicit
    function theorem applied to
    $q_{\rm in}(x)q_{\rm out}(x)=\delta$ shows that $x(\delta)$
    is analytic near zero. This fact, along with analyticity of the functions in (\ref{eqn:sumssums}) and positivity of the denominator in (\ref{eqn:cdelt}), allows one to infer analyticity of $C_{\rm clamped}(1-\delta)$. 
    It follows that
    \begin{align}
    C_{\rm clamped}(1-\delta)&=C_{\rm clamped}(1) +  K_{n}\delta+O(\delta^{2})\nonumber \\
        \frac{dC_{\rm clamped}}{d\epsilon}
        &=-K_n+O(1-\epsilon).
        \label{eqn:dderiv}
    \end{align}
    Differentiating at $\delta=0$ gives
    \begin{align}
        K_n
        ={}&\mathcal F_n(\bar{x})
        \left[
            1+\frac{1}{p_n^2}
            -\frac{2R_n(\bar{x})}{S_n(\bar{x})p_n}
        \right]-\frac{\mathcal F_n'(\bar{x})}
        {p_nq'_{\rm out}(\bar{x})}.
        \label{eqn:kn}
    \end{align}
    Hence $K_n>0$ implies that the derivative (\ref{eqn:dderiv}) is negative in
    a left neighborhood of $\epsilon=1$. It remains
    to establish positivity of $K_n$ for sufficiently large $n$.

    All large-$n$ estimates below are taken with the coupling
    constants fixed. The trigonometric sums above imply
    \begin{equation}
        \mathcal F_n(x)
        =
        \frac{\left({\sin x\over x}\right)^2}
        {\frac12+\frac{\sin(2x)}{4x}}
        +O(n^{-2}),
    \end{equation}
    uniformly for $x$ in a fixed neighborhood of $\pi/2$.
    Since $\bar{x}=x(0)=\pi/2+O(n^{-1})$, it follows that
    \begin{align}
        \mathcal F_n(\bar{x})
        &=\frac{8}{\pi^2}+O(n^{-1}), \\
        \mathcal F_n'(\bar{x})&=O(1), \\
        \frac{R_n(\bar{x})}{S_n(\bar{x})}
        &=\frac{2}{\pi}+O(n^{-1}).
    \end{align}
    Together with the previously obtained estimates,
    \[
        p_n^{-1}=O(n),
        \qquad
        p_n^{-2}=\frac{16n^2}{\pi^2d^2}+O(n),
        \qquad
        \frac{1}{p_nq'_{\rm out}(\bar{x})}=O(n),
    \]
    these show that the only contribution of order $n^2$
    to $K_n$ in (\ref{eqn:kn}) is $\mathcal F_n(\bar{x})/p_n^2$.
    The $n^{2}$ scaling establishes eventual positivity of $K_{n}$. Specifically,
    \begin{align}
        K_n
        &=\frac{\mathcal F_n(\bar{x})}{p_n^2}+O(n) \nonumber\\
        &=\left(\frac{8}{\pi^2}+O(n^{-1})\right)
        \left(\frac{16n^2}{\pi^2d^2}+O(n)\right)+O(n)
        \nonumber\\
        &=\frac{128}{\pi^4d^2}\,n^2+O(n).
    \end{align}
\end{proof}

The existence of a critical value of $\epsilon$ above which the gradients of $\epsilon(\theta)$ and $C_{\rm clamped}(\theta)$ are antiparallel stands in contrast to the statement of parallel gradients for the $J_{\rm out}=J_{\rm in}=1$ setting shown in Theorem 1 of the main text. The fact that the gradients are parallel for that particular value of the dynamical parameters is easily verified by computing ${dC_{\rm clamped}\over d\epsilon}$. It turns out that the gradients are parallel for all $J_{\rm out}=J_{\rm in}=J\ge 0$, and here we report this much richer technical result.
\begin{lemma}
\label{lem:llll}
    For $J_\mathrm{in} = J_\mathrm{out}=: J$, the overlap cost function is given by \begin{align}
    C_{\rm clamped}(\vartheta) 
    & =  1 - \frac{1}{T+1} 
    \frac{\left[\sin((T+1)\vartheta)(1-a^2\tan^2(\frac{\vartheta}{2})) - 2a\tan(\frac{\vartheta}{2})\cos{((T+1)\vartheta)}\right]}{4\left[(T+1)\sin((T+1)\vartheta) - a \tan(\frac{\vartheta}{2})\left((T+1)\cos((T+1)\vartheta) + \frac{\sin((T+1)\vartheta)}{\sin(\vartheta)}\right) \right]}
    \frac{\sin^2((T+1)\vartheta)}{\sin^2(\frac{\vartheta}{2})}. \label{eq:JinEqualJoutCostFunctionOnevariable}
\end{align}
\end{lemma}
\begin{proof}
When $J_\mathrm{in} = J_\mathrm{out}=: J$, we use the secular equation \eqref{eqn:genjsecular} to simplify the general ground state \eqref{eq:GS_general_Jin_Jout} to
\begin{align}
    \ket{\Psi_\clamped} &\propto  -\sqrt{\epsilon}\sum_{t=0}^{T}\cos \left( (t+{1\over 2}) \vartheta \right) \ket{\gamma_{t}} + \sqrt{1-\epsilon} \sum_{t=0}^{T}\cos\left( (t+{1\over 2})\vartheta\right)\ket{\eta_{t}} 
    - \sum_{t=0}^{T}{\cos\left( (T-t+{1\over 2})\vartheta \right) }\ket{\eta_{t}}.
    \label{eqn:jstate}
\end{align}
    Further, taking $J_\mathrm{in} = J_\mathrm{out}=:J$ in the secular equation (\ref{eqn:genjsecular}) gives the simplified equation
\begin{align}
    1-J\left(1 + \cot\left( {\vartheta\over 2} \right)\cot\left( (T+1)\vartheta \right) \right) &= J\sqrt{1-\epsilon}{\cot\left( {\vartheta\over 2} \right) \over \sin\left( (T+1)\vartheta \right) } \sigma \label{eqn:huhu0} \\
    {\sin\left( {\vartheta\over 2} \right) \sin\left( (T+1)\vartheta \right)\over \cos\left( {\vartheta\over 2} \right) }\left(1-J\right) - J\cos\left((T+1)\vartheta\right)&= J\sigma \sqrt{1-\epsilon}
    \label{eqn:huhu}
\end{align}
where $\sigma = \pm 1$. But the left hand side of (\ref{eqn:huhu}) is continuous on $[0,{\pi \over T+1}]$ on which it varies from $[-J,J]$. Therefore, the smallest positive solution occurs when taking $\sigma=-1$. Using (\ref{eqn:huhu0}) with $\sigma=-1$ in (\ref{eq:GS_general_Jin_Jout}) gives (\ref{eqn:jstate}).  By using the fact that the variables $a_{\text{in}}, a_{\text{out}}$ from Theorem \ref{thm:gap_gen_j_in_j_out} satisfy $a_{\text{in}}=a_{\text{out}}=a = {J-1\over J}$ in the present case, we further simplify the secular equation to $q(\vartheta)= \sqrt{1-\epsilon}$ where
\begin{align}
    q(\vartheta) &:= \cos((T+1)\vartheta) + a\tan(\frac{\vartheta}{2})\sin((T+1)\vartheta) \label{eqn:secsimple} 
\end{align}

We compute the normalization of the above state (\ref{eqn:jstate}) as follows. Let $c_t = \cos((t+\frac12)\vartheta)$. Then 
\begin{align}
   \mathcal{N} & = \epsilon\sum_{t}c_t^2 + \sum_t \left(\sqrt{1-\epsilon}c_t - c_{T-t}\right)^2 \nonumber \\
    & = 2\sum_t c_t^2 - 2\sqrt{1-\epsilon}\sum_t c_t c_{T-t} \nonumber \\
    & = T+1 + \frac{\sin(2(T+1)\vartheta)}{2\sin(\vartheta)} - \sqrt{1-\epsilon} \left[(T+1)\cos((T+1)\vartheta)+\frac{\sin((T+1)\vartheta)}{\sin(\vartheta)}\right] \label{eq:normalizationGSEqualJ}
\end{align}
where we used $\sum_t c_t^2 = \sum_t c_{T-t}^2$ and some basic trigonometric simplifications.  This gives the overlap cost function as
\begin{align}
    C_{\rm clamped}(\epsilon,\vartheta) & = 1 - \frac{1}{\mathcal{N}}\frac{\epsilon}{T+1} \left(\sum_{t=0}^T \cos((t+\frac{1}{2})\vartheta)\right)^2  = 1 - \frac{1}{\mathcal{N}}\frac{\epsilon}{T+1} \frac{\sin^2((T+1)\vartheta)}{4\sin^2({\vartheta\over 2})} \label{eq:OverlapCost_Hclampd_JinJoutequal}
\end{align}
Using the secular equation (\ref{eqn:secsimple}) gives
\begin{align}
    \mathcal{N} & = \sin((T+1)\vartheta)\left[(T+1)\sin((T+1)\vartheta) - a \tan(\frac{\vartheta}{2})\left((T+1)\cos((T+1)\vartheta) + \frac{\sin((T+1)\vartheta)}{\sin(\vartheta)}\right) \right] \nonumber \\
    & = -\sin((T+1)\vartheta) q'(\vartheta) 
    \end{align}
    which gives the overlap cost function
    \begin{align}
    C(\vartheta) 
    & = 1-\frac{1}{T+1}\frac{q^2(\vartheta)-1}{q'(\vartheta)} \frac{\sin((T+1)\vartheta)}{4\sin^2(\frac{\vartheta}{2})} 
\end{align}
which is equal to (\ref{eq:JinEqualJoutCostFunctionOnevariable}).
\end{proof}

We can decompose the cost function for general $a:={J-1\over J}$ as 
$ C(\vartheta,a) = 1- A(\vartheta)B(\vartheta)$, where
\begin{align}
    A(\vartheta) &= \left(\frac{1}{\sin(\frac{\vartheta}{2})}\right)\frac{\left(1-\left(\cos(n\vartheta)+a\tan\left(\frac{\vartheta}{2}\right)\sin(n\vartheta)\right)^{2}\right)}{n\sin(n\vartheta)-a\tan(\frac{\vartheta}{2})\left(n\cos(n\vartheta)+\frac{\sin(n\vartheta)}{\sin(\vartheta)}\right)} \nonumber \\
    B(\vartheta) &= \left(\frac{\sin\left(n\vartheta\right)}{4n\sin\left(\frac{\vartheta}{2}\right)}\right).
\end{align}
However, we must analyze the situations of $0\le a \le 1$ and $-\infty < a < 0$ separately in Lemma \ref{lem:GradientsOverlapClampedJin=JoutPosa} and Lemma \ref{lem:GradientsOverlapClampedJin=JoutNega} respectively, as the parts in this factorization have different properties in each case.
Note that $B(\vartheta) > 0$ and $B'(\vartheta) < 0$ for $\vartheta \in (0,\pi/n)$. As for $A(\vartheta,a)$, note that
\begin{align}
    A(\vartheta) = \frac{q^2(\vartheta) -1 }{q'(\vartheta)} \frac{1}{\sin(\frac{\vartheta}{2})} = \frac{1-q^2(\vartheta)}{-q'(\vartheta)} \frac{1}{\sin(\frac{\vartheta}{2})} .
\end{align}
For $0\leq a \leq 1$, we have $q(0) = 1$ and $q'(\vartheta) < 0$, which implies $1-q^2(\vartheta) < 0$ and hence $A(\vartheta) > 0$ for $\vartheta \in (0,\pi/n)$.

\begin{lemma}\label{lem:GradientsOverlapClampedJin=JoutPosa}
For $0\leq a \leq 1$
    \begin{align}
        \frac{d A(\vartheta)}{d \vartheta} < 0 
    \end{align}
    where $\vartheta\in (0,\pi/n)$. 
    \label{lem:lem0a1}
\end{lemma}
\begin{proof}
Introduce the variables
\begin{align}
    t:=\tan\left(\frac{\vartheta}{2}\right),
\qquad
r:=\tan\left(\frac{n\vartheta}{2}\right),
\qquad
m:=\frac{r}{t}.
\end{align}
Since $0<\vartheta<\pi/n$, we have $0<n\vartheta/2<\pi/2$. Moreover, because
\[
y\mapsto \frac{\tan y}{y}
\]
is strictly increasing on $(0,\pi/2)$, we have
\[
m=\frac{\tan(n\vartheta/2)}{\tan(\vartheta/2)}>n.
\]
In particular, $m>n\ge 2>a.$
Using the half-angle identities
\[
\sin(n\vartheta)=\frac{2r}{1+r^2},
\qquad
\cos(n\vartheta)=\frac{1-r^2}{1+r^2},
\]
we obtain
\[
q(\vartheta) = \cos(n\vartheta)+a\tan\left(\frac{\vartheta}{2}\right)\sin(n\vartheta)
=
\frac{1-r^2+2atr}{1+r^2}.
\]
Therefore
\begin{align}
1-q(\vartheta)^2
&=
1-
\left(
\frac{1-r^2+2atr}{1+r^2}
\right)^2 \nonumber \\
&=
\frac{(1+r^2)^2-(1-r^2+2atr)^2}{(1+r^2)^2} \nonumber \\
&=
\frac{
\left(1+r^2-1+r^2-2atr\right)
\left(1+r^2+1-r^2+2atr\right)
}{
(1+r^2)^2
} \nonumber \\
&=
\frac{4r(r-at)(1+atr)}{(1+r^2)^2}.
\end{align}
Using $\sin(\vartheta) = \frac{2t}{1+t^2}$
\begin{align}
D(\vartheta)
:= -q'(\vartheta)& =
n\sin(n\vartheta)
-a\tan\left(\frac{\vartheta}{2}\right)
\left(
n\cos(n\vartheta)+\frac{\sin(n\vartheta)}{\sin(\vartheta)}
\right) \nonumber \\
&=
\frac{2nr}{1+r^2}
-at
\left(
n\frac{1-r^2}{1+r^2}
+
\frac{r(1+t^2)}{t(1+r^2)}
\right) \nonumber \\
&=
\frac{
2nr-ant(1-r^2)-ar(1+t^2)
}{
1+r^2
}.
\end{align}
Substituting $r=mt$ gives $D(\vartheta) 
=
\frac{tK}{1+m^2t^2},$
where
\begin{align}
    K
:=
2mn-a(m+n)+am(mn-1)t^2.
\end{align}

Since
\[
\sin(\vartheta/2)=\frac{t}{\sqrt{1+t^2}},
\]
we obtain
\begin{align}
A(\vartheta)
&=
\frac{
4r(r-at)(1+atr)
}{
(1+r^2)^2
}
\frac{1+r^2}{tK}
\frac{\sqrt{1+t^2}}{t} \nonumber \\
&=
\frac{
4m(m-a)(1+amt^2)\sqrt{1+t^2}
}{
(1+m^2t^2)K
}.
\end{align}

We now differentiate this expression. The variables $t$ and $m$ both depend on $\vartheta$. We have
\[
\frac{dt}{d\vartheta}
=
\frac{1+t^2}{2},
\quad 
\frac{dr}{d\vartheta}
=
\frac{n(1+r^2)}{2}
=
\frac{n(1+m^2t^2)}{2}.
\]
Therefore
\[
\frac{dm}{d\vartheta}
=
\frac{1}{t}\frac{dr}{d\vartheta}
-
\frac{m}{t}\frac{dt}{d\vartheta}
=
\frac{
n(1+m^2t^2)-m(1+t^2)
}{
2t
}.
\]
We compute the derivative by logarithmic differentiation. 
 \begin{align} 
 \frac{A'(\vartheta)}{A(\vartheta)} &= \frac{m'}{m} + \frac{m'}{m-a} + \frac{a t^2 m' + 2amt t'}{1+amt^2} + \frac{t t'}{1+t^2} - \frac{2m t^2 m' + 2m^2 t t'}{1+m^2t^2} - \frac{K_m m' + K_t t'}{K}, \label{eqn:alogder}\end{align} 
 where 
 \[ t'=\frac{dt}{d\vartheta} = \frac{1+t^2}{2}, \qquad m'=\frac{dm}{d\vartheta} = \frac{ n(1+m^2t^2)-m(1+t^2) }{ 2t }, \] 
 and 
 \[ K_m = \frac{\partial K}{\partial m} = 2n-a+a(2nm-1)t^2, \] 
 \[ K_t = \frac{\partial K}{\partial t} = 2am(nm-1)t. \] 
 Note that the fourth and fifth terms of (\ref{eqn:alogder}) are polynomials in $t$, e.g., the fifth term being equal to $-mnt$, and that the first three terms and the sixth term can be converted to polynomials in $t$ by multiplying (\ref{eqn:alogder}) by $Ktm(m-a)(1+amt^{2})$. The factor of $t$ comes from the fact that $m'$ has an $O(t^{-1})$ term. It follows that multiplying both sides of (\ref{eqn:alogder}) by $A(\vartheta)$ gives
 \begin{equation}
     A'(\vartheta)={4\sqrt{1+t^{2}} \over (1+m^{2}t^{2})K^{2}t}\mathcal{Q}_{a,n}(t,m)
 \end{equation}
 where the coefficients of $t^{2j}$, $j=0,1,2,3$, are obtained by computer algebra system, and given by ${-P_{j}\over 2}$ with the $P_{j}$ given below. Substituting the $K_{m}$ and $K_{t}$ expressions into the logarithmic derivative, multiplying by $A$, and clearing denominators gives 
 \begin{align} 
    A'(\vartheta) = - \frac{ 2\sqrt{1+t^2} }{ t^4(1+m^2t^2)K^2 } \mathcal P_{a,n}(t,m), \label{eq:derivativeA}
 \end{align} 
 where \[ \mathcal P_{a,n}(t,m) = t^3\left(P_0+P_1t^2+P_2t^4+P_3t^6\right). \] 
 The coefficients are as follows. First, 
 \[ P_0 = (m-n)\left(a^2n-am^2-2amn+2m^2n\right). \] 
 Next, 
 \begin{align} 
    P_1 &= m\Big[ -a^3m^2-a^3mn-2a^3n^2 +4a^2m^2n+5a^2mn^2+a^2m \nonumber \\ &\qquad +am^3n-8am^2n^2-3am^2+2amn +2m^3n^2 \Big]. \nonumber \\
    & = m \Big[-(a^3 (m^2+m n+2 n^2))+a^2 m (4 m n+5 n^2+1)+a m (m^2n-8 mn^2-3m+2 n)+2 m^3 n^2\Big]
 \end{align} 
 Next, 
 \begin{align} 
    P_2 &= am^2\Big[ a^2m^2n-a^2m-2a^2n +am^3n-5am^2n^2-am^2 +8amn+a+2m^3n^2-2m^2n-2m \Big].  \\
     P_3 &= a^2m^4(mn-1)(mn+1-2an).
\end{align} 
For the positivity argument it is useful to rewrite these coefficients in terms of \[ u:=m-n > 0. \]
In terms of $u$, 
\begin{align}
P_0 = u\left[ (2n-a)u^2 + 4n(n-a)u + n(2n^2-3an+a^2) \right] =   u\left[ (2n-a)u^2 + 4n(n-a)u + n(2n-a)(n-a) \right]  
\end{align} Each term in brackets is positive because $n\ge 2$ and $0<a<1$. Hence
\[
P_0>0.
\]
Next, \[ P_1 = m(A_3u^3+A_2u^2+A_1u+A_0), \] where 
\begin{align}
    A_3 & =an+2n^2 > 0, \\
    A_2 &= -a^3+4a^2n-5an^2-3a+6n^3,\\
    A_1 &= -3a^3n+13a^2n^2+a^2-13an^3-4an+6n^4,\\
    A_0 &= -4a^3n^2+9a^2n^3+a^2n-7an^4-an^2+2n^5.
\end{align} 

For $A_2$, we use $0<a<1$ and $n \geq 2$ to get
\begin{align}
    A_2
>
6n^3-5n^2-4 = n^2(6n-5) - 4 \geq 7n^2-4 >0.
\end{align}
For $A_1$, note that
\[
\frac{dA_1}{da}
=
-9a^2n+26an^2+2a-13n^3-4n.
\]
This derivative is increasing in $a$ on $(0,1)$, since
\[
\frac{d^2A_1}{da^2}
=
-18an+26n^2+2> 26n^2-18n>0.
\]
Thus
\[
\frac{dA_1}{da}
\le
\left.\frac{dA_1}{da}\right|_{a=1}
=
-13n^3+26n^2-13n+2
=
-13n(n-1)^2+2<0
\]
for $n\ge 2$. Hence $A_1$ is decreasing in $a$, and therefore
\[
A_1
>
A_1|_{a=1}
=
(n-1)(6n^3-7n^2+6n-1)>0.
\]

Similarly,
\[
\frac{dA_0}{da}
=
-12a^2n^2+18an^3+2an-7n^4-n^2.
\]
This derivative is increasing in $a$ on $(0,1)$, because
\[
\frac{d^2A_0}{da^2}
=
-24an^2+18n^3+2n>0.
\]
Thus
\[
\frac{dA_0}{da}
\le
\left.\frac{dA_0}{da}\right|_{a=1}
=
-7n^4+18n^3-13n^2+2n
=
-n(n-1)(7n^2-11n+2)<0.
\]
Therefore $A_0$ is decreasing in $a$, and hence
\[
A_0
>
A_0|_{a=1}
=
n(n-1)^3(2n-1)>0.
\]
Thus
\[
P_1>0.
\]

Next,
\[
P_2
=
am^2(B_3u^3+B_2u^2+B_1u+B_0),
\]
where
\begin{align}
    B_3&=an+2n^2>0,\\
    B_2&=a^2n-2an^2-a+6n^3-2n,\\
    B_1 &=2a^2n^2-a^2-7an^3+6an+6n^4-4n^2-2,\\
    B_0 &=a^2n^3-3a^2n-4an^4+7an^2+a+2n^5-2n^3-2n.
\end{align}
For $B_2$,
\[
B_2
>
6n^3-2n^2-2n-1>0
\]
for all $n\ge 2$. For $B_1$,
\[
B_1
>
6n^4-7n^3-4n^2-2 > 6n^4 - (3.5n^4 + n^4 + 0.5n^4) = n^4>0
\]
for all $n\ge 2$. Finally,
\[
\frac{dB_0}{da}
=
2a(n^3-3n)-4n^4+7n^2+1.
\]
Since $n\ge 2$, this derivative is increasing in $a$. Therefore
\[
\frac{dB_0}{da}
\le
\left.\frac{dB_0}{da}\right|_{a=1}
=
2n^3-6n-4n^4+7n^2+1
=
-(n-1)(4n^3+2n^2-5n+1)<0.
\]
Thus $B_0$ is decreasing in $a$, and hence
\[
B_0
>
B_0|_{a=1}
=
(n-1)^3(2n^2+2n-1)>0.
\]
Therefore
\[
P_2>0.
\]
The last coefficient is
\[
P_3
=
a^2m^4(mn-1)(mn+1-2an).
\]
Since $m>n\ge 2$, we have $mn>1$. Also,
\[
mn+1-2an
>
n^2+1-2n
=
(n-1)^2>0.
\]
Hence
\[
P_3>0.
\]

We have shown that
\[
P_0,P_1,P_2,P_3>0.
\]
Since $t>0$, it follows that
\[
\mathcal P_{a,n}(t,m)>0.
\]

Moreover,
\[
t^4>0,
\qquad
1+m^2t^2>0,
\qquad
K^2>0,
\qquad
\sqrt{1+t^2}>0.
\]
Therefore, from \eqref{eq:derivativeA}
\begin{align}
    A'(\vartheta)  =
-
\frac{
2\sqrt{1+t^2}
}{
t^4(1+m^2t^2)K^2
}
\mathcal P_{a,n}(t,m)
<0.
\end{align}
Hence $A(\vartheta)$ is strictly decreasing on $\left(0,\frac{\pi}{n}\right)$.\end{proof}

\begin{lemma}\label{lem:GradientsOverlapClampedJin=JoutNega}
For $-\infty < a < 0$ and $J_{\rm in}=J_{\rm out}=J$,
    \begin{align}
        \frac{d C_{\rm clamped}(\vartheta)}{d \vartheta} > 0 
    \end{align}
    where $\vartheta\in (0,\pi/2n)$. 
\end{lemma}
\begin{proof}
We use $n:=T+1$ and the variables $r$ and $t$ and $m$ from the proof of Lemma \ref{lem:lem0a1}. Note that for $a<0$, the lowest positive solution of the secular equation (\ref{eqn:secsimple}) is in $(0,{\pi\over 2n})$ for $0<\epsilon <1$, so we consider this interval for $\vartheta$. Define also $\alpha:= -a >0$. Use of double angle formulas gives
\begin{align}
    C_{\rm clamped}(\vartheta)&= 1-S(\vartheta) \nonumber \\
    S(\vartheta)&:= {2r^{2}(1+t^{2})(\alpha t + r)(1-\alpha rt) \over nt^{2}(1+r^{2})^{2}D(\vartheta)}\nonumber \\
    D(\vartheta)&:= 2nr + n\alpha t(1-r^{2}) + \alpha r(1+t^{2}).
\end{align}
Therefore, it is sufficient that $S'(\vartheta)<0$ on $(0,\pi/2n)$ to prove the theorem.
Noting that $t'={1+t^{2}\over 2}$ and $r'={n(1+r^{2})\over 2}$, the logarithmic derivative of $S$ is
\begin{align}
    {S'\over S}&= {n(1+r^{2})\over r} + t + {{\alpha\over 2}(1+t^{2}) + {n\over 2}(1+r^{2}) \over \alpha t + r} \nonumber \\
    &-{\alpha \left( {nt\over 2}(1+r^{2}) + {r\over 2}(1+t^{2}) \right) \over 1-\alpha r t} - {1+t^{2}\over t} - 2nr - {D'\over D}.
    \label{eqn:dddd}
\end{align}
Note that the product $2rtD(\alpha t+r)(1-\alpha rt)$ clears denominators of $S'/S$, with the factor of 2 just used to simplify coefficients. It follows that
\begin{equation}
    S'= {r(1+t^{2})\over nt^{3}(1+r^{2})^{2}D^{2}}\mathcal{P}(\alpha,n,t,r)
    \label{eqn:sp}
\end{equation}
with $\mathcal{P}(\alpha, n , t,r)$ a polynomial in $\alpha, n , t,r$ which is inferred from the expression (\ref{eqn:dddd}) and 
\begin{align}
    D'&= n^{2}(1-\alpha rt)(1+r^{2}) + \alpha(n+rt)(1+t^{2}).
\end{align}
Specifically,
\begin{align}
    \mathcal{P}(\alpha, n , t,r) &= 2tnD(\alpha t + r)(1-\alpha r t)(1+r^{2}) \nonumber \\
    &-2rD(\alpha t + r)(1-\alpha r t)\nonumber \\
    &+ trD(1-\alpha r t)\left(\alpha(1+t^{2})+n(1+r^{2})\right) \nonumber \\
    &-trD\alpha(\alpha t + r)\left( nt(1+r^{2}) + r(1+t^{2})\right) \nonumber \\
    &- 4ntr^{2}D(\alpha t+r)(1-\alpha rt)\nonumber \\
    &- 2trD'(\alpha t+r)(1-\alpha r t).
\end{align}
Note that the function multiplying $\mathcal{P}(\alpha, n , t,r)$ in (\ref{eqn:sp}) is positive, so one only needs to prove that $\mathcal{P}(\alpha, n , t,r)<0$ in order to prove the theorem. We introduce the scalar $0<x<1$ that scales $m$ to $n$ according to $n=xm$. Then $t={xr\over n}$, and a computer algebra system is used to compute the $\alpha$ monomials in the function
\begin{equation}
    \mathcal{P}\left( \alpha,n,t={xr\over n},r \right)= {r^{3}\over n^{5}}\left( Q_{0}+\alpha Q_{1} + \alpha^{2}Q_{2}+\alpha^{2}Q_{3}\right),
\end{equation}
where $Q_{j}$ are  polynomials in $n$. One finds that (with inequalities following directly from $0<r^{2} < 1$ and $0<x<1$ except where noted)
\begin{align}
    Q_{0} &= -4n^{6}(1-x+r^{2}x) < 0 
\end{align}
and
\begin{align}
    Q_{1}&= -n(n^{4}A_{1} + n^{2}A_{2} + 2r^{4}x^{4}) \nonumber \\
    A_{1}&=2+3(1-r^{2})x -(5r^{4}-14r^{2}+5)x^{2} >0 \nonumber \\
    A_{2}&= r^{2}x^{2}(4-3(1-r^{2})x^{2}) >0
\end{align}
and
\begin{align}
    Q_{2}&= x\left( n^{4}B_{1} + n^{2}B_{2} + B_{3}\right) \nonumber \\
    B_{1}&= -2r^{6}x^{2} + 10r^{4}x^{2}-r^{4}x-10r^{2}x^{2} \nonumber \\
    &{} -2r^{2}x+r^{2}+2x^{2}-x-1 \nonumber \\
    B_{2}&= r^{2}x^{2}(r^{4}x-10r^{2}x + 2r^{2} + x-2) <0 \nonumber \\
    B_{3}&=r^{4}x^{4}(r^{2}-1)<0
\end{align}
with $B_{1}<0$ following from considering it as function of $x$ and finding no minima on the interval $0<x<1$ while it is negative on the endpoints. Lastly,
\begin{align}
    Q_{3}&= nr^{2}x^{3}\left[ n^{2}(-1-3x+r^{2}(1+2x) -3r^{4}x) + 3r^{2}x^{2}(s-1) \right],
\end{align}
where $3r^{2}x^{2}(s-1)<0$ and $-1-3x + r^{2}(1+2x)-3r^{4}x < -1-3x + 1+2x = -x <0$, so $Q_{3}<0$. 
\end{proof}

\begin{theorem}
\label{thm:jin_jout_parallel}
    For $J_{\mathrm{in}} = J_{\mathrm{out}}=J$, the overlap cost function for $H_{\rm clamped}$  \eqref{eq:JinEqualJoutCostFunctionOnevariable} has a non-negative derivative for any $J\geq 0$, i.e., $\frac{\partial C_{\rm clamped}}{\partial \epsilon} \geq 0$, with equality only at $\epsilon = 0$.
\end{theorem}
\begin{proof}
    The proof follows by using the results of Lemma \ref{lem:GradientsOverlapClampedJin=JoutPosa} and Lemma \ref{lem:GradientsOverlapClampedJin=JoutNega}.
\end{proof}

The final theorem gives a constant lower bound for the fidelity between the free and clamped ground states, which allows the history state (\ref{eq:historystate}) to be utilized as an input state for quantum phase estimation.

\begin{theorem}\label{thm:LowerBoundOverlapLargeTClampedGeneralJ}
    The fidelity between the free and clamped ground states has the lower bound
\begin{equation}
    1- C_{\rm clamped}(\vartheta,\epsilon) \ge {1\over 4}
    \end{equation}
    on $\vartheta \in (0,{\pi\over 2n-1})$ containing the smallest positive solution of the secular equation \eqref{eq:SecGenJinJout} for $0<J_{\text{out}}\le J_{\text{in}}$.
\end{theorem}
\begin{proof}
    We use the overlap cost function in \eqref{eq:OverlapCost_Hclampd_GenJ}. 
    From (\ref{eqn:qinqout}), we have $q_{\text{out}}\le q_{\text{in}}$ and $q_{y}< \cos(n\vartheta) + \tan{\vartheta\over 2}\sin(n\vartheta) = {\cos\left( {1\over 2}(2n-1)\vartheta\right) \over \cos{\vartheta\over 2}}$ for $y=\text{in},\text{out}$. Further, on  $(0,\vartheta_{*}]$, $q_{y}\ge 0$ due to the characteristic equation. It follows that the smallest positive solution of the characteristic equation is $\vartheta_{*}\in (0,{\pi\over 2n-1})$. On the larger interval $(0,{\pi \over n})$, $q_{y}<1$. The normalization $\mathcal{N}$ satisfies $\mathcal{N}\ge 0$ on $(0,{\pi \over 2n-1})$ due to the expression (\ref{eq:normalizationGeneralJ}).

    Calling $A:=n+T_{n}U_{n-1}$ and $B=nT_{n}+U_{n-1}$ for Chebyshev functions $T_{n}:=\cos(n\vartheta)$, $U_{n-1}:= {\sin(n\vartheta)\over\sin\vartheta}$, we have that
    \begin{align}
        \mathcal{N} &= {1\over 2}\left( A-q_{\text{out}}B\right) + {1\over 2}{q_{\text{out}}\over q_{\text{in}}}\left( A-q_{\text{in}}B\right) .
    \end{align}
    But \begin{equation}A-q_{y}B=(1-T_{n})(n-U_{n-1}) + (1-q_{y})B\label{eqn:abeq}\end{equation} with the first term being positive on $\vartheta\in (0,{\pi\over 2n-1})$ and $1-q_{y}>0$ on the same. To prove that $B\ge 0$ on $(0,{\pi\over 2n-1})$, we introduce the main trick that will pervade the rest of the proof (we state the trick without proof): for $n\ge 2$ and  $0\le t < {\pi \over n-1}$, then
    \begin{equation}
        {\sin(nt)\over n\sin t} \ge \cos\left( t\sqrt{ {n^{2}-1 \over 3}}\right).
    \end{equation}
    But \begin{align}
        B/n&=\cos(n\vartheta) + {\sin(n\vartheta)\over n\sin\vartheta} \nonumber \\
        &\ge \cos(n\vartheta) + \cos\left(\vartheta \sqrt{{n^{2}-1\over 3}}\right) \nonumber \\
        &= 2\cos \left( {n+\sqrt{{n^{2}-1\over 3}} \over 2}\vartheta \right)\cos \left( {n-\sqrt{{n^{2}-1\over 3}} \over 2}\vartheta \right)
    \end{align}
    because $\vartheta \in (0,{\pi \over 2n-1})$ satisfies the assumptions of the trick. Since $n+\sqrt{{n^{2}-1\over 3}} \le 2n-1$ and $0<n+\sqrt{{n^{2}-1\over 3}} \le n-1$, $B/n\ge 0$  on $(0,{\pi \over 2n-1})$. Positivity of $B$ allows to conclude that $A-q_{y}B\ge 0$ via (\ref{eqn:abeq}), and therefore, that 
    \begin{align}
        \mathcal{N}  &\le {1\over 2}\left( A-q_{\text{out}}B\right) + {1\over 2}\left( A-q_{\text{in}}B\right)\nonumber \\
        &\le  A-q_{\text{out}}B.
        \label{eqn:neq}
    \end{align}

    To complete the proof of the theorem, we will show that ${\epsilon \sin^{2}(n\vartheta) \over 4n\mathcal{N}\sin^{2}\left( {\vartheta \over 2}\right)}\ge {1\over 4}$ on $(0,{\pi \over 2n-1})$. Define $M(\vartheta):={\cos\left( {1\over 2}(2n-1)\vartheta\right) \over \cos{\vartheta\over 2}}$ so that from the first line of the proof we have $1-q_{y}\ge 1-M$. Then (\ref{eqn:abeq}) gives 
    \begin{equation}
        A-q_{y}B \le (1-q_{y})\left( B + { (1-T_{n})(n-U_{n-1})\over 1-M} \right).
        \label{eqn:otherform}
    \end{equation}
     where $M < 1$ on $\vartheta \in (0,\pi/(2n-1))$.
    From (\ref{eqn:neq}) and (\ref{eqn:otherform}), we note that if we can show that 
    \begin{equation}
        B + { (1-T_{n})(n-U_{n-1})\over 1-M} \le {\sin^{2}(n\vartheta) \over n\sin^{2}\left( {\vartheta \over 2}\right)}
        \label{eqn:hypineq}
    \end{equation}     
    it will follow that
    \begin{equation}
        \mathcal{N}\le A-q_{\text{out}}B \le (1-q_{\text{out}}){ \sin^{2}(n\vartheta) \over n\sin^{2}\left( {\vartheta \over 2}\right)} \le (1-q_{\text{in}}q_{\text{out}}){ \sin^{2}(n\vartheta) \over n\sin^{2}\left( {\vartheta \over 2}\right)}
    \end{equation}
    which implies the desired inequality
    \begin{equation}
        {\epsilon \sin^{2}(n\vartheta) \over 4n\mathcal{N}\sin^{2}\left( {\vartheta \over 2}\right)} \ge  {1\over 4}
    \end{equation}
    on $(0,{\pi\over 2n-1})$.

We are left to prove (\ref{eqn:hypineq}). Introducing $x:= {\vartheta\over 2}$ one has $0<x<{\pi \over 4n-2}$ and a series of elementary trigonometric identities and clearing of denominators brings inequality (\ref{eqn:hypineq}) to the form
\begin{align}
    &{}\sin^{2}(2nx)\cos(x)\sin((n-1)x) \nonumber \\
    &{} - n^{2}\cos(2nx)\sin^{2}(x)\cos(x)\sin((n-1)x) \nonumber \\
    &{} - n\sin(nx)\sin^{2}(x)\left(n\cos^{2}(x) - \cos^{2}(nx)\right) \ge 0.
    \label{eqn:hypsimp}
\end{align}
    This is a technical trigonometric inequality, but it can be proven using elementary identities and the (non-elementary) trick above. The elementary part is to factor out the trigonometric functions with $n$-independent argument, because these are positive on $x\in (0,{\pi \over 4n-2})$. To this end, introduce the symbolic dictionary $s=\sin x$, $c=\cos x$, $S=\sin(nx)$, $C=\cos(nx)$, $p={S\over ns}$, $q={C\over c}$ so that $\sin((n-1)x) = Sc-Cs$, $S=nsp$, $C=qc$, $Sc-Cs = sc(np-q)$, $\sin^{2}(2nx)=4S^{2}C^{2}$, $\cos(2nx)=C^{2}-S^{2}$ to write the left hand side of (\ref{eqn:hypsimp}) as
    \begin{align}
        &{} c\sin((n-1)x)\left( 4S^{2}C^{2} - n^{2}s^{2}(C^{2}-S^{2}) \right) - ns^{2}S(nc^{2}-C^{2})\nonumber \\
        &= sc^{2}(np-q)\left( 4S^{2}C^{2} - n^{2}s^{2}(C^{2}-S^{2}) \right) - n^{2}s^{3}p(nc^{2}-C^{2}) \nonumber \\
        &= sc^{2}(np-q)\left( 4n^{2}s^{2}p^{2}C^{2} - n^{2}s^{2}(C^{2}-n^{2}s^{2}p^{2}) \right) - n^{2}ps^{3}(nc^{2}-C^{2})\nonumber \\
        &= n^{2}s^{3}c^{2}\left[ (np-q)\left( (4p^{2}-1)C^{2}+n^{2}s^{2}p^{2} \right) - p(n-q^{2}) \right]\nonumber \\
        &= n^{2}s^{3}c^{2}\left[ (np-q) (4p^{2}-2)C^{2} -q(1-pq)  \right]
    \end{align}
    where the last line used $n^{2}s^{2}p^{2}=1-C^{2}$.
    The trick gives the fact that $p \ge \cos \left( x \sqrt{ {n^{2}-1\over 3}}\right) > {\sqrt{3}\over 2}$, i.e., $4p^{2}-3 >0$, because $x\sqrt{ {n^{2}-1\over 3}} < {\pi \over 4n-2}\sqrt{ {n^{2}-1\over 3}} \le {\pi \over 6}$, and one further has $np-q = {Sc-Cs\over sc} = {\sin((n-1)x)\over sc}>0$, so one is compelled to write the above expression as
    \begin{equation}
        n^{2}s^{3}c^{2}\left[ (np-q) (4p^{2}-3)C^{2} +(np-q)C^{2}-q(1-pq)  \right]
    \end{equation}
    and it becomes sufficient to show that $(np-q)C^{2}-q(1-pq)\ge 0$, i.e.,
    \begin{equation}
        {\sin((n-1)x)C^{2}\over sc} - {C\over c}\left(1-{SC \over nsc} \right) \ge 0.
    \end{equation}
    Clearing the denominators (which are positive on $0<x<{\pi \over 4n-2}$), and introducing a convenient factor of 4 gives the equivalent inequalities, with the second obtained by dividing the first by $n\sin(2x)$:
    \begin{align}
        (n+2)\sin(2nx) + n\sin((2n-2)x) - 3n\sin(2x) &\ge 0\nonumber \\
        (n+2){\sin(n\vartheta)\over n\sin\vartheta} + (n-1){\sin((n-1)\vartheta)\over (n-1)\sin\vartheta} - 3 &\ge 0.
    \end{align}
    For $n=2$, no trick is needed; one directly verifies the second inequality because $\vartheta < {\pi \over 3}$. For $n\ge 3$, use the fact that on $\vartheta \in (0,{\pi\over 2n-1})$, both $\vartheta \sqrt{{n^{2}-1\over 3}}$ and $\vartheta \sqrt{{(n-1)^{2}-1\over 3}}$ are $<{\pi \over 3}$, so the second inequality becomes ${n+2\over 2} + {n-1\over 2} - 3 >0$, which is clear.
    
\end{proof}

\subsection{Analysis of the cyclic energy cost function $E_{\rm cyclic}$}
We first derive the characteristic function of $P_{\mathcal{D}}H_{\rm cyclic}P_{\mathcal{D}}$, resulting in the expression in Theorem 2 of the main text in the special case of $J_{\rm cyc}=1$.
\begin{theorem}
\label{thm:cyclicsec}
    For $0<J_{\rm cyc}$, the ground state energy of $P_{\mathcal{D}}H_{\rm cyclic}P_{\mathcal{D}}$ is given by $E_{\rm cyclic}(\epsilon)=1-\cos\vartheta^*(\epsilon)$, where $\vartheta^*$ is the smallest non-negative solution of
\begin{align}\label{eqn:flosec_simple}
    \frac{\cos(\frac{(T+2)\vartheta}{2})}{\cos(\frac{T\vartheta}{2})} = 1- J_{\rm cyc}(1-\sqrt{1-\epsilon}).
\end{align}
$E_{\rm cyclic}(\epsilon)$  takes values in $\mathbb{R}_{\ge 0}$ and is equal to zero if and only if the circuit parameters $\theta$ are such that $\epsilon(\theta) =0$.
\end{theorem}
\begin{proof}
    Working in $\text{span}\mathcal{D}$ corner of $CS$, one finds that $P_{\mathcal{D}}H_{\rm cyclic}P_{\mathcal{D}}$ has the form of $H$ in (\ref{eq:modified_H}) with
\begin{align}
    H_0 &= H_\prop \nonumber \\
    a_1 &= {J_{\rm cyc}\over 2} \left(1-\sqrt{1-\epsilon}\right)\nonumber \\
    \ket{\xi_1} &=  {1\over \sqrt{2}}\left(\ket{\gamma_{0}} +\ket{\gamma_{T}} \right) \nonumber \\
    a_2 &= {J_{\rm cyc}\over 2} \left(1+\sqrt{1-\epsilon}\right) \nonumber \\
    \ket{\xi_2} &= {1\over \sqrt{2}}\left(\ket{\gamma_{0}} -\ket{\gamma_{T}} \right).
    \label{eqn:cyclicidents}
\end{align}
In $\text{span}\mathcal{D}$, the resolvent of $H_{\rm prop}$ is
\begin{equation}
    R_{0}(\lambda)=\sum_{k=0}^{T}{\ket{\Psi_{k}}\bra{\Psi_{k}}\over \omega_{k}-\lambda},
\end{equation}
with $\omega_{k}$ in (\ref{eqn:h0eigs}).
From (\ref{eq:eig_equation}) and Lemma \ref{lem:clem}, and taking $\lambda = 1-\cos\vartheta$ as before, we have
\begin{align}
    0&= \left( 1+{J_{\rm cyc}\over 2}\left(1-\sqrt{1-\epsilon}\right)\langle \xi_{1}\vert R_{C}(\cos\vartheta) \vert \xi_{1}\rangle \right)\left( 1+{J_{\rm cyc}\over 2}\left(1+\sqrt{1-\epsilon}\right)\langle \xi_{2}\vert R_{C}(\cos\vartheta) \vert \xi_{2}\rangle\right) \nonumber \\
    &{} - {J_{\rm cyc}^{2}\epsilon\over 4}\langle \xi_{1}\vert R_{C}(\cos\vartheta)\vert \xi_{2}\rangle^{2}
    \label{eqn:cycsec1}
\end{align}
where the matrix $C$ from Lemma \ref{lem:clem} appears.
The second line is equal to zero by the symmetry of $R_{C}(\cos\vartheta)$. Using the matrix elements of $R_{C}(\cos\vartheta)$ from Theorem \ref{thm:clampedsec}, we have
\begin{align}
    \langle \xi_{1}\vert R_{C}(\cos\vartheta) \vert \xi_{1}\rangle &= -{\cos\left( {T\vartheta \over 2} \right) \over \sin \left( {\vartheta\over 2}\right)\sin\left( {(T+1)\vartheta\over 2}\right)} \nonumber \\
    \langle \xi_{2}\vert R_{C}(\cos\vartheta) \vert \xi_{2}\rangle &= {\sin\left( {T\vartheta \over 2} \right) \over \sin \left( {\vartheta\over 2}\right)\cos\left( {(T+1)\vartheta\over 2}\right)}.
\end{align}
The second factor in the first line of (\ref{eqn:cycsec1}) can therefore not be set to zero to obtain the characteristic equation for the ground state eigenvalue because the resulting equation does not have a solution as $\vartheta\rightarrow 0$. The appropriate secular equation comes from setting the first factor in the first line of (\ref{eqn:cycsec1}) to zero
\begin{equation}
    {2\sin \left( {\vartheta\over 2}\right)\sin\left( {(T+1)\vartheta\over 2}\right) \over \cos\left( {T\vartheta \over 2} \right)} = J_{\rm cyc}\left( 1 - \sqrt{1-\epsilon}\right),
\end{equation}
which is (\ref{eqn:flosec_simple}).

$P_{\mathcal{D}}H_{\rm cyclic}P_{\mathcal{D}} \succeq 0$ because $H_{\rm prop}\succeq 0$ in ${\rm span}\mathcal{D}$ and the perturbations are positive semidefinite. 
If $\vartheta_{*}(\epsilon) = 0$, i.e., $E_{\rm cyclic}(\epsilon(\theta))=0$ then substituting in \eqref{eq:secEquationFloquent} gives $\epsilon(\theta) = 0$. Conversely, if $\epsilon(\theta) = 0$, then $\vartheta = 0$ is a solution. Since $\vartheta(\epsilon)$ is defined as the smallest non-negative solution, $E_{\rm cyclic}(\epsilon(\theta))= 0.$ 
\end{proof}

We now show that the gradient of $E_{\rm cyclic}(\epsilon(\theta))$ with respect to circuit parameters $\theta$ is parallel to the gradient of $\epsilon(\theta)$.

\begin{theorem}
    The cost function $E_{\rm cyclic}(\epsilon(\theta))$ corresponding to the ground state energy of $P_{\mathcal{D}}H_{\rm cyclic}P_{\mathcal{D}}$ for all $J_{\rm in}, J_{\rm cyc}>0$ satisfies ${dE_{\rm cyclic}(\epsilon)\over d\epsilon } \ge 0$ for $\epsilon\in (0,1)$, and $T> 1$. \label{thm:GSE_H_cyclic_gradient}
\end{theorem}
\begin{proof}
    Note that the characteristic equation (\ref{eqn:flosec_simple}) is independent of $J_\mathrm{in}$. We show that $\frac{d\epsilon}{d\vartheta} \geq 0$ on the smallest-root branch of \eqref{eqn:flosec_simple} for any $J_{\rm cyc} >0$. Define     \begin{align}
        \beta(\vartheta) := 
        \frac{
        \cos\!\left(\frac{(T+2)\vartheta}{2}\right)
        }{
        \cos\!\left(\frac{T\vartheta}{2}\right)
        }. \label{eq:betaDef}
    \end{align} Differentiating the secular equation \eqref{eqn:flosec_simple} given by $\beta(\vartheta) = 1-J_{\rm cyc}(1-\sqrt{1-\epsilon})$ implicitly with respect to $\epsilon$ gives
    \begin{align}
        \beta'(\vartheta)\frac{d\vartheta}{d\epsilon} = -\frac{J_{\rm cyc}}{2\sqrt{1-\epsilon}} \implies \frac{d\vartheta}{d\epsilon} = -\frac{1}{\beta'(\vartheta)}\frac{J_{\rm cyc}}{2\sqrt{1-\epsilon}}.
    \end{align}
    In the relevant branch, we claim $\beta'(\vartheta) \leq 0$.
    Clearly, the denominator is positive. 
As for the numerator, when \((T+1)\vartheta\le \pi\), then
 the claim follows. Otherwise, suppose 
 ${\pi \over T+1}< \vartheta \le {\pi \over T}$
 and set $\delta:=(T+1)\vartheta-\pi.$
Then \(0<\delta\le \vartheta\le \pi/T\). For \(T\ge 2\), \(\vartheta\le \pi/2\), and hence \(\sin x\) is increasing on
\([0,\vartheta]\). Therefore $\sin\delta\le \sin\vartheta,$
and
\begin{align}
    \sin((T+1)\vartheta)
=
\sin(\pi+\delta)
=
-\sin\delta \ge
-\sin\vartheta.
\end{align}
Hence,
\begin{align}
    \sin((T+1)\vartheta)+(T+1)\sin\vartheta
\ge
T\sin\vartheta>0.
\end{align}
Therefore 
\begin{align}
    \beta'(\vartheta)<0 \label{eq:betaPositive}
\end{align} 
on ${\pi \over T+1}< \vartheta \le {\pi \over T}$. Noting  $J_{\rm cyc} >0$ and $\epsilon \in (0,1)$ completes the proof.
\end{proof}

\begin{theorem}
\label{thm:cycliclargeTgap}
The clock reflection $W\ket{\gamma_{t}}_{CS}:=\ket{\gamma_{T-t}}_{CS}$ commutes with the Hamiltonian $P_{\mathcal{D}}H_{\rm cyclic}P_{\mathcal{D}}$, with the modifying vectors in (\ref{eqn:cyclicidents}) being in the symmetric and antisymmetric sector, respectively. Let
\[
\Delta_{\pm}:=J_{\mathrm{cyc}}\bigl(1\pm \sqrt{1-\epsilon}\bigr),
\qquad
\rho_{\pm}:=1-\Delta_{\pm} .
\]
The positive low-lying roots of the  secular equations
\begin{align}
\frac{\cos\!\left(\frac{(T+2)\vartheta}{2}\right)}
{\cos\!\left(\frac{T\vartheta}{2}\right)}
&=
\rho_{-} \; , \; \text{(symmetric sector)} \nonumber \\
\frac{\sin\!\left(\frac{(T+2)\vartheta}{2}\right)}
{\sin\!\left(\frac{T\vartheta}{2}\right)}
&=
\rho_{+} \; , \; \text{(antisymmetric sector)}
\end{align}
satisfy
\begin{align}
    \vartheta_k
 &= \frac{(2k-1)\pi}{T} \left[1- \frac{2}{\Delta_{-} T} + O(T^{-2})\right]
\qquad k=1,2,\dots \text{ (symmetric sector) }\nonumber \\
\vartheta_k
 &= \frac{2k\pi}{T} \left[1- \frac{2}{\Delta_{+} T} + O(T^{-2})\right]
\qquad k=1,2,\dots \text{ (antisymmetric sector) }
\label{eqn:rootexp}
\end{align}
The spectral gap of $P_{\mathcal{D}}H_{\rm cyclic}P_{\mathcal{D}}$ in the symmetric sector is given by
\begin{align}
    \Delta E
:=
E_2-E_1
= \frac{4\pi^2}{T^2}\left[1-\frac{4}{\Delta_{-} T} + O(T^{-2})\right].
\end{align}
Note that the QPE dynamics is reflection-sector preserving and that $\ket{\Psi_{\rm free}}$ belongs to the symmetric sector.
\end{theorem}

\begin{proof}
The $W$-symmetry of $P_{\mathcal{D}}H_{\rm cyclic}P_{\mathcal{D}}$ and that of its its perturbations (\ref{eqn:cyclicidents}) is clear from inspection. The fact that the first (second) factor in (\ref{eqn:cycsec1}) gives the secular equation for the symmetric (antisymmetric) sector follows from (\ref{eqn:eigeq}) and the $W$-symmetry of the modifying vectors.

    We prove the results for the  symmetric sector. We will take  $n:= T+1$ and  write the  symmetric secular equation as $B_{n}(\vartheta) = \rho_{-}$. Since $\Delta_{-} > 0$, one has $\rho_{-} \leq 1$. Note that where it is defined, $B_{n}'(\vartheta)<0$. Label the poles $p_{j}={(2j-1)\pi\over n-1}$, $j=1,\ldots,$ on the positive axis. On the first branch $(0,p_{1})$, $B_{n}$ decreases from 1 to 0.  On the other branches, $B_{n}$ decreases from $\infty$ to $-\infty$. Since $\rho_{-}\le 1$, there is exactly one root in each branch. We will take $n$ sufficiently large that the eigenvalue indices of interest satisfy $k<{n\over 2}$. Define  $a_{k}:={(2k-2)\pi\over n}$, $b_{k}={(2k-1)\pi\over n}$ (note that $a_{1}=0$, $b_{1}<p_{1}$ and for $k\ge 2$, $a_{k},b_{k}\in (p_{k-1},p_{k})$) such that
    \begin{equation}
        B_{n}(a_{k})=1 \; , \, B_{n}(b_{k})=-1.
    \end{equation}
    If $0<\Delta_{-}<2$, then $-1<\rho_{-}<1$ and the root is in $(a_{k},b_{k})=({(2k-2)\pi\over n},{(2k-1)\pi\over n})$. If $\Delta_{-}=2$, then the root is at $b_{k}$. If $\Delta_{-}>2$, then the root is in $(b_{k},p_{k})$. Therefore, regardless of $J_{\rm in}>0$, the $k$-{th} positive root lies in the interval $\left( \frac{(2k-2)\pi}{n}, \frac{(2k-1)\pi}{n-1}\right)$.

    We now define scaled roots $x_{k}={\vartheta_{k}\over 2}$ and parametrize these roots by
    \begin{equation}
        2(n-1)x_{k} = q_{k} - \delta_{k}
    \end{equation}
    where $q_{k}:=(2k-1)\pi$. From the branches $(0,p_{1})$, $(p_{1},p_{2})$, $\ldots$, one finds that $0<\delta_{1}<\pi$ and $0<\delta_{k}<2\pi$ for $k\ge 2$. In fact, substituting the $x_{k}$ into the secular equation gives
    \begin{equation}
\left(\Delta_{-} - 2\sin^{2} x_k\right)\sin\frac{\delta_k}{2}
= \sin(2x_k)\cos\frac{\delta_k}{2}
\label{eqn:subsub}
\end{equation}
and for sufficiently large $T$, the left hand side and $\sin(2 x_{k})>0$, so for large $T$, $0<\delta_{k}<\pi$ and we can invert the tangent
\begin{align}
\frac{\delta_k}{2}
&= \arctan\left(
\frac{\sin(2x_k)}{\Delta_{-} - 2\sin^{2} x_k}
\right)\nonumber \\
&\le \frac{\sin(2x_k)}{\Delta_{-} - 2\sin^{2} x_k}\nonumber \\
&\leq \frac{4x_k}{\Delta_{-}}.
\end{align}
Since $q_{k}=O(1)$, $\delta_{k}=O(T^{-1})$. Expanding (\ref{eqn:subsub}) for large $T$ gives
\begin{align}
    \delta_k &= \frac{4}{\Delta_{-}} x_k + O(T^{-3})\nonumber \\
    &= \frac{2q_k}{\Delta_{-} T}
- \frac{4q_k}{\Delta_{-}^{2} T^{2}}
+ O(T^{-3}).
\end{align}
Using $\vartheta_{k}={2x_{k}}$ gives
\begin{align}
\vartheta_k
&= \frac{q_k}{T}
- \frac{2q_k}{\Delta_{-} T^2}
+ \frac{4q_k}{\Delta_{-}^2 T^3}
+ O(T^{-4})
\end{align}
which is the symmetric sector entry in (\ref{eqn:rootexp}).
Subsequently using $E_{k}:=1-\cos\vartheta_{k}$  gives the asymptotic expansion for the eigenvalue and the gap.

\end{proof}

\subsection{Analysis of the cyclic overlap cost function $C_{\rm cyclic}$}

We start by deriving the $\vartheta^{*}$ dependence of the ground state of $P_{\mathcal{D}}H_{\rm cyclic}P_{\mathcal{D}}$.
\begin{theorem}
\label{thm:cyclicgs}
    The ground state of $P_{\mathcal{D}}H_{\rm cyclic}P_{\mathcal{D}}$ is a $CS$-valued function of the circuit parameters $\theta$ through its dependence on $\epsilon(\theta)$, and is given by
    \begin{align}
    \ket{\Psi_{\rm cyclic}}
    &\propto \sum_{t=0}^{T}\left(  \vphantom{\sum} \sin((t+1)\vartheta^{*}) +\sin((T+1-t)\vartheta^{*}) \vphantom{\sum} \right)\ket{\gamma_{t}} .
        \label{eqn:cycgs0}
\end{align}
where $\vartheta^{*}$ is the smallest non-negative solution of (\ref{eqn:flosec_simple}).
\end{theorem}
\begin{proof}
    Starting from (\ref{eqn:resolvprop}), one has 
\begin{equation}\ket{\Psi_{\rm cyclic}}\propto R_{0}(1-\cos\vartheta^{*})(c_{1}a_{1}\ket{\xi_{1}}+ c_{2}a_{2}\ket{\xi_{2}})
    \end{equation}
    with variables defined in (\ref{eqn:cyclicidents}) and $\vec{c}\in\text{ker}M(1-\cos\vartheta^{*})$ according to (\ref{eq:c}). Therefore,
    \begin{equation}
        \begin{pmatrix}
            1 + a_{1}R_{C}(\cos\vartheta^{*})_{1,1} & 0 \\
            0 & 1 + a_{2}R_{C}(\cos\vartheta^{*})_{2,2}
        \end{pmatrix}\begin{pmatrix}
            c_{1}\\c_{2}
        \end{pmatrix}=0.
    \end{equation}
    But $1 + a_{1}R_{C}(\cos\vartheta^{*})_{1,1}=0$ is exactly the characteristic equation (\ref{eqn:flosec_simple}), so $\vec{c}=(c_{1},0)^{\intercal}$ gives the ground state
    \begin{align}
        \ket{\Psi_{\rm cyclic}}&\propto R_{0}(1-\cos\vartheta^{*})\ket{\xi_{1}} \nonumber \\
        &\propto \sum_{t=0}^{T}\left( R_{C}(\cos\vartheta^{*})_{t,0} + R_{C}(\cos\vartheta^{*})_{t,T} \right)\ket{\gamma_{t}}\nonumber \\
        &= \sum_{t=0}^{T}\left(\cos\left( (T-t+{1\over 2})\vartheta^{*} \right) + \cos \left( (t+{1\over 2})\vartheta^{*}\right) \right)\ket{\gamma_{t}} \nonumber \\
        & \propto \sum_{t=0}^{T}\left(  \vphantom{\sum} \sin((t+1)\vartheta^{*}) +\sin((T+1-t)\vartheta^{*}) \vphantom{\sum} \right)\ket{\gamma_{t}},
    \end{align}
    where the third line uses the matrix elements derived in Lemma \ref{lem:clem}.
\end{proof}

In the following theorem, we compute the normalization constant of $\ket{\Psi_{\rm cyclic}}$ to obtain the overlap cost function, which is immediately seen to agree with the expression in Theorem 2 of the main text.
\begin{theorem}
    \label{thm:overlapcostcyclic}
    The overlap cost function $C_{\rm cyclic}(\epsilon(\theta))$ is given by
    \begin{equation}
C_{\rm cyclic}(\epsilon)=1-\frac{
2\sin^2\!\left(\frac{(T+1)\vartheta^{*}}{2}\right)
}{
(T+1)\sin^2\!\left(\frac{\vartheta^{*}}{2}\right)
\left(
T+1+\frac{\sin((T+1)\vartheta^{*})}{\sin\vartheta^{*}}
\right)
}. \label{eqn:cycgs0}
    \end{equation}
\end{theorem}
\begin{proof}
    The amplitudes in (\ref{eqn:cycgs0}) are 
\begin{align}
a_t:=\sin((t+1)\vartheta^{*})+\sin((T+1-t)\vartheta^{*}),
\qquad t=0,\dots,T.
\end{align}
Then 
\begin{align}
\ket{\Psi_{\mathrm{cyclic}}}
=
\frac{1}{\sqrt{\mathcal N}}
\sum_{t=0}^T a_t \ket{\gamma_t},\quad \mathcal N=\sum_{t=0}^T a_t^2.
\label{eqn:cycgsnorm}
\end{align}
We know $\ket{\Psi_{\free}}=\frac{1}{\sqrt{T+1}}\sum_{t=0}^T \ket{\gamma_t}.$
Therefore,
\begin{align}\label{eq:costFuncOverlap}
\langle \Psi_{\rm free}\vert \Psi_{\rm cyclic}\rangle
&=
\frac{1}{\sqrt{T+1}}\frac{1}{\sqrt{\mathcal N}}
\sum_{t=0}^T a_t, \qquad \left|\langle \Psi_{\rm free}\vert \Psi_{\rm cyclic}\rangle\right|^2
=
\frac{\left(\sum_{t=0}^T a_t\right)^2}{(T+1)\sum_{t=0}^T a_t^2}.
\end{align}
This simplifies 
\begin{align}
\sum_{t=0}^T a_t
&=
2\sum_{k=1}^{T+1}\sin(k\vartheta^{*}) = \frac{
2\sin\!\left(\frac{(T+1)\vartheta^{*}}{2}\right)
\sin\!\left(\frac{(T+2)\vartheta^{*}}{2}\right)
}{
\sin\!\left(\frac{\vartheta^{*}}{2}\right)
}.
\end{align}
Using
\begin{align}
a_t
=
2\sin\left(\frac{(T+2)\vartheta^{*}}{2}\right)
\cos\left(\frac{(2t-T)\vartheta^{*}}{2}\right),
\end{align}
and so
\begin{align}
\sum_{t=0}^T a_t^2
&=
4\sin^2\!\left(\frac{(T+2)\vartheta^{*}}{2}\right)
\sum_{t=0}^T \cos^2\!\left(\frac{(2t-T)\vartheta^{*}}{2}\right).
\end{align}
Using
$\cos^2 x=\frac{1+\cos(2x)}{2},$
we obtain
\begin{align}
\sum_{t=0}^T \cos^2\left(\frac{(2t-T)\vartheta^{*}}{2}\right)
&=
\frac{T+1}{2}
+\frac12\sum_{t=0}^T \cos((2t-T)\vartheta^{*}).
\end{align}
Now
\begin{align}
\sum_{t=0}^T \cos((2t-T)\vartheta^{*})
=
\sum_{t=0}^T \cos(T\vartheta^{*}-2t\vartheta^{*})
=
\frac{\sin((T+1)\vartheta^{*})}{\sin\vartheta^{*}},
\end{align}
so
\begin{align}
\sum_{t=0}^T a_t^2
=
2\sin^2\!\left(\frac{(T+2)\vartheta^{*}}{2}\right)
\left(
T+1+\frac{\sin((T+1)\vartheta^{*})}{\sin\vartheta^{*}}
\right).
\end{align}
Substituting these expressions in the overlap formula gives
\begin{align}
    1-\left|\langle \Psi_{\rm free}\vert \Psi_{\rm cyclic}\rangle\right|^2
&=1-\frac{
2\sin^2\!\left(\frac{(T+1)\vartheta^{*}}{2}\right)
}{
(T+1)\sin^2\!\left(\frac{\vartheta^{*}}{2}\right)
\left(
T+1+\frac{\sin((T+1)\vartheta^{*})}{\sin\vartheta^{*}}
\right)
}. 
\end{align}
\end{proof}

\begin{theorem}
    The overlap cost function $C_{\rm cyclic}(\epsilon)$ corresponding to the fidelity of the ground state of $H_{\rm cyclic}$ with the ground state of $H_{\rm free}$ for all $J_{\rm cyc}>0$ satisfies ${dC_{\rm cyclic}\over d\epsilon} \ge 0$ for $\epsilon\in (0,1)$ and $T\geq 2.$ 
    \label{thm:GSO_H_cyclic2}    
\end{theorem}
\begin{proof}
    We first write
    \begin{align}
        \frac{dC_{\rm cyclic}}{d\epsilon} = \frac{dC_{\rm cyclic}}{d\vartheta}/\frac{d\epsilon}{d\vartheta}.
    \end{align}
    The denominator being positive is immediate from Theorem \ref{thm:GSE_H_cyclic_gradient}.
    We show the positivity of the numerator as follows. Note that the LHS of \eqref{eqn:flosec_simple} equals $1$ at $\vartheta=0$, is continuous and decreasing until $\vartheta = \frac{\pi}{T},$ where it approaches $-\infty$ for $T\geq 2$. Hence the smallest solution for $\vartheta$ with $\epsilon\in (0,1)$ lies in $(0, \frac{\pi}{T}).$ 
    Define
\begin{align}
    S_1(\vartheta) &:= \frac{\sin((T+1)\frac\vartheta2)}{\sin(\frac{\vartheta}{2})} = \sum_{j=0}^T\cos((T-2j)\frac{\vartheta}{2}), \nonumber \\
    S_2(\vartheta) &:= \sum_{j=0}^T\cos((T-2j)\frac{\vartheta}{2})^2 = \frac{T+1}{2} + \frac12\frac{\sin((T+1)\vartheta)}{\sin({\vartheta})}.
\end{align}
This gives
\begin{align}
    1-C_{\rm cyclic}(\vartheta) = \frac{S_1(\vartheta)^2}{(T+1)S_2(\vartheta)}
\end{align}
\begin{align}
    \frac{d}{d\vartheta}(1-C_{\rm cyclic}) = \frac{1}{(T+1)} \frac{S_1}{S_2^2}(2S_1'S_2 - S_1S_2')
\end{align}
For \(j=0,1,\dots,T\), define $m_j:=T-2j,$ $c_j(\vartheta):=\cos(m_j\frac\vartheta2),$
\begin{align}
    w_{j}\left( {\vartheta\over 2}\right):=-c_{j}'\left( {\vartheta\over 2}\right)=\frac{|m_j|}{2}\sin\left( {|m_j|\vartheta\over 2}\right)\ge 0.
\end{align}
Then
\begin{align}
    S_1'=-\sum_{j=0}^{T}w_j,
\quad
S_2' = 2\sum_{j=0}^{T} c_j c_j'
      = -2\sum_{j=0}^{T} c_j w_j.
\end{align}
Hence
\begin{equation}
\frac{d}{d\vartheta}(1-C_{\rm cyclic})
=
-\frac{2S_1}{(T+1)S_2^2}
\left[
\Big(\sum_j w_j\Big)S_2
-
S_1\Big(\sum_j c_jw_j\Big)
\right].
\label{eq:d1minusC}
\end{equation}
Thus it suffices to prove
\begin{equation}
\Big(\sum_j w_j\Big)S_2
\ge
S_1\Big(\sum_j c_jw_j\Big).
\label{eq:key-ineq}
\end{equation}
In the smallest root interval, $|m_j|\frac\vartheta2 \in [0,\pi/2)$, and hence $c_j(\vartheta) > 0$. Also, $c_j$ is an decreasing function of $|m_j|$ and $w_j$ is an increasing function. Applying Chebyshev' weighted sum inequality to sequences $1\over c_j$ and $w_j$ with weights as $c_j$, we get
\begin{align}
    \sum_j w_j \geq \frac{(T+1) \sum_j c_j w_j}{\sum_j c_j} \implies \frac{1}{T+1}S_1 \sum_j w_j \geq \sum_j c_jw_j
\end{align}
From Cauchy-Schwarz, $S_1^2 \leq (T+1)S_2$. Combing these, we get
\begin{align}
    S_1 \sum_j c_j w_j \leq \frac{S_1^2}{T+1}\sum_j w_j \leq S_2 \sum_j w_j.
\end{align}
Therefore, $\frac{dC_{\rm cyclic}}{d\vartheta} \geq 0$. 

\end{proof}

\begin{theorem}
\label{thm:worstOverlapCyclicGeneral}
Fix \(T\ge 2\). We have 
\begin{align}
    &\inf_{\epsilon\in (0,1), J_{\rm cyc}\in \mathbb{R}^+}|\langle\Psi_\mathrm{cyclic}|\Psi_{\rm free}\rangle|^2 = \frac{8}{\pi^2} - \frac{8}{\pi^2(T+1)}+ O(T^{-2}) 
\end{align}
Moreover,
\begin{align}
    \frac{1}{3} \leq \inf_{\epsilon\in (0,1), J_{\rm cyc}\in \mathbb{R}^+}|\langle\Psi_{\rm cyclic}|\Psi_{\rm free}\rangle|^2 \leq \frac{8}{\pi^2}
\end{align}
Since the boundary point \(\epsilon=1\), \(J_{\rm cyc}=\infty\) is not included in the admissible set, this value is an infimum.
\end{theorem}
\begin{proof}
From \eqref{eqn:cycgs0}, $C_{\rm cyclic}$  depends on \(\epsilon\) and \(J_{\mathrm{cyc}}\) only through
\(\vartheta\). The chain rule gives
\begin{align}
    \frac{\partial C_{\rm cyclic}}{\partial J_{\mathrm{cyc}}}
=
\frac{dC_{\rm cyclic}}{d\vartheta}
\frac{\partial \vartheta}{\partial J_{\mathrm{cyc}}},
\qquad
\frac{\partial C_{\rm cyclic}}{\partial \epsilon}
=
\frac{dC_{\rm cyclic}}{d\vartheta}
\frac{\partial \vartheta}{\partial \epsilon}.
\end{align}
We begin by arguing the positivity of the above gradients.
From Theorem \ref{thm:GSO_H_cyclic2}, $\frac{dC_{\rm cyclic}}{d\epsilon}>0$ for any $J_{\rm cyc}>0$. The proof also shows $\frac{dC_{\rm cyclic}}{d\vartheta}>0$. It remains to show $\frac{d\vartheta}{dJ_{\rm cyc}}>0$. 
Let $\alpha:=1-\sqrt{1-\epsilon}$.
Since \(\epsilon\in(0,1)\), we have \(\alpha>0\). 

The secular equation \eqref{eqn:flosec_simple} is $\beta(\vartheta)=1-J_{\mathrm{cyc}}\alpha,$ where $\beta(\vartheta)$ is defined in \eqref{eq:betaDef}. Differentiating implicitly with respect to $J_{\rm cyc}$ gives
\begin{align}
    \beta'(\vartheta)\frac{d\vartheta}{dJ_{\rm cyc}}
=
-\alpha,
 \implies
    \frac{d\vartheta}{dJ_{\rm cyc}}
=
-\frac{\alpha}{\beta'(\vartheta)}.
\label{eqn:implicitder}
\end{align}
It remains to show that \(\beta'(\vartheta)<0\). A direct differentiation and using basic trigonometric identities yields
\begin{align}
    \beta'(\vartheta)
=
-
\frac{
\sin((T+1)\vartheta)+(T+1)\sin\vartheta
}{
2\cos^2\!\left(\frac{T\vartheta}{2}\right)
}.
\end{align}
On the relevant branch \(0<\vartheta\le \pi/T\), we note that $\beta'(\vartheta)<0$. 
Since \(\alpha>0\), it follows that in (\ref{eqn:implicitder})
\begin{align}
    \frac{d\vartheta}{dJ_{\rm cyc}}>0.
\end{align}
 Hence \(C_{\rm cyclic}\) is increasing in both
\(\epsilon\) and \(J_{\rm cyc}\). Therefore its supremum is obtained
in the limiting regime $\vartheta\to \frac{\pi}{T}.$
 Substituting
\(\vartheta=\pi/T\), we have
\begin{align}
    \sup C_{\rm cyclic}
=
1-
\frac{
2\cos^2\!\left(\frac{\pi}{2T}\right)
}{
T(T+1)\sin^2\!\left(\frac{\pi}{2T}\right)
}
=
1-
\frac{2}{T(T+1)}
\cot^2\!\left(\frac{\pi}{2T}\right).
\end{align}
\noindent Since the squared overlap is $1-C_{\rm cyclic}$, we obtain the result.

\end{proof}

\subsection{Low energy dynamical subspaces\label{sec:leds}}

In the complement of $\text{span}\mathcal{B}$, the matrix of $H_{\rm clamped}$ is the same as that of $H_{\prop}+(\ket{T}\bra{T}_{C}+\ket{0}\bra{0}_{C})\otimes \mathbb{I}_{S}$. In this complementary sector, the energy is larger than the second lowest eigenvalue of $H_{\clamped}$ in the subspace spanned by $\mathcal{B}$.
\begin{theorem}
\label{thm:clampedparams}
    Consider a subspace $\mathcal{K}$ of $CS$ that is invariant under $H_{\mathrm{clamped}}$ (with $J_{\mathrm{in}}=J_{\mathrm{out}}=1$) and orthogonal to $\mathrm{span}\mathcal{B}$. Then  
    \begin{equation}
        H_{\mathrm{clamped}}\big\vert_{\mathcal{K}} \succeq 1-\cos \vartheta^{*}_{1}
    \end{equation}
    where $1-\cos\vartheta_{1}^{*}$ is the second lowest eigenvalue of $H_{\mathrm{clamped}}$ in $\mathrm{span}\mathcal{B}$.
    \end{theorem}
\begin{proof}
    Without loss of generality, take the subspace $\mathcal{K}$ to have the minimal dimension $2(T+1)$. Because $\ket{T}\ket{\psi_{f}}\in \mathcal{B}$, we have $\ket{T}\ket{\psi_{f}}\in \mathcal{K}^{\perp}$, and the matrix in $\mathcal{K}$ has the form
    \begin{align}
        H_{\text{clamped}}\big\vert_{\mathcal{K}}  &= H_{\prop}+(\ket{T}\bra{T}_{C}+\ket{0}\bra{0}_{C})\otimes \mathbb{I}_{S} \big\vert_{\mathcal{K}} \nonumber \\
        & =  \begin{pmatrix}
{3\over 2} &-{1\over 2} &             &             &                    \\
-{1\over 2}&  1         &-{1\over 2}&  &             &                     \\
           &-{1\over 2} &1 &-{1\over 2}&            &                      \\
           &            &  &\ddots     &            &                       \\
           &            &  &-{1\over 2}& 1          &    -{1\over 2}    \\
           &            &  &           &-{1\over 2} & {3\over 2} 
\end{pmatrix}.
    \end{align}

But this matrix has eigenvalues $1-\cos \left( {(k+1)\pi \over T+1}\right)$, $k=0,\ldots, T$ \cite{Losonczi1992}. The second lowest eigenvalue of $H_{\text{clamped}}$ in the subspace spanned by $\ket{\gamma_{t}}$, $\ket{\eta_{t}}$ is $1-\cos \vartheta^{*}_{1}$ with $\vartheta^{*}_{1} = {\pi - \sin^{-1}\sqrt{\epsilon} \over T+1}$.
\end{proof}

The projected dynamics of $H_{\rm cyclic}$ in $\text{span}\mathcal{D}$ can be separated from its projected dynamics in an orthogonal subspace spanned by $\ket{t}_{C}\otimes U_{t}\cdots U_{1}\ket{\phi_{\rm in}}_{S}$, $t=0,\ldots, T$ with $\langle \phi_{\rm in}\vert \psi_{\rm in} \rangle=0$ by choosing $J_{\rm in}$ sufficiently large.
\begin{theorem}
\label{thm:cyclicparams}
    Assume that $\epsilon=1$ in the matrix for $P_{\mathcal{D}}H_{\rm cyclic}P_{\mathcal{D}}$ (i.e., the circuit parameters $\theta$ give $\bra{\gamma_{0}}\ket{0}\bra{T}_{C}\otimes V^{\dagger}\ket{\gamma_{T}}=  \langle \psi_{\rm in}\vert V^{\dagger}U_{T}\cdots U_{1}\ket{\psi_{\rm in}}=0$).   Consider an orthogonal subspace  $\mathcal{K}\subset CS$ given by $\mathcal{K}=\text{span}\lbrace \ket{\xi_{0}},\ldots,\ket{\xi_{T}}\rbrace$ with $\ket{\xi_{t}}_{CS}=\ket{t}_{C}\otimes U_{t}\cdots U_{1}\ket{\phi_{\rm in}}_{S}$ and $\langle \phi_{\rm in}\vert \psi_{\rm in}\rangle = 0$, but with $P_{\mathcal{K}}H_{\rm cyclic}P_{\mathcal{K}}$ having $\epsilon=0$ at circuit parameters $\theta$. For large $T$, $J_{\rm cyc}<{J_{\rm in}\over 2}$ is a sufficient condition that $\lambda_{\rm min}\left( P_{\mathcal{D}}H_{\rm cyclic}P_{\mathcal{D}}\right) < \lambda_{\rm min}\left( P_{\mathcal{K}}H_{\rm cyclic}P_{\mathcal{K}}\right)$.
\end{theorem}
\begin{proof}
In the case of $H_{\rm cyclic}$, consider a $n:= T+1$ dimensional subspace $\ket{\zeta_{0}},\ldots, \ket{\zeta_{T}}$ in which the circuit perfectly compiles, but $\ket{\zeta_{0}}$ is orthogonal to $\ket{\psi_{\text{in}}}$, thereby incurring a $J_{\text{in}}$ penalty. In this subspace, $H_{\text{cyclic}}$ is
\begin{equation}
\begin{pmatrix}
\frac12+J_{\text{in}}+\frac{J_{\text{cyc}}}{2} & -\frac12 & 0 & \cdots & 0 & -\frac{J_{\text{cyc}}}{2} \\
-\frac12 & 1 & -\frac12 & \cdots & 0 & 0 \\
0 & -\frac12 & 1 & \ddots & 0 & 0 \\
\vdots & \vdots & \ddots & \ddots & -\frac12 & \vdots \\
0 & 0 & 0 & -\frac12 & 1 & -\frac12 \\
-\frac{J_{\text{cyc}}}{2} & 0 & 0 & \cdots & -\frac12 & \frac{1+J_{\text{cyc}}}{2}
\end{pmatrix}
\end{equation}
The secular equation is
\begin{equation}
\left(\cos\vartheta + J_{\mathrm{in}} + J_{\mathrm{cyc}} - 1\right)
U_{n-1}(\cos\vartheta)
+
\left[J_{\mathrm{in}}\left(J_{\mathrm{cyc}} - 1\right) - J_{\mathrm{cyc}}\right]
U_{n-2}(\cos\vartheta)
-
J_{\mathrm{cyc}}
= 0.
\label{eqn:overlapexpl}
\end{equation}
with $U_n(\cos\vartheta)
=
\frac{\sin((n+1)\vartheta)}{\sin\vartheta}$ the Chebyshev polynomial of the second kind.

We will assume the opposite situation to occur in the $\ket{\gamma_{t}}$ sector (i.e., in $\text{span}\mathcal{D}$): the circuit totally fails to compile ($\epsilon=1$). We know that $H_{\rm cyclic}$ does not incur a $J_{\rm in}$ penalty in this sector and the secular equation is (\ref{eqn:flosec_simple}) with $\epsilon=1$. Take the smallest positive solution $\vartheta^{*}$ of that equation, and use it to solve for $J_{\rm in}$ in (\ref{eqn:overlapexpl}). 
In other words, at this value of $J_{\mathrm{in}}$, $\vartheta^{*}$ is also a solution of (\ref{eqn:overlapexpl}) 
(below this value of $J_{\text{in}}$, the $\ket{\zeta_{t}}$ sector will have a lower energy state than the $\ket{\gamma_t}$ sector because $J_{\text{in}}$ has been reduced while the other couplings remain fixed). The critical value of $J_{\text{in}}$ is 
\begin{align}
    J_{\text{in}} &= 
J_{\text{cyc}}+
\frac{
J_{\text{cyc}}^2\left(J_{\text{cyc}}-1+\cos\vartheta^{*}\right)
}{
J_{\text{cyc}}^2+2(1-J_{\text{cyc}})\left(1-\cos\vartheta^{*}\right)
} \nonumber \\
&= 2J_{\text{cyc}} + O\left( {1\over T^{2}} \right).
\end{align}
\end{proof}

\subsection{Simulation of time dynamics of $P_{\mathcal{D}}H_{\rm cyclic}P_{\mathcal{D}}$\label{sec:qpesupp}}

Consider $H_{\rm free}$ with $J_{\rm in}=J_{\rm prop}=1$. With 
\begin{equation}
    W_{t}(\theta):= \sum_{t=0}^{T}\ket{t}\bra{t}_{C}\otimes U_{t}(\theta_{t})\cdots U_{1}(\theta_{1})
\end{equation}
one finds that
\begin{align}
    W_{t}(\theta)^{\dagger}H_{\rm free}W_{t}(\theta) &= L\otimes \mathbb{I}_{S} + \ket{0}\bra{0}_{C}\otimes \left( \mathbb{I}_{S} - \ket{\psi_{\rm in}}\bra{\psi_{\rm in}} \right) \\
    &= L\otimes \ket{\psi_{\rm in}}\bra{\psi_{\rm in}} + \left( L+\ket{0}\bra{0}_{C} \right)\otimes \left( \mathbb{I}_{S} - \ket{\psi_{\rm in}}\bra{\psi_{\rm in}} \right)
\end{align}
where $L:={1\over 2}\sum_{t=1}^{T}\left( \ket{t}-\ket{t-1}\right)\left( \bra{t}-\bra{t-1} \right)$. There are two spectral sectors of $W_{t}(\theta)^{\dagger}H_{\rm free}W_{t}(\theta)$ with interlacing eigenvalues: the sector with $\ket{\psi_{\rm in}}\bra{\psi_{\rm in}}$ has eigenvalues $1-\cos {k\pi \over T+1}$  and the sector with $\mathbb{I}_{S} - \ket{\psi_{\rm in}}\bra{\psi_{\rm in}}$ has eigenvalues $1-\cos {\left(k + {1\over 2} \right)\pi \over T+1}$, $k=0,\ldots, T$. The sectors are separated by $O(T^{-2})$. But
\begin{equation}
    W_{t}(\theta)L\otimes \ket{\psi_{\rm in}}\bra{\psi_{\rm in}}W_{t}(\theta)^{\dagger} = P_{\mathcal{D}}H_{\rm free}P_{\mathcal{D}},
\end{equation}
so doing QPE with respect to $H_{\rm free}$ to precision $O(T^{-2})$ is enough to coherently implement the reflection operator $R_{\mathcal{D}}:=2P_{\mathcal{D}} - \mathbb{I}_{S}$ by giving a $-1$ phase to the $1-\cos {\left(k + {1\over 2} \right)\pi \over T+1}$ eigenvalues. 
By writing $H_{\rm cyclic}$ as a block matrix according to the decomposition $CS=\text{span}\mathcal{D} \oplus \text{span}\mathcal{D}^{\perp}$, one finds that 
\begin{equation}
    {H_{\rm cyclic}+R_{\mathcal{D}}H_{\rm cyclic}R_{\mathcal{D}} \over 2} = P_{\mathcal{D}}H_{\rm cyclic}P_{\mathcal{D}} + P_{\mathcal{D}}^{\perp}H_{\rm cyclic}P_{\mathcal{D}}^{\perp}.
\end{equation}
It follows that for an initial state in $\text{span}\mathcal{D}$, e.g., the history state, the alternating dynamics gives
\begin{equation}
    \lim_{m\rightarrow \infty}\left( e^{-i{t\over 2m}H_{\rm cyclic}}R_{\mathcal{D}}e^{-i{t\over 2m}H_{\rm cyclic}}R_{\mathcal{D}}\right)^{m}\ket{\Psi_{\rm free}} = e^{-itP_{\mathcal{D}}H_{\rm cyclic}P_{\mathcal{D}}}\ket{\Psi_{\rm free}},
\end{equation}
where $H_{\rm cyclic}$ is not necessarily implemented with $J_{\rm in}=1$. For fixed $m$, the error is obtained from the first order Trotter expansion.

\section{Numerical optimization methods}
\subsection{Optimization of cost functions and trainability} 

For randomly initialized circuits, the cost functions above do not have a straightforward guarantee on the convergence of gradient descent due to their non-convexity. However, our analytical results indicate that for $\epsilon(\theta)$ near 1, the curvature of these cost functions at local maxima is much greater than for $\epsilon(\theta)$ itself, suggesting that descent to local minima should converge faster. To substantiate this intuition, we have identified short depth circuits for which the cost functions have a geometric convergence rate ${\mu_{\text{PL}}\over L}$ in a large neighborhood of the global minimum due to locally possessing the Polyak-Lojasiewicz property ${1\over 2}\Vert \nabla C\Vert^{2} \ge \mu_{\text{PL}} C$, where $\mu_{\text{PL}}>0$ is the PL constant \cite{karimi}, and having $L$-Lipschitz gradient. Table \ref{tab:prl_example} shows guarantees for gradient descent on two short-depth identity compiling tasks: Task A-- compiling the $n$-qubit identity from a single-layer ($T=1$) Pauli rotation on $[0,\pi]$ (on $[0,\pi-\delta]$ for $\epsilon(\theta)$), and Task B-- compiling the one-qubit identity from an $SU(2)$ $KAK$ decomposition $e^{-i\theta_{3}J_{z}}e^{-i\theta_{2} J_{y}}e^{-i\theta_{1}J_{z}}$ (so $T=3$) on the domain $S_{\delta}=\lbrace \theta \in \mathbb{R}^{3} \, : \, 0\le \theta_{1}+\theta_{3} \le \pi - \delta \, \wedge \, 0 \le \theta_{2} \le \pi - \delta\rbrace$. The latter task exhibits a 5 order-of-magnitude difference in PL constants on the stated neighborhood. 

\begin{table}[t]
\caption{Gradient descent rates $\mu_{\text{PL}}/L$ for compiling Tasks A and B of the main text. For Task A, $\epsilon(\theta)$ is considered on the domain $\theta\in [0,\pi - \delta]$, whereas other cost functions are defined on $[0,\pi]$ (the overlap cost for $H_{\text{cyclic}}$ is not defined for $T=1$). For Task B, $\epsilon(\theta)$ is defined on $S_{\delta}$ defined in the main text, and the other cost functions are defined on $S_{0.1}$. All coupling constants in the Hamiltonians are equal to 1. The overlap cost function for $H_{\text{cyclic}}$ is identically zero for $T=1$.}
\label{tab:prl_example}
\begin{ruledtabular}
\begin{tabular}{cccc}
Cost function & Hamiltonian & Rate A& Rate B \\
\hline
$\epsilon(\theta)$ & -- & $\sin^{2}\left( {\delta \over 2} \right)$ & $\mathcal{O}\left( \sin^{6}\left( {\delta\over 2}\right)\right)$\\
$\lambda(\theta)$ & $H_{\text{clamped}}$ & $\cos^{2}\left( {\pi \over 8}\right)$ & $3.1 \times 10^{-3}$ \\
$1-\vert\langle \Psi_{\text{free}}\vert \Psi\rangle\vert^{2}$  & $H_{\text{clamped}}$ & $0.37$ & $1.7 \times 10^{-3}$\\
$\lambda(\theta)$ & $H_{\text{cyclic}}$ & ${1\over 4}$ & $1.0 \times 10^{-3}$\\
$1-\vert\langle \Psi_{\text{free}}\vert \Psi\rangle\vert^{2}$  & $H_{\text{cyclic}}$ & -- & Not PL 
\end{tabular}
\end{ruledtabular}
\end{table}

In Task B of Table \ref{tab:prl_example}, the Polyak-Lojasiewicz (PL) constant and Lipschitz constant of the cost function and the cost function gradients, respectively, on the domain $S_{\delta}:=\lbrace \theta \, : \, \theta_{2}\in[0,\pi-\delta] \wedge \theta_{1}+\theta_{3} \in [0,\pi-\delta]\rbrace$ were calculated using
\begin{align}
    \mu_{\delta}&:= \text{min}_{\theta \in S_{\delta}}{\Vert \nabla C(\theta) \Vert^{2} \over 2C(\theta)} \nonumber \\
    L_{\delta}&:=\max_{\theta\in S_{\delta}}\Vert C''(\theta)\Vert,
\end{align}
where $C''$ is the Hessian, whose operator norm gives the Lipschitz constant of $\nabla C$ because the cost functions are smooth.

Despite the fact that the cost functions scale non-linearly with the fidelity and have a range that is determined by $T$, we expect that their asymptotic trainability properties are inherited from those of $\epsilon(\theta)$ (e.g., barren plateau phenomenon \cite{McClean2018BarrenPlateaus, Cerezo2021}), although the actual scaling must be considered carefully. For example, if $\epsilon(\theta)$ probabilistically concentrates according to $\text{Var}_{\theta}\epsilon(\theta)$, the ${1\over 2}$-H\"{o}lder property of cost function $\lambda(\epsilon(\theta))$ (note that this cost function is not Lipschitz on $[0,1]$)
\begin{equation}
    \vert \lambda(\epsilon) - \lambda(\epsilon')\vert \le {\pi^{2}\over 4(T+1)^{2}}\vert \epsilon - \epsilon ' \vert^{1\over 2} 
\end{equation}
implies that $\text{Var}_{\theta}\lambda(\epsilon(\theta)) \le {\pi^{4}\over 16(T+1)^{4}}\sqrt{\text{Var}_{\theta}\epsilon(\theta)}$.

\subsection{Gradients and sweeps}
We evaluate exact Hellmann--Feynman coordinate gradients of the FK energy using the analytic $\pi$-shift identity and apply them in fixed-order coordinate-descent sweeps.  
A training sweep counts one full optimizer pass over all trainable parameters. In QAQC, each parameter gradient is obtained from two evaluations of the circuit fidelity, at $\theta_j\pm\pi/2$, giving $2p$ fidelity evaluations per sweep. In the sequential FK implementation used in our simulations, for $p$ parameters, a sweep consists of $p$ clamped FK ground state preparations, plus one preparation of the free FK ground state used to calculate the fidelity. Thus the observed sweep reduction should be interpreted as improved optimization conditioning. The potential hardware advantage of clamped methods would require a substrate in which the FK/clamped equilibrium state is prepared natively or efficiently, so that equilibrium relaxation replaces digital ground state preparation. 

 Implemented sequentially, one sweep uses $p$ ground state preparations for the energy costs and $4p$ for the finite-difference overlap cost, compared with $2p$ shifted-circuit fidelity evaluations for direct QAQC; Sec.~\ref{sec:twoGS} reduces the energy-cost count to two preparations per sweep using the clock-edge structure.

Gradients are determined analytically using the exact $\pi$-shift identity.  
In the numerical implementation, the relevant FK Hamiltonian is assembled at the current parameter vector and diagonalized to obtain the required ground state. For E2 this is the clamped Hamiltonian; for Floquet training it is the corresponding cyclic Hamiltonian; and for O2, the free and clamped ground states are computed at $\theta_j\pm\delta$ and the central difference of the overlap cost gives the coordinate gradient, so an O2 sweep uses $4p$ ground state solves. The $\pi$-shift is used only to assemble the analytic operator $\partial_{\theta_t}H_{\rm prop}$: for a trainable gate, the code replaces the local propagation gate by $U_t(\theta_t+\pi)$ in the derivative of the corresponding FK propagation edge.  It then evaluates the expectation value of this derivative operator on the computed ground state.  Thus the simulation does  not finite-difference the FK energy with respect to $\theta_t$.

After each coordinate update in the fully sequential implementation, the parameter vector is immediately updated before moving to the next coordinate in the sweep.  Thus the reported sequential FK curves use one fresh equilibrium solve per trainable coordinate.  In the two ground state variants, the same analytic shifted-gate derivative is used, but the ground state is reused across one parity block of coordinates, respectively.  Fixed nontrainable gates are skipped.  No stochastic sampling,  or hardware measurement noise are included in these simulations. 
All clamped-Hamiltonian solves use $J_{\rm out}=J_{\rm in}=J_{\rm prop}=1$. All cyclic-Hamiltonian solves, including the fully-sequential implementation and the two ground state energy descent and the incremental cost $C_{\rm inc}$, use $J_{\rm in}=J_{\rm prop}=1$ and $J_{\rm cyc}=0.2$, satisfying the dynamical-sector condition $2J_{\rm cyc}<J_{\rm in}$ of Theorem~\ref{thm:cyclicparams}. Learning rates were calibrated at these couplings.

\paragraph{Numerical training protocol.} To compare the cost functions directly for deeper circuits, we numerically calculate the fidelity $F=1-\epsilon(\theta)$ under the update rule given by the respective cost function gradients.
All compiling benchmarks were performed in noiseless state-vector simulation.  For each task and optimizer, circuit parameters were initialized randomly as $\theta_j(0)\sim \mathrm{Unif}[-\pi,\pi]$.  The same initialization seeds were used across all methods, so that training times could be compared seed-by-seed.  Convergence was defined as reaching target fidelity $F\geq 0.99$.  Runs that did not reach this threshold within the fixed training budget were assigned the maximum sweep count when computing median training times and paired differences.

For the fixed-task benchmark suite, final performance statistics were computed using $20$ random initializations per task and method.  The fixed-task suite consisted of two-qubit Bell-state preparation, four-qubit GHZ-state preparation, Choi-state identity compiling, and Choi-state CNOT compiling tasks.  For the random-generator identity-compiling benchmark, we used $5$ independently sampled generator instances and $20$ random initializations per instance, giving $100$ paired optimization trajectories per method.  Learning-rate calibration was performed using a separate set of $10$ random initializations that were not included in the final reported statistics.

\paragraph{Gradients and objective functions.}
Direct QAQC used the standard $\pm\pi/2$ parameter-shift rule on the output fidelity.  
 At each training step we diagonalized the clamped Feynman-Kitaev Hamiltonian, obtained its ground state $|\Psi_0(\theta)\rangle$, and evaluated
$\partial_{\theta_j}E_0=\langle \Psi_0(\theta)|\partial_{\theta_j}H_{\rm FK}(\theta)|\Psi_0(\theta)\rangle$.
The derivative $\partial_{\theta_j}H_{\rm FK}$ was constructed analytically. For Pauli rotations $U_j(\theta_j)=\exp(-i\theta_j G_j/2)$ with $G_j^2=I$, the gate derivative can be written using the exact $\pi$-shift identity $\partial_{\theta_j}U_j(\theta_j)=U_j(\theta_j+\pi)/2$, which gives the corresponding exact derivative of the propagation Hamiltonian. This is distinct from applying the ordinary $\pm\pi/2$ parameter-shift rule to the history-state energy itself. The ordinary parameter-shift rule does not apply directly because the FK cost is evaluated on an equilibrium state \(\ket{\Psi_0(\theta)}\) that changes with the same Hamiltonian \(H(\theta)\) being differentiated.

The E2 and Floquet methods use exact state-vector Hellmann-Feynman gradients assembled from the analytic $\pi$-shift propagation derivative. The O2 method performs coordinate descent on the scalar overlap cost $C_{\rm overlap}$, with each coordinate gradient estimated by central finite differences ($\delta=10^{-3}$) of the cost evaluated from the free and clamped ground states at $\theta_j\pm\delta$; because $C_{\rm overlap}$ depends on the ground state eigenvectors, no Hellmann-Feynman identity applies. Sampling noise and hardware-measurement error were excluded from all simulations in order to isolate the training dynamics.

\paragraph{Learning-rate calibration and fair comparison.}
Learning rates were calibrated independently for each method and task using logarithmic learning-rate scans on the calibration initializations.  For each method/task pair, all candidate learning rates were evaluated on the same calibration seeds.  The learning rate used in the final benchmark was selected from this scan by prioritizing successful convergence and then minimizing the median number of sweeps to reach $F\geq0.99$ on the calibration set.  After selection, the learning rate was held fixed for all final runs of that method and task.  Thus the reported direct-QAQC baseline is a tuned fixed-learning-rate direct-fidelity optimizer, not an adaptive or line-search method.

For the fixed-task benchmark suite, learning rates were selected by a two-stage procedure: an initial logarithmic scan over method-dependent ranges was performed on held-out calibration seeds, followed by a local refinement of $17$ logarithmically spaced values $\eta=\eta_0,10^x$ with $x\in[-0.75,0.75]$ around the best initial value, evaluated on $10$ calibration seeds per method/task pair. 

To check that the comparison was not limited by under-tuning of the direct-QAQC baseline,  we performed a broader audit using $41$ logarithmically spaced learning rates over $\eta\in[0.05,100]$, again on held-out calibration seeds.  The selected QAQC learning rates from this audit were then evaluated on the same held-out final seeds used for the other methods.  We also verified that the selected learning rates were not artifacts of being forced to the boundary of the scanned range.

For the random-generator identity benchmark, each method used a $25$-point logarithmic learning-rate scan over a method-appropriate four-decade range, evaluated on $10$ calibration seeds for each generator instance before final evaluation on held-out optimization seeds. 
The breadth of this audit indicates that the slower QAQC convergence is not an artifact of learning-rate selection. 

\paragraph{Paired performance statistics.}
Final comparisons were made using paired initialization seeds.  For each task and seed, we compared the number of training sweeps required by each method to the corresponding direct-QAQC run.  We report absolute median sweeps to reach $F\geq0.99$ and paired sweep differences $\Delta N=N_{\mathrm{method}}-N_{\mathrm{QAQC}}$.  Negative $\Delta N$ indicates that the method reaches the target fidelity in fewer sweeps than direct QAQC.  Win, tie, and loss counts were computed from these paired differences over identical task and initialization instances.

\section{Graph Laplacian}
\label{sec:graph-laplacian}
Here we formulate the resistor-network and FK Hamiltonian as special cases of a weighted connection Laplacian. Consider a directed graph $G=(V,E)$ with nodes $V=\{0,\dots,T\}$, positive edge weights $g_e$, and  a unitary voltage assignment $\sigma:E\to U(d)$.  For each edge $e:a\to b$, let $S_e:=\ketbra{b}{a}\otimes\sigma(e)$ act on $\mathcal{H}_{\text{clock}}\otimes\mathcal{H}_{\text{sys}}=\mathbb{C}^{|V|}\otimes\mathbb{C}^d$.   The connection Laplacian of $(G,\sigma)$ is
\begin{equation}
  L_\sigma := \frac{1}{2}\sum_{e:a\to b} g_e \Big[\ketbra{a}{a}\otimes I +\ketbra{b}{b}\otimes I - S_e - S_e^\dagger\Big]\,.
  \label{eq:connection_laplacian}
\end{equation}
On the path graph $P_{T+1}$ with voltages $\sigma(e_t)=U_t(\theta_t)$, this reduces to the FK propagation Hamiltonian, $H_{\text{prop}}(\theta)=L_\sigma$. 

We generalize the resistor-network Laplacian form, $L_{\rm res}$, by replacing scalar node voltages $v_t$ with clock-slice amplitudes $\phi_t\in\mathcal H_{\rm sys}$, and scalar edge differences $v_b-v_a$ with differences $\phi_b-\sigma(e)\phi_a$; the scalar conductivity $g_e$ is the classical edge weight, while the circuit gate $U_t$ is the unitary connection used to compare neighboring clock slices:
\begin{align}
L_{\rm res}
&=
\frac12\sum_{e:a\to b} g_e\,
\bigl(\mathbf e_b-\mathbf e_a\bigr)
\bigl(\mathbf e_b-\mathbf e_a\bigr)^\dagger ,
\nonumber\\
L_{\rm FK}(\theta)
&=
\frac12\sum_{t=1}^{T}
\Bigl(|t\rangle\otimes I-|t-1\rangle\otimes U_t^\dagger(\theta_t)\Bigr)
\Bigl(\langle t|\otimes I-\langle t-1|\otimes U_t(\theta_t)\Bigr).
\label{eq:resistor_to_fk_laplacian}
\end{align}
Here $\mathbf e_a$ denotes the standard basis vector associated with vertex $a$. Thus $L_{\rm res}$ is the scalar specialization of Eq.~\eqref{eq:connection_laplacian}, with $d=1$, $\sigma(e)=1$, and vertex field $\phi_a=v_a$, whereas $L_{\rm FK}$ is its clock-path specialization, with $g_{e_t}=1$, $\sigma(e_t)=U_t(\theta_t)$, and vertex field $\phi_t\in\mathcal H_{\rm sys}$. 

Having identified \(H_{\rm prop}\) as the clock-network Laplacian, we now add the analogue of terminal loads, which provides the basis of our formulation of  equilibrium learning.  In classical loaded networks, sources and loads enter the variational principle as boundary terms~\cite{Meixner_JMathPhys_1963,Meixner_1973,BraytonMoser1964}.  In the FK clock network, the corresponding boundary variables are the endpoint components which fail to lie in the prescribed input and target subspaces:
\begin{equation}
    \delta_{\rm in}\phi_0:=(I-P_{\rm in})\phi_0,
    \qquad
    \delta_{\rm out}\phi_T:=(I-P_f)\phi_T,
    \label{eq:boundary_residuals}
\end{equation}
where \(P_{\rm in}=\ketbra{\psi_{\rm in}}{\psi_{\rm in}}\) and \(P_f=\ketbra{\psi_f}{\psi_f}\). 
Consider the clock path $P_{T+1}$ with vertices $\{0,\dots,T\}$ and directed edges $e_{t}\colon(t{-}1)\to t$. Decompose a state as $\ket{\Psi}=\sum_{t=0}^{T}\ket{t}\otimes\ket{\phi_{t}}$ with $\phi_{t}\in\mathcal{H}_{\mathrm{sys}}$. The gate $U_{t}(\theta_{t})$ acts as the transport operator on edge~$e_{t}$, and we define the \emph{edge mismatch}
\begin{equation}
    \delta_U\phi_t \;:=\;\phi_{t}\;-\;U_{t}(\theta_{t})\,\phi_{t-1}\,,
    \qquad t=1,\dots,T\,.
    \label{eq:constitutive}
\end{equation}
This is a Hilbert space-valued constitutive relation: it relates the ``current'' $\delta_U\phi_t$ on each edge to the ``potentials'' $\phi_{t},\phi_{t-1}$ at  adjacent vertices.
The covariant voltage drop across the edge \( (t-1) \to t\) is \(\phi_t-U_t(\theta_t)\phi_{t-1}\), the voltage assignment \(U_t(\theta_t)\) plays the corresponding edge-local role: it specifies how the adjacent clock-slice amplitudes are compared, and the resulting mismatch is \(\delta_U\phi_t\).  Thus \(U_t\) is a unitary, lossless analogue of a trainable admittance, i.e., conductivity, while \(\delta_U\phi_t\) is the current-like residual which enters the quadratic network energy, shown below. 
The loaded FK functional is
\begin{equation}
     \mathcal E[\phi;\theta,J_{\rm out}]
    =
    \frac{J_{\rm prop}}{2}\sum_{t=1}^{T}\|\delta_U\phi_t\|^2
    +J_{\rm in}\|\delta_{\rm in}\phi_0\|^2
    +J_{\rm out}\|\delta_{\rm out}\phi_T\|^2 .
    \label{eq:loaded_fk_functional}
\end{equation}
Equivalently,
\begin{align}
    &\mathcal E[\phi;\theta,J_{\rm out}]
    =
    \bra{\Psi}\big(J_{\rm prop} H_{\rm prop}(\theta)+H_{\rm in}+H_{\rm out}\big)\ket{\Psi},
    \label{eq:loaded_fk_hamiltonian}\\
    &H_{\rm in} = J_{\rm in}\ketbra{0}{0}\otimes(I-P_{\rm in}),
    \quad
    H_{\rm out} = J_{\rm out}\ketbra{T}{T}\otimes(I-P_f).
\end{align}
 For fixed circuit parameters \(\theta\), the normalized minimizer of \(\mathcal E\) is the equilibrium state of this loaded clock network.  Setting \(J_{\rm out}=0\) gives the free, input-clamped equilibrium; setting \(J_{\rm out}>0\) gives the clamped equilibrium, where target mismatch appears as terminal load energy.  
 Accordingly, the $J_{\rm out}=0$ and $J_{\rm out}>0$ cases of Eq.~\eqref{eq:loaded_fk_functional} are the FK counterparts of the free and clamped resistor-network functionals in Eqs.~\eqref{eq:resistor_dirichlet} and \eqref{eq:resistor_clamped_energy}, respectively. 
Training is the outer problem: the circuit parameters are updated using the dependence of these ground energies, or free-clamped contrasts, on \(\theta\).

In the FK circuit, the edge mismatch, eq.~\eqref{eq:constitutive}, is exactly a gate mismatch which 
 defines the circuit parallel transport, precisely as Ohm's law relates branch current to node voltages via $i_{e}=G_{e}(v_{\mathrm{start}}-v_{\mathrm{end}})$ in a resistive circuit~\cite{Meixner_JMathPhys_1963,BraytonMoser1964}.
When the transport is perfect ($\phi_{t}=U_{t}\phi_{t-1}$), the edge carries zero mismatch. 

\section{Benchmark circuit definitions}
\label{subsec:benchmark-circuits}

We specify here the trainable circuits used in the benchmark suite.  
All benchmarks whose trainable subsystem consists of two qubits use the same two-qubit circuit block.  
The block has fifteen trainable rotation angles and three fixed entangling gates:
\begin{align}
U_2(\theta)
={}&
{\cal R}_{z}^{(2)}(\theta_{14},\theta_{15})
{\cal R}_{y}^{(2)}(\theta_{12},\theta_{13})
{\cal R}_{z}^{(2)}(\theta_{10},\theta_{11})
C_{21}
R_y^{(2)}(\theta_9)
C_{12}
R_y^{(2)}(\theta_8)
R_z^{(1)}(\theta_7)
\nonumber\\
&\times
C_{21}
{\cal R}_{z}^{(2)}(\theta_5,\theta_6)
{\cal R}_{y}^{(2)}(\theta_3,\theta_4)
{\cal R}_{z}^{(2)}(\theta_1,\theta_2).
\label{eq:two-qubit-benchmark-ansatz}
\end{align}
The rightmost factor acts first.  For a single-qubit rotation on qubit $j$,
\begin{equation}
R_z^{(j)}(\alpha)
=
\exp\left(-\frac{i}{2}\alpha Z_j\right),
\qquad
R_y^{(j)}(\alpha)
=
\exp\left(-\frac{i}{2}\alpha Y_j\right).
\end{equation}
We denote by $C_{jk}$ the controlled-NOT gate with control qubit $j$ and target qubit $k$.  For compactness, we also write
\begin{equation}
{\cal R}_{\alpha}^{(n)}(a_1,\ldots,a_n)
=
\prod_{j=1}^{n} R_{\alpha}^{(j)}(a_j),
\qquad
\alpha\in\{y,z\},
\end{equation}
where the factors commute because they act on different qubits.

\begin{figure}[ht!]
\centering
\begin{adjustbox}{max width=\linewidth}
\begin{quantikz}[row sep=0.45cm, column sep=0.38cm]
\lstick{$q_1$}
& \gate{R_z(\theta_1)}
& \gate{R_y(\theta_3)}
& \gate{R_z(\theta_5)}
& \targ{}
& \gate{R_z(\theta_7)}
& \ctrl{1}
& \qw
& \targ{}
& \gate{R_z(\theta_{10})}
& \gate{R_y(\theta_{12})}
& \gate{R_z(\theta_{14})}
& \qw \\
\lstick{$q_2$}
& \gate{R_z(\theta_2)}
& \gate{R_y(\theta_4)}
& \gate{R_z(\theta_6)}
& \ctrl{-1}
& \gate{R_y(\theta_8)}
& \targ{}
& \gate{R_y(\theta_9)}
& \ctrl{-1}
& \gate{R_z(\theta_{11})}
& \gate{R_y(\theta_{13})}
& \gate{R_z(\theta_{15})}
& \qw
\end{quantikz}
\end{adjustbox}
\caption{
Two-qubit trainable block $U_2(\theta)$.  This same block is used as the active circuit in the Bell state preparation,  and unitary compiling benchmarks.
}
\label{fig:supp-v2-block}
\end{figure}
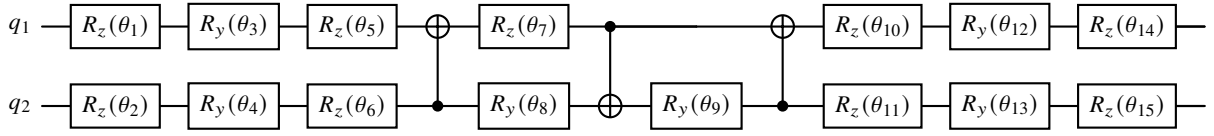

Equivalently, the two-qubit trainable block is shown in Fig.~\ref{fig:supp-v2-block}.

We separate the benchmark tasks as state preparation, Bell state and GHZ state preparation, Sec.~\ref{subsec:2bell} and Sec.~\ref{subsec:GHZ}, and unitary compiling tasks, Sec.~\ref{subsec:unitary}.

For a state-preparation task, the target operation $V$ may be chosen as any fixed unitary extension satisfying $V\ket{\psi_{\rm in}}=\ket{\psi_f}$. Consequently,
\begin{equation}
\left| \bra{\psi_{\rm in}}V^\dagger U(\theta)\ket{\psi_{\rm in}} \right|^2
= \left| \bra{\psi_f}U(\theta)\ket{\psi_{\rm in}} \right|^2,
\end{equation}
so the unitary-overlap formulation reduces to the desired state fidelity.

\subsection{Two-qubit Bell-state preparation}
\label{subsec:2bell}
The Bell-state preparation task uses the input and target states
\begin{equation}
\ket{\psi_{\rm in}}
=
\ket{00},
\qquad
\ket{\psi_{\rm tar}}
=
\ket{\Phi^+}
=
\frac{1}{\sqrt{2}}
\left(\ket{00}+\ket{11}\right).
\end{equation}
The trained output is
\begin{equation}
\ket{\psi_{\rm out}(\theta)}
=
U_2(\theta)\ket{00},
\end{equation}
and the reported fidelity is
\begin{equation}
F_{\rm Bell}(\theta)
=
\left|
\left\langle \Phi^+ \right|
U_2(\theta)
\left|00\right\rangle
\right|^2 .
\end{equation}

\subsection{Four-qubit GHZ-state preparation}
\label{subsec:GHZ}
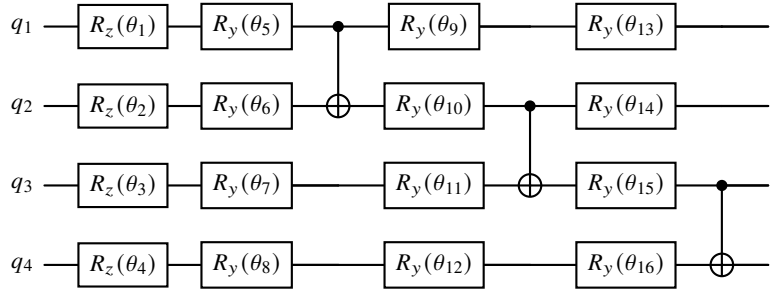
\begin{figure}[ht!]
\centering
\begin{adjustbox}{max width=\linewidth}
\begin{quantikz}[row sep=0.45cm, column sep=0.45cm]
\lstick{$q_1$}
& \gate{R_z(\theta_1)}
& \gate{R_y(\theta_5)}
& \ctrl{1}
& \gate{R_y(\theta_9)}
& \qw
& \gate{R_y(\theta_{13})}
& \qw
& \qw \\
\lstick{$q_2$}
& \gate{R_z(\theta_2)}
& \gate{R_y(\theta_6)}
& \targ{}
& \gate{R_y(\theta_{10})}
& \ctrl{1}
& \gate{R_y(\theta_{14})}
& \qw
& \qw \\
\lstick{$q_3$}
& \gate{R_z(\theta_3)}
& \gate{R_y(\theta_7)}
& \qw
& \gate{R_y(\theta_{11})}
& \targ{}
& \gate{R_y(\theta_{15})}
& \ctrl{1}
& \qw \\
\lstick{$q_4$}
& \gate{R_z(\theta_4)}
& \gate{R_y(\theta_8)}
& \qw
& \gate{R_y(\theta_{12})}
& \qw
& \gate{R_y(\theta_{16})}
& \targ{}
& \qw
\end{quantikz}
\end{adjustbox}
\caption{
Four-qubit GHZ-state preparation ansatz.  This is a state-preparation benchmark on four active qubits, not a universal four-qubit unitary compilation ansatz.
}
\label{fig:supp-ghz-block}
\end{figure}

The GHZ benchmark is not a two-qubit unitary-compiling task.  It instead uses a separate four-qubit state-preparation ansatz in which all four qubits are active.  The input and target states are
\begin{equation}
\ket{\psi_{\rm in}}
=
\ket{0000},
\qquad
\ket{\psi_{\rm tar}}
=
\ket{\mathrm{GHZ}_4}
=
\frac{1}{\sqrt{2}}
\left(\ket{0000}+\ket{1111}\right).
\end{equation}
The trainable circuit has sixteen rotation angles and a nearest-neighbor CNOT chain:
\begin{align}
U_{\rm GHZ}(\theta)
={}&
C_{34}
{\cal R}_{y}^{(4)}
(\theta_{13},\theta_{14},\theta_{15},\vartheta_{16})
C_{23}
{\cal R}_{y}^{(4)}
(\theta_9,\theta_{10},\theta_{11},\vartheta_{12})
\nonumber\\
&\times
C_{12}
{\cal R}_{y}^{(4)}
(\theta_5,\theta_6,\theta_7,\theta_8)
{\cal R}_{z}^{(4)}
(\theta_1,\theta_2,\theta_3,\theta_4).
\label{eq:ghz-benchmark-ansatz}
\end{align}
Thus $ \ket{\psi_{\rm out}(\theta)} = U_{\rm GHZ}(\theta)\ket{0000}$, with fidelity
\begin{equation}
F_{\rm GHZ}(\theta)
=
\left|
\left\langle \mathrm{GHZ}_4 \right|
U_{\rm GHZ}(\theta)
\left|0000\right\rangle
\right|^2 .
\end{equation}
The circuit is given in Fig.~\ref{fig:supp-ghz-block}.

The four-qubit GHZ benchmark is a state-preparation task on four active qubits, not a universal four-qubit unitary-compilation task.  

\subsection{Identity and  \texttt{CNOT} compiling}
\label{subsec:unitary}

This reference-register encoding is the Choi-Jamio{\l}kowski isomorphism; the fidelity of Eq.~\eqref{eq:ChoiFidelity} coincides with the Hilbert-Schmidt-test cost used in QAQC~\cite{khatri2019quantum}. 
The unitary compiling benchmarks test two-qubit unitary synthesis.  We write $A=A_1A_2$ for the active two-qubit register and $R=R_1R_2$ for the two-qubit reference register.  The code ordering is $A_1,A_2,R_1,R_2$. The input  state is
\begin{equation}
\ket{\Omega}_{AR}
=
\frac{1}{2}
\sum_{x\in\{0,1\}^2}
\ket{x}_{A}\ket{x}_{R}.
\label{eq:AR_state}
\end{equation}
The trainable circuit acts only on the active register:
\begin{equation}
\ket{\Omega_{U_2(\theta)}}
=
\left(U_2(\theta)\otimes I_R\right)\ket{\Omega}_{AR}.
\end{equation}
For a target two-qubit unitary $V$, the target  state is
\begin{equation}
\ket{\Omega_{V}}
=
\left(V\otimes I_R\right)\ket{\Omega}_{AR}.
\end{equation}
The  fidelity is
\begin{equation}
F_{\rm Choi}(\theta;V)
=
\left|
\left\langle
\Omega_{V}
\middle|
\Omega_{U_2(\theta)}
\right\rangle
\right|^2
=
\frac{1}{16}
\left|
\operatorname{Tr}
\left[
V^{\dagger}U_2(\theta)
\right]
\right|^2 .
\label{eq:ChoiFidelity}
\end{equation}
This reference-register encoding is the Choi-Jamio{\l}kowski isomorphism; the fidelity of Eq.~\eqref{eq:ChoiFidelity} coincides with the Hilbert-Schmidt-test cost used in QAQC~\cite{khatri2019quantum}. 
For identity compiling,
\begin{equation}
V=I,
\qquad
\ket{\Omega_{V}}=\ket{\Omega}_{AR}.
\end{equation}
For  \texttt{CNOT} compiling, the target is $V=C_{12}$.

The reference-register preparation and the active trainable block are represented as
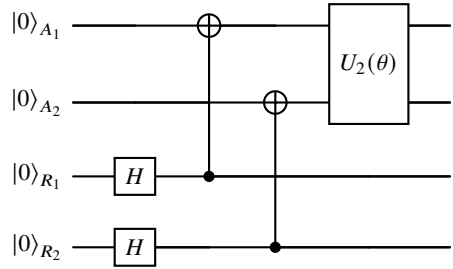
\begin{figure}[ht!]
\centering
\begin{adjustbox}{max width=\linewidth}
\begin{quantikz}[row sep=0.45cm, column sep=0.55cm]
\lstick{$\ket{0}_{A_1}$}
& \qw
& \targ{}
& \qw
& \gate[2]{U_2(\theta)}
& \qw \\
\lstick{$\ket{0}_{A_2}$}
& \qw
& \qw
& \targ{}
&
& \qw \\
\lstick{$\ket{0}_{R_1}$}
& \gate{H}
& \ctrl{-2}
& \qw
& \qw
& \qw \\
\lstick{$\ket{0}_{R_2}$}
& \gate{H}
& \qw
& \ctrl{-2}
& \qw
& \qw
\end{quantikz}
\end{adjustbox}
\caption{
Unitary compiling  benchmark circuit in the implementation ordering $A_1,A_2,R_1,R_2$.
The Bell-pair preparation produces $\ket{\Omega}_{AR}$, after which the trainable two-qubit circuit acts only on the active register $A$. Replacing $U_2(\theta)$ by $V$ gives the target  state.
}
\label{fig:supp-choi-circuit}
\end{figure}
Here $H$ denotes the Hadamard gate,
\begin{equation}
H
=
\frac{1}{\sqrt{2}}
\begin{pmatrix}
1 & 1 \\
1 & -1
\end{pmatrix}.
\end{equation}

\subsection{Random-Pauli-generator identity compiling}
In this benchmark, each circuit generator is an independently sampled two-qubit Pauli string. The term ``random generator'' refers to this sampling of the circuit generators; the quantum input state remains fixed. The random-generator benchmark is a short identity-compiling task, akin to Fig~\ref{fig:supp-choi-circuit}.  In these tasks, the quantum input state is fixed and is not sampled over random states.  In the implementation ordering $A_1,A_2,R_1,R_2$, the active register $A=A_1A_2$ is entangled with the reference register $R=R_1R_2$ in the double-Bell state of Eq.~\eqref{eq:AR_state}, which takes the explicit computational-basis form 
\begin{equation}
\ket{\psi_{\rm in}}
=
\ket{\Omega}_{AR}
=
\frac{1}{2}
\sum_{a=0}^{3}
\ket{a}_{A}\ket{a}_{R}
=
\frac{1}{2}
\left(
\ket{0000}
+
\ket{0101}
+
\ket{1010}
+
\ket{1111}
\right).
\end{equation}
The target operation is the identity on the active register, so
\begin{equation}
\ket{\psi_{\rm tar}} = \ket{\Omega}_{AR}.
\end{equation}
For generator seed $s$, the circuit instance is defined by sampled two-qubit Pauli generators 
\begin{equation}
G_t^{(s)}
=
P_{a_t}^{(s)}\otimes P_{b_t}^{(s)},
\qquad
P_{a_t}^{(s)},P_{b_t}^{(s)}
\in
\{I,X,Y,Z\}.
\end{equation}
The trainable active-register circuit is
\begin{equation}
U_{\rm rg}^{(s)}(\phi)
=
\exp\left(
-\frac{i}{2}\phi_T G_T^{(s)}
\right)
\cdots
\exp\left(
-\frac{i}{2}\phi_2 G_2^{(s)}
\right)
\exp\left(
-\frac{i}{2}\phi_1 G_1^{(s)}
\right),
\end{equation}
with the rightmost factor acting first.  In the reported short-identity benchmark, $T=4$, and the full four-qubit output state is
\begin{equation}
\ket{\psi_{\rm out}^{(s)}(\phi)}
=
\left(
U_{\rm rg}^{(s)}(\phi)\otimes I_R
\right)
\ket{\Omega}_{AR}.
\end{equation}
The fidelity is therefore
\begin{equation}
F_{\rm rgid}^{(s)}(\phi)
=
\left|
\left\langle
\Omega
\right|
\left(
U_{\rm rg}^{(s)}(\phi)\otimes I_R
\right)
\left|
\Omega
\right\rangle
\right|^2
=
\frac{1}{16}
\left|
\operatorname{Tr}
\left[
U_{\rm rg}^{(s)}(\phi)
\right]
\right|^2 .
\end{equation}
The production benchmark varied the generator seed $s$ and the optimization seed for the initial circuit parameters; it did not vary the input quantum state.

\subsection{Initial states}
In all fixed benchmark tasks, the quantum input states and target states are deterministic.  Randomness enters through the initial trainable parameters, which are sampled independently across optimization seeds. For the Bell task,
\begin{equation}
\ket{\psi_{\rm in}}=\ket{00},
\qquad
\ket{\psi_{\rm tar}}=\ket{\Phi^+}
=
\frac{1}{\sqrt{2}}
\left(
\ket{00}+\ket{11}
\right).
\end{equation}
For the GHZ task,
\begin{equation}
\ket{\psi_{\rm in}}=\ket{0000},
\qquad
\ket{\psi_{\rm tar}}=\ket{\mathrm{GHZ}_4}
=
\frac{1}{\sqrt{2}}
\left(
\ket{0000}+\ket{1111}
\right).
\end{equation}
For the identity compiling task,
\begin{equation}
\ket{\psi_{\rm in}}
=
\ket{\psi_{\rm tar}}
=
\ket{\Omega}_{AR},
\qquad
\ket{\Omega}_{AR}
=
\frac{1}{2}
\sum_{x\in\{0,1\}^2}
\ket{x}_{A}\ket{x}_{R}.
\end{equation}
For the  \texttt{CNOT} compiling task, the input remains $\ket{\Omega}_{AR}$, while the target is
\begin{equation}
\ket{\psi_{\rm tar}}
=
\left(
C_{12}\otimes I_R
\right)
\ket{\Omega}_{AR} .
\end{equation}
In the random-generator identity benchmark, every optimization run starts from the same fixed input state; the quantum input state is not sampled over Haar-random states.  In the code, the active two-qubit register $A=A_1A_2$ and reference register $R=R_1R_2$ are ordered as $A_1,A_2,R_1,R_2$, and the initial state is
\begin{equation}
\ket{\psi_{\rm in}}
=
\ket{\Omega}_{AR}
=
\frac{1}{2}
\sum_{a=0}^{3}
\ket{a}_{A}\ket{a}_{R}
=
\frac{1}{2}
\left(
\ket{0000}
+
\ket{0101}
+
\ket{1010}
+
\ket{1111}
\right).
\end{equation}
Equivalently, this is the product of Bell pairs $\ket{\Phi^+}_{A_1R_1}\ket{\Phi^+}_{A_2R_2}$, written in the implementation ordering $A_1,A_2,R_1,R_2$.

\section{Additional sequential-update benchmarks}

\begin{figure}[t]
\centering
\includegraphics[width=.8\columnwidth]{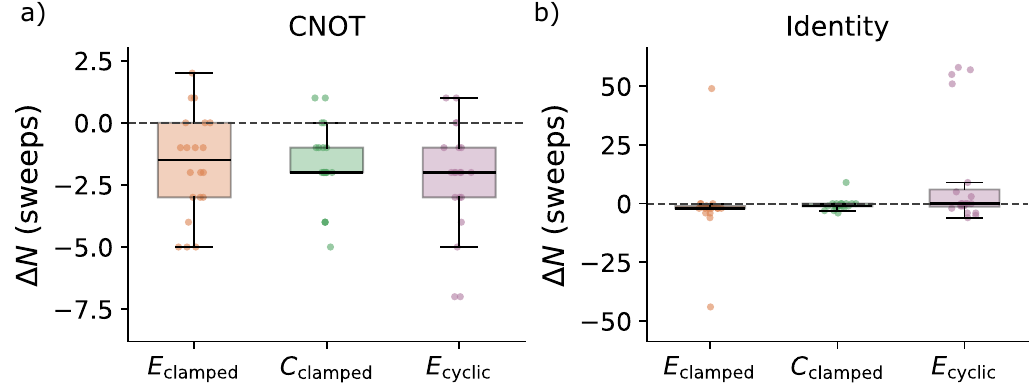}
\caption{Unitary compiling benchmarks for $V=\mathrm{CNOT}_{12}$ (a) and $V=I$ (b).  Paired sweep differences to convergence, $\Delta N$ relative to tuned QAQC.}
\label{fig:SpecialTasks_performance}
\end{figure}

Figure~\ref{fig:SpecialTasks_performance} reports the paired sweep differences for the unitary compiling benchmarks of Fig.~\ref{fig:Fig1_Compiling}(c,d) of the main text.

\begin{figure}[t]
\centering
\includegraphics[width=.8\textwidth]{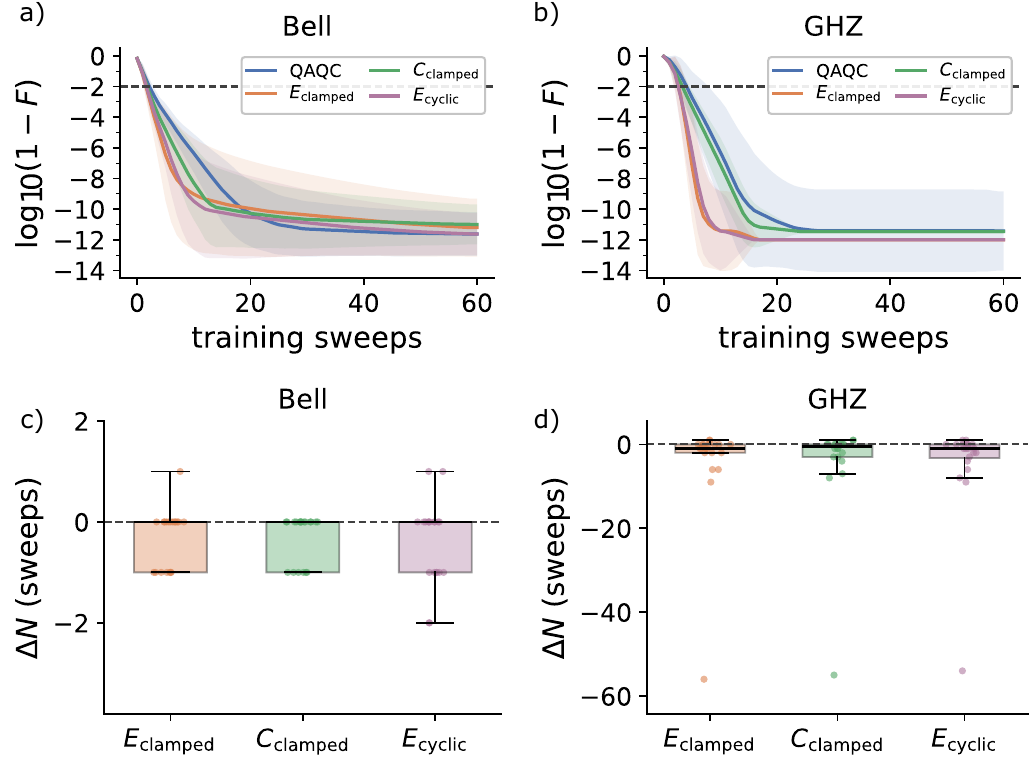}
\caption{
Average training curves and paired sweep differences for the Bell- and GHZ-state-preparation benchmarks using fully sequential coordinate updates. Negative \(\Delta N\) indicates convergence in fewer sweeps than the corresponding tuned QAQC run.}
\label{fig:SpecialTasks_training}
\end{figure}

Figure~\ref{fig:SpecialTasks_training} reports results for the Bell- and GHZ-state preparation, discussed above. 
On the Bell task the equilibrium costs are competitive with tuned QAQC, with median $\Delta N$ within about one sweep for $E_{\rm clamped}$ and $C_{\rm clamped}$ and a modest positive median for $E_{\rm cyclic}$. On the GHZ task all median $\Delta N$ are zero, with $E_{\rm clamped}$ converging fastest in the mean training curves and a small number of initializations gaining tens of sweeps.

 \section{Phase diagrams} 
For the learning-rate and clamp-strength scans, each grid point was evaluated over the calibration initializations.  At each point we recorded the median number of sweeps to threshold and the success probability.  Iteration-time phase diagrams display the median sweep count relative to the best point for the same method and task, while success-basin diagrams display the fraction of calibration runs reaching $F\geq0.99$.  These scans were used to assess robustness and select stable operating points; the final benchmark statistics were computed separately on the held-out final initializations.

\begin{figure*}[t]
\centering
\includegraphics[width=\textwidth]{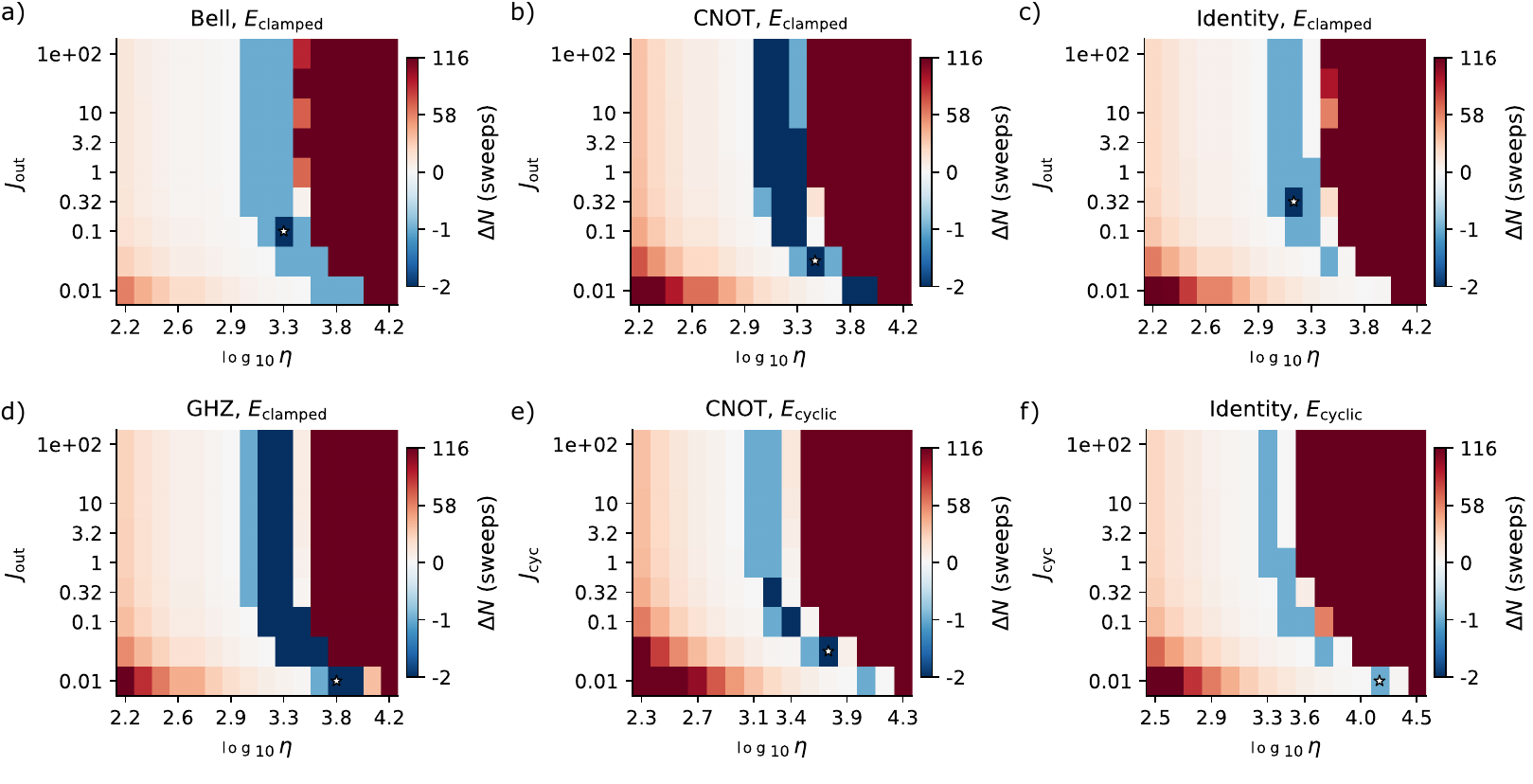}
\caption{Iteration-time phase diagrams for the special compiling tasks, comparing Feynman-Kitaev energy-based training methods with direct-QAQC training over learning-rate and clamp-strength scans.}
\label{fig:phasediagram_iterations}
\end{figure*}

\begin{figure*}[t]
\centering
\includegraphics[width=\textwidth]{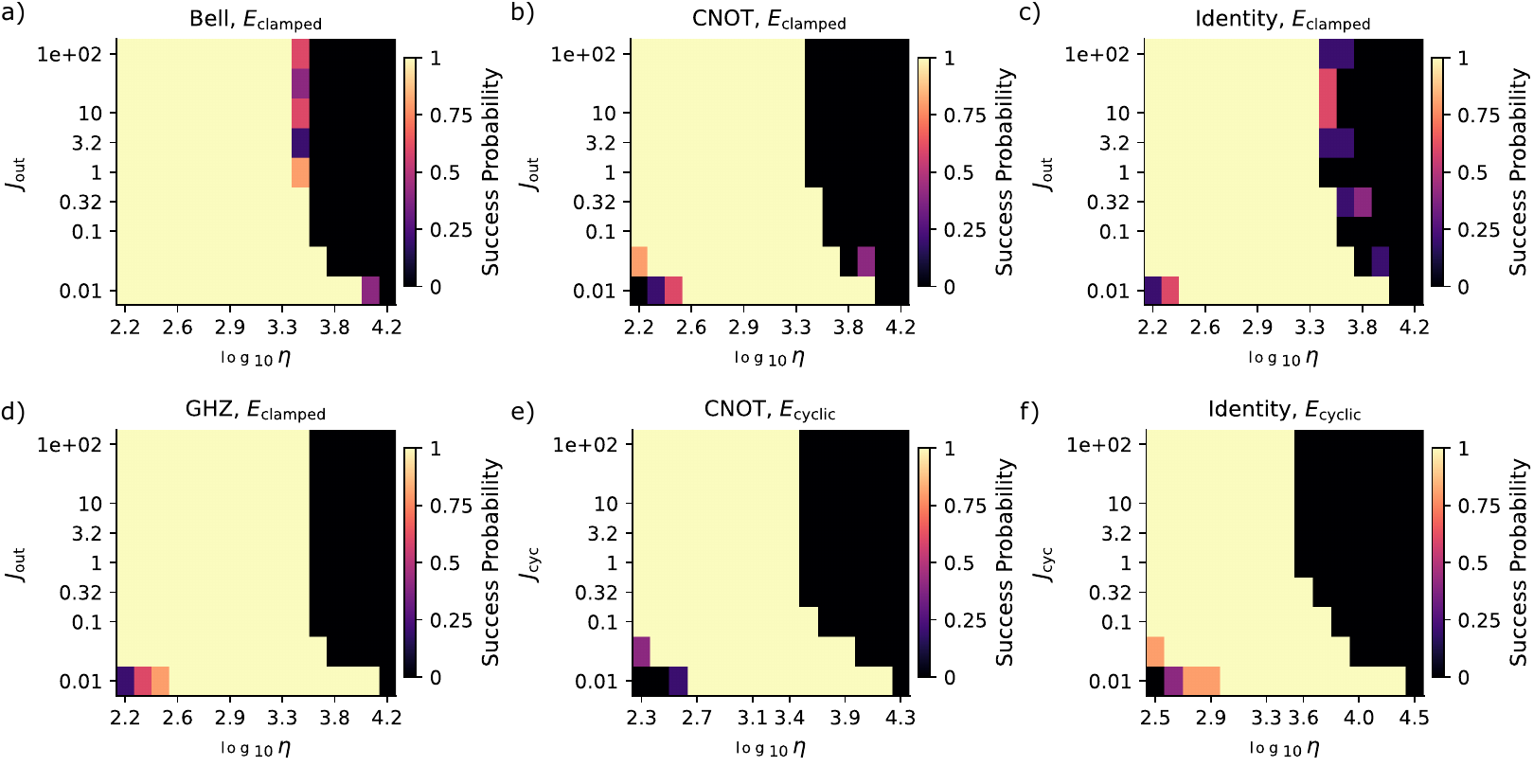}
\caption{Success-basin phase diagrams for the same special compiling tasks, showing the robustness of convergence across learning-rate and clamp-strength scans.}
\label{fig:phasediagram_success}
\end{figure*}

Figures~\ref{fig:phasediagram_iterations} and~\ref{fig:phasediagram_success} separate optimization speed from convergence robustness. At each hyperparameter pair, we evaluate the method on the held-out calibration initializations. Figure~\ref{fig:phasediagram_iterations} reports the median paired sweep difference $\Delta N=N_{\rm FK}-N_{\rm QAQC}$, so negative values identify regions that converge in fewer sweeps than tuned QAQC. Figure~\ref{fig:phasediagram_success} reports the fraction of runs reaching $F\geq0.99$.

In Figure~\ref{fig:phasediagram_success}, the success diagrams exhibit broad connected regions of near-unit success probability, whereas the sweep reductions occupy narrower bands within those regions. Thus the favorable performance is not confined to a single isolated hyperparameter point, although overly aggressive learning rates produce a sharp loss of convergence. 

\section{ Incremental cost function using ground state of $H_{\rm cyclic}$}
We also tested a cyclic incremental cost based on nearest-neighbor clock coherences. Let $|\Psi_{\rm free}\rangle$ denote the  history state for some input state $\ket{\psi_{\rm in}}_{S}$, and let $|\Psi_{\rm cyclic}\rangle\in \text{span}\mathcal{D}$ denote the ground state of the $H_{\rm cyclic}$ in the low-energy dynamical sector.  
With cyclic indexing of the clock labels, we define
\begin{equation}
C_{\rm inc}(\theta) =1-\frac{1}{T+1}\sum_{t=0}^{T} \frac{\big\langle \ket{t-1}\!\bra{t}_{C}\big\rangle_{\ket{\Psi_{\rm cyclic}}} }     {\big\langle \ket{t-1}\!\bra{t}_{C}\big\rangle_{\ket{\Psi_{\rm free}}}}.
\label{eq:cyclic_incremental_ratio}
\end{equation}
Like the overlap cost function $C_{\rm cyclic}$, $C_{\rm inc}$ contrasts an equilibrium with the free history state, but through local clock observables rather than a global state overlap.  However,  in the clamped setting $C_{\rm inc}$ does not depend on the circuit parameters through $\epsilon(\theta)$ alone, so we apply this cost function only in the cyclic setting, and for $T\geq 2$ (it vanishes identically for $T=1$). It is faithful in the dynamical parameter range $0<2J_{\rm cyc}\lesssim J_{\rm in}$ that we identified in Theorem \ref{thm:cyclicparams}. The fact that $\epsilon=0$ implies $C_{\rm in}=0$ follows from the fact that $\lim_{\vartheta_{*}\rightarrow 0 }\ket{\Psi_{\rm cyclic}} = \ket{\Psi_{\rm free}}$. To see the other direction, namely the fact that $C_{\rm in}=0$ implies $\epsilon=0$, first
call the $\ket{\gamma_{t}}$ amplitudes of the state $\ket{\Psi_{\rm cyclic}}$ in Theorem \ref{thm:cyclicgs} by ${a_{t}\over \sqrt{\sum_{t}a_{t}^{2}}}$. It follows that 
\begin{equation}
    C_{\rm in}(\epsilon)=1-{\sum_{t=0}^{T}a_{t-1}a_{t}\over \sum_{t=0}^{T}a_{t}^{2}}.
    \label{eqn:incrementamps}
\end{equation} Since the amplitudes are real, $C_{\rm in}(\epsilon)$ implies that $(a_{0},\ldots, a_{T})$ is proportional to the all-ones vector. Simplifying the equation $a_{t-1}-a_{t}=0$ gives
\begin{equation}
        4\sin\left( {\vartheta^{*}\over 2} \right) \sin\left( {(T+2)\vartheta^{*} \over 2}\right)\sin\left( {(T+1-2t)\vartheta^{*} \over 2}\right) = 0.
        \label{eqn:samevec}
    \end{equation}
But on the interval $\vartheta \in [0,{\pi\over T+2}]$ containing the lowest solution of the secular equation (\ref{eq:secEquationFloquent}), the only way that  (\ref{eqn:samevec}) can be true is if is $\vartheta^{*} = 0$. The expression (\ref{eqn:incrementamps}) reveals why it is necessary to take the quotient in defining $C_{\rm in}$, namely, if the cost function were defined by an additive difference in expectation values of the clock hopping operator, the overlap $\langle \psi_{t-1}\vert \psi_{t}\rangle$ would appear, which depends on $\theta$ not only through $\epsilon(\theta)$.

\begin{figure}[t]
\centering
\includegraphics[width=.8\columnwidth]{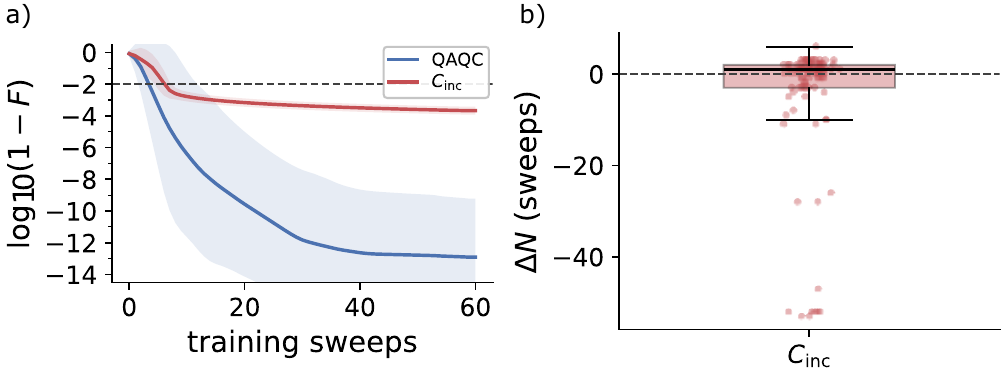}
\caption{Performance and training curves for the random-generator identity-compiling benchmark using the incremental cyclic objective \(C_{\rm inc}\). }
\label{fig:IncrementalCompiling}
\end{figure}

Figure~\ref{fig:IncrementalCompiling} compares training with $C_{\rm inc}$ against QAQC on the random-generator benchmark. 
These runs use $J_{\rm in}=J_{\rm prop}=1$ and $J_{\rm cyc}=0.2$, inside the faithful range identified above. 

 \section{ Two ground states per sweep}
 \label{sec:twoGS}

In a local FK realization, Eq.~\eqref{eq:HF_gradient} expresses each coordinate gradient as the ground-state expectation value of an operator supported on a single clock edge, and these operators have disjoint clock support for nonadjacent edges. We therefore partition the trainable propagation edges into even and odd edges: one prepared equilibrium state supplies gradients for the approximately $p/2$ parameters of one class of edges, e.g., even edges, that class is updated, and a second equilibrium state is prepared for the complementary class of edges, so one full sweep uses two ground state preparations. This reuse applies to the energy costs $E_{\rm clamped}$ and $E_{\rm cyclic}$; for $C_{\rm clamped}$ only the block-update schedule is shared, since its finite-difference gradient still requires the free and clamped ground states at $\theta_j\pm\delta$, four solves per coordinate.

\begin{figure}[t]
\centering
\includegraphics[width=.8\columnwidth]{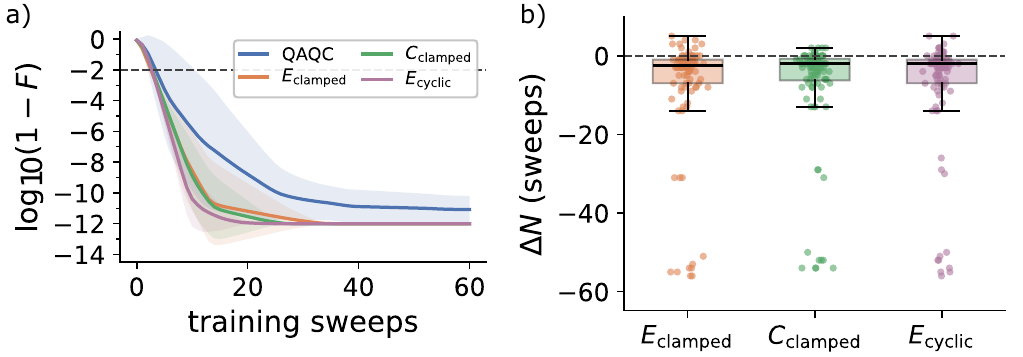}
\caption{Performance and training curves for the random-generator identity-compiling benchmark with two ground state preparations per sweep. One optimized learning rate per random generator. For $C_{\rm clamped}$ only the block-update schedule is shared; its finite-difference gradient uses four ground state solves per coordinate.}
\label{fig:IdentityCompiling-1GS}
\end{figure}

\begin{figure}[t]
\centering
\includegraphics[width=.8\columnwidth]{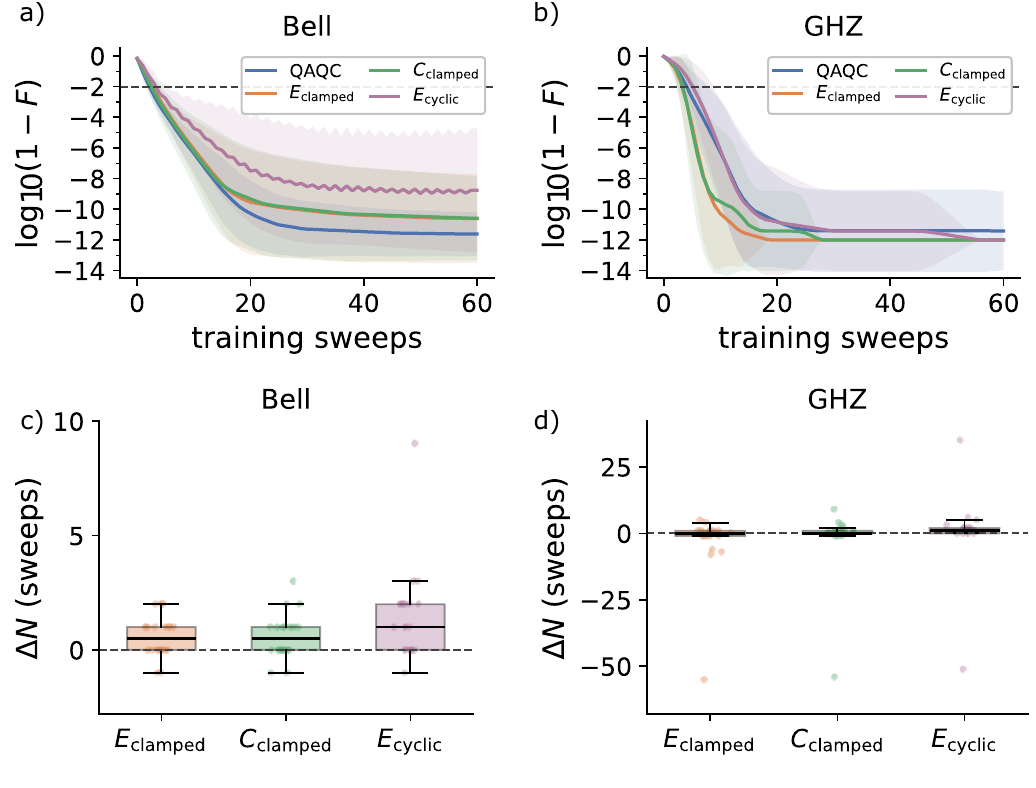}
\caption{Bell- and GHZ-state-preparation benchmarks with two ground state preparations per sweep.  For $C_{\rm clamped}$ only the block-update schedule is shared.}
\label{fig:SpecialTasks_trainig-1GS}
\end{figure}

\begin{figure}[t]
\centering
\includegraphics[width=.8\columnwidth]{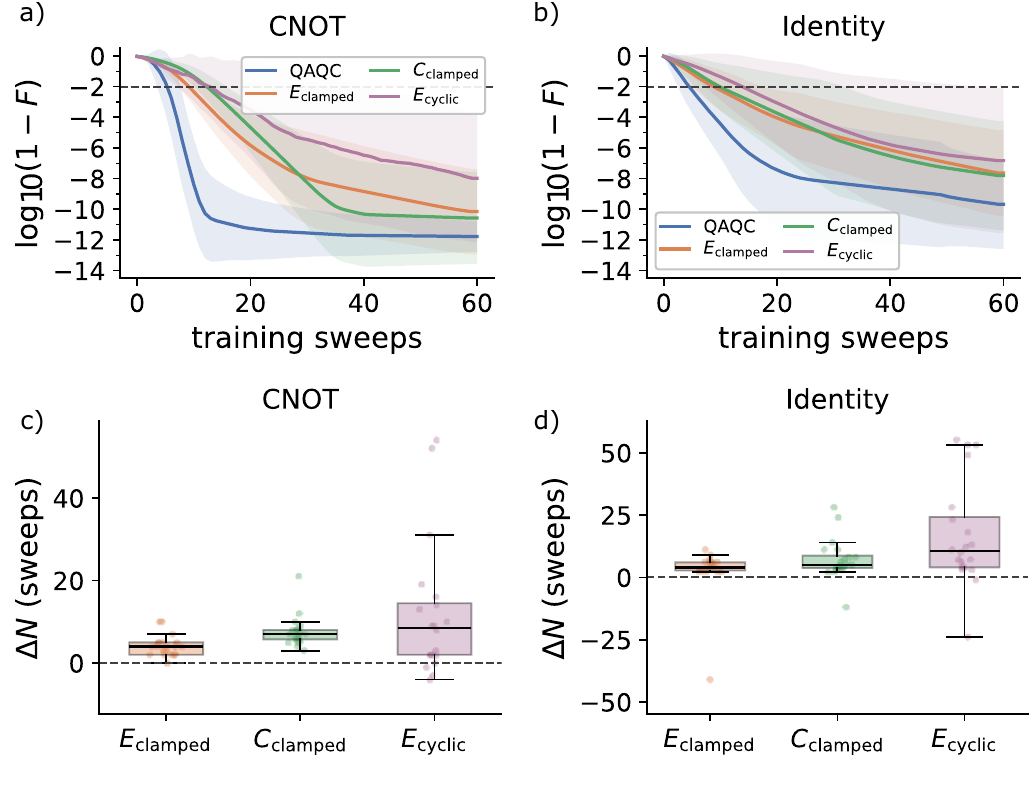}
\caption{Unitary compiling benchmarks ($V=\mathrm{CNOT}_{12}$ and $V=I$) with two ground state preparations per sweep. For $C_{\rm clamped}$ only the block-update schedule is shared.}
\label{fig:SpecialTasks_performance-1GS}
\end{figure}

Figures~\ref{fig:IdentityCompiling-1GS}-\ref{fig:SpecialTasks_performance-1GS} repeat the benchmark suites under this schedule, at the couplings stated in the numerical-methods section:  random-generator identity compiling (Fig.~\ref{fig:IdentityCompiling-1GS}), Bell- and GHZ-state preparation (Fig.~\ref{fig:SpecialTasks_trainig-1GS}), and fixed-ansatz \texttt{CNOT} and identity compiling (Fig.~\ref{fig:SpecialTasks_performance-1GS}). On the random-generator suite, median $\Delta N$ remains at or below zero with the same long tails of large gains ($\Delta N\lesssim-50$) as the fully sequential implementation, and on the state-preparation tasks the medians stay within about one sweep of QAQC. On the fixed-ansatz unitary tasks, the reduced number of preparations trades away sweep efficiency: median $\Delta N$ is positive for all three costs, a few sweeps for $E_{\rm clamped}$ and $C_{\rm clamped}$ and largest for $E_{\rm cyclic}$.

\end{document}